\documentclass[usegraphicx,usenatbib,useAMS]{mn2e}\usepackage{times}
  \let\url\relax
\def\apj{{ApJ}}
\def\apjs{{ApJS}} 
\def\apjl{{ApJL}}

\def\aj{{AJ}}
\def\mnras{{MNRAS}}

\def\prd{{Phys Rev D}}

\def\physrep{{Physics Reports}}
\newcommand{\be}{\begin{equation}}
\newcommand{\ba}{\begin{eqnarray}}
\newcommand{\ee}{\end{equation}}
\newcommand{\ea}{\end{eqnarray}}  

\def\lesssim{\mathrel{\hbox{\rlap{\hbox{\lower4pt\hbox{$\sim$}}}\hbox{$<$}}}}
\def\gtrsim{\mathrel{\hbox{\rlap{\hbox{\lower4pt\hbox{$\sim$}}}\hbox{$>$}}}}

\def\gtsima{$\; \buildrel > \over \sim \;$}
\def\ltsima{$\; \buildrel < \over \sim \;$}
\def\gsim{\lower.5ex\hbox{\gtsima}}
\def\lsim{\lower.5ex\hbox{\ltsima}}
\def\simgt{\lower.5ex\hbox{\gtsima}}
\def\simlt{\lower.5ex\hbox{\ltsima}}
\def\simpr{\lower.5ex\hbox{\prosima}}

\def\simless{\mathbin{\lower 3pt\hbox
   {$\rlap{\raise 5pt\hbox{$\char'074$}}\mathchar''7218$}}}   
\def\simgreat{\mathbin{\lower 3pt\hbox
   {$\rlap{\raise 5pt\hbox{$\char'076$}}\mathchar''7218$}}}   

\begin{document}

\title  [Large-scale reionization] {Simulating cosmic reionization: How 
large a volume is large enough?}
 
\author[I. T. Iliev, et al.]{Ilian~T.~Iliev$^{1}$\thanks{e-mail: 
I.T.Iliev@sussex.ac.uk}, Garrelt Mellema$^{2}$, Kyungjin Ahn$^3$,
Paul R. Shapiro$^{4}$, Yi Mao$^{5}$, \newauthor 
Ue-Li Pen$^{6}$ 
\\
$^1$ Astronomy Centre, Department of Physics \& Astronomy, Pevensey II 
Building, University of Sussex, Falmer, Brighton BN1 9QH, United Kingdom\\    
$^2$ Department of Astronomy \& Oskar Klein Centre, Stockholm University, 
AlbaNova, SE-10691 Stockholm, Sweden\\
$^3$ Department of Earth Sciences, Chosun University, 
Gwangju 501-759, Korea\\
$^{4}$ Department of Astronomy, University of Texas, Austin, 
TX 78712-1083, U.S.A.\\
$^5$ UPMC Univ Paris 06, CNRS, Institut Lagrange de Paris, 
Institut d'Astrophysique de Paris, UMR7095, 98 bis, boulevard 
Arago, F-75014, Paris, France\\
$^6$ Canadian Institute for Theoretical Astrophysics, University
  of Toronto, 60 St. George Street, Toronto, ON M5S 3H8, Canada
}
\date{\today} \pubyear{2013} \volume{000}
\pagerange{1} \twocolumn \maketitle
\label{firstpage}

\begin{abstract}
We present the largest-volume ($425\, h^{-1}\rm Mpc=607\,$Mpc on a side) 
full radiative transfer simulation of cosmic reionization to date. We
show that there is significant additional power in density fluctuations 
at very large scales. Although the halo power spectra are unaffected by this,
there is strong modulation of the halo abundance reaching very large scales.  
We systematically investigate the effects this additional power has on 
the progress, duration and features of reionization, as well as on selected
reionization observables. We find that comoving simulation volume of 
$\sim100\,h^{-1}$~Mpc per side is sufficient for deriving a convergent 
mean reionization history, but that the reionization patchiness is 
significantly underestimated. In the 
large-scale volume the isolated ionized regions reach volumes up to 4-10 
times larger (depending on the measure being used) than they do in a 
smaller, $114\,h^{-1}=163$~Mpc volume with the same source properties and 
reionization history, though the abundance of smaller ionized patches 
agrees well in the two cases. We use jackknife splitting of the large 
simulation volume to quantify the convergence of reionization properties 
with simulation volume for both mean-density and variable-density 
sub-regions. We find that sub-volumes of $\sim100\,h^{-1}$~Mpc per 
side or larger yield convergent reionization histories, except for the 
earliest times (corresponding to $z\gtrsim20-25$ for our parameters), but 
smaller volumes of $\sim50\,h^{-1}$~Mpc or less are not well converged at 
any redshift. Reionization history milestones, defined here as the redshifts 
at which the ionized fraction by mass reaches 10\%, 50\%, 90\% and 99\%, 
show significant scatter between the sub-volumes, of $\Delta z=0.6-1$ for 
$\sim50\,h^{-1}$Mpc volumes, decreasing to $\Delta z=0.3-0.5$ for 
$\sim\,100h^{-1}$Mpc volumes, and $\Delta z\sim0.1$ for $\sim200\,h^{-1}$Mpc 
volumes. If we only consider mean-density sub-regions the scatter decreases, 
but remains at $\Delta z\sim0.1-0.2$ for the different size sub-volumes. 
Consequently, many potential reionization observables like 21-cm rms, 
21-cm PDF skewness and kurtosis all show good convergence for volumes of 
$\sim200\,h^{-1}$Mpc, but retain considerable scatter for smaller volumes. 
In contrast, the three-dimensional 21-cm power spectra at large scales 
($k\lesssim0.25\,h\,\rm Mpc^{-1}$) do not fully converge for any sub-volume 
size. These additional large-scale fluctuations significantly 
enhance the 21-cm fluctuations. At the rough beam- and bandwidth resolution 
expected for the LOFAR EoR experiment (3' and 440 kHz) and for our simulation 
parameters, the peak value of the rms 21-cm brightness temperature
fluctuations as a function of frequency, derived from the large volume is 
$\sim10\%$ higher than for a $\sim100\,h^{-1}$Mpc volume. At late
times (high frequency), close to the overlap epoch,  the 
signal derived from the large volume is up to $2.5$ times larger, which 
should improve the prospects of detection considerably, given the lower 
foregrounds and greater interferometer sensitivity at higher frequencies. 
\end{abstract}


\section{Introduction}

The first billion years of cosmic evolution remain the only period in the 
history of the universe still largely unconstrained by direct observations.
While we now have fairly detailed data on the Cosmic Microwave Background
originating from the last scattering surface at redshift $z\sim1100$ and 
a wealth
of multiwavelength observations at later times, $z<6$, the intermediate 
period remains largely uncharted. A number of ongoing observational programs 
aim to provide observations of this epoch in e.g. high-$z$ Ly-$\alpha$ 
\citep[e.g.][]{2012ApJ...745..122K}, CMB secondary anisotropies 
\citep[e.g.][]{2012ApJ...756...65Z} and redshifted 21-cm 
\citep[e.g.][]{2010MNRAS.405.2492H,2009IEEEP..97.1497L,2010AJ....139.1468P}. 
Improved observational constraints could provide a wealth of information 
about the nature of the first stars and galaxies, their properties, abundances 
and clustering, the timing and duration of the reionization transition and the 
complex physics driving in this process.

However, most observations will yield only statistical measures, e.g. power 
spectra and auto- and cross-correlations, whose interpretation requires 
detailed simulations and modelling in order to extract the quantities of 
interest. In recent years there has been significant improvement in terms 
of the dynamic range of such simulations and their sophistication in 
implementing the variety of relevant physical processes \citep[see e.g.][for 
a recent review]{2011ASL.....4..228T}. 

One question that has not received much attention is what simulation volume 
is required for obtaining reliable results on the different aspects of cosmic 
reionization - for example the mean reionization history, the scales and 
distribution of the patchiness, and the various corresponding observational 
features. Due to limited dynamic range and the necessity to resolve the 
low-mass galaxies driving reionization, the earliest numerical simulations 
of this period inevitably followed relatively small volumes, typically just 
a few comoving Mpc per side \citep{1997ApJ...486..581G,2000MNRAS.314..611C,
2000ApJ...535..530G,2002ApJ...575...33R}. Such small-volume simulations 
yield a rather late, sharp, and fast reionization occurring over a short 
interval in redshift. Subsequent simulations managed to increase the 
studied volumes to $10-20\, {h}^{-1}{\rm Mpc}$ per side \citep{2003MNRAS.344L...7C,
2003MNRAS.344..607S}. This resulted in more extended reionization histories, 
in better agreement with the then current constraints on the CMB optical 
depth for electron scattering \citep[for a review see][]{2003ApJS..148..175S}.

\citet{2004ApJ...609..474B} used theoretical calculations based on analytical
halo mass functions and the extended Press-Schechter conditional mass function
to show that local abundances of galaxies vary significantly on up to tens of
Mpc scales. Consequently, periodic cosmological volumes of sizes less than this
significantly underestimate the abundance of rare haloes (which at high 
redshifts are most or all of the star-forming haloes), with corresponding 
effects on the reionization
observables. In particular, the scatter in the local reionization is greatly
reduced in the latter case, resulting in quicker reionization than larger
volumes would predict. The global mean halo collapsed fraction was shown to 
be largely converged for $\sim100\,$Mpc volumes. However, the question what 
simulation volume is required for a particular reionization feature to be 
derived reliably was left open, to be tested and quantified by future 
simulations and observations.  

The first cosmological reionization simulations at large scales (specifically,
$100\,h^{-1}\rm Mpc=143\,\rm $Mpc per side) were presented in 
\citet{2006MNRAS.369.1625I}. By 
sub-dividing this large volume into sub-volumes of different sizes and 
calculating their reionization history, they demonstrated that a mean-density 
volume of a few tens of Mpc per side is sufficient to derive the mean 
reionization history with a reasonable accuracy. However, the same work also
showed that there is considerable scatter in the reionization histories for 
all (i.e. not just mean-density) volumes. This scatter becomes very large 
for $5-10\,h^{-1}$Mpc volumes, to the point of making even the mean 
reionization history unreliable. The convergence of other properties with 
volume, e.g. the size distributions of the ionized regions, remained an open 
question since individual ionized regions can reach tens of Mpc across, 
similar to the size of the sub-divided volumes, and thus even such 
$\sim 100\,h^{-1}$~Mpc sized volumes are insufficient for such a study. 
Furthermore, the answer to this question likely depends on the property 
being studied. In the currently-favoured $\Lambda$CDM universe density 
fluctuations occur on all scales, albeit with diminishing amplitude for 
larger sizes, so any result should be quantified based on the level of 
convergence required.

Previously, reionization at large scales has only been approached by 
semi-numerical models \citep{2007ApJ...657...15K,2010MNRAS.406.2421S}. 
\citet{2007ApJ...657...15K} used small-scale reionization simulations 
as sub-grid prescription in a very large, 1~Gpc, volume to study the 
regions around luminous QSOs at the end of reionization. More recently, 
the results in \citet{2010MNRAS.406.2421S,2011MNRAS.411..955M} indicated 
that the additional large-scale density power does result in larger 
ionized patches and increased 21-cm fluctuations. The semi-numerical 
approach, whereby the 
reionization transition is modelled based on analytical prescriptions 
for the local halo abundances and clustering combined with smoothed 
density fields generated as initial conditions for N-body simulations, 
has the advantage that it is computationally much cheaper than full 
simulations. This allows for variation of the unknown parameters and 
thus testing a variety of models. However, the available semi-numerical 
and semi-analytical models generally do not include many important 
reionization features, such as the suppression of low-mass sources due to 
Jeans-mass filtering, recombinations, or nonlinear halo clustering, which 
limits their applicability. Furthermore, to date the semi-numerical models 
have only been tested against simulations at smaller scales, below 
$\sim100$~Mpc, for lack of larger-scale simulations. Such larger-scale 
simulations would therefore be also useful for verification and further 
development of semi-numerical modelling.  

In addition to being able to probe very large scales, simulations of 
at least a few hundred comoving Mpc per side also have the advantage 
of matching better wide-field surveys of high-redshift structures such as 
the one currently conducted by LOFAR, with field-of-view (FOV) of a few 
tens of square degrees. The first detections are likely to be statistical, 
measuring the redshifted 21-cm power spectra and variance
\citep{2009MNRAS.393.1449H,2012MNRAS.423.2222I}, although a limited form 
of imaging might be possible in some cases \citep{2012MNRAS.424..762D,
2012MNRAS.425.2964Z}. Another important advantage of large simulation 
volumes is that they provide a wider range of environments (e.g. larger, 
deeper voids, higher density peaks, large-scale bulk motions) and much 
richer statistics of objects, particularly very rare ones which might 
not be found at all in a smaller volume \citep[e.g.][]{2013arXiv1305.1976W}, 
for example luminous high-redshift QSOs as found by e.g. SDSS 
\citep{2001AJ....122.2833F}. In fact the simulation we present here
has already been used to study the impact and detectability of a rare
QSO \citep{2012MNRAS.424..762D}.

In this work we present the first direct, radiative transfer (``RT'')
simulation of reionization at very large scales, 
in a comoving volume of $425\,h^{-1}{\rm Mpc}=607\,$Mpc per side, 
and investigate the effects of the previously ignored long wavelength 
fluctuations on the formation of early structures and on the progress 
and features of cosmic reionization. The structure of this paper is as
follows: in \S~\ref{sect:sims} we present our simulations; in 
\S~\ref{sect:early_struct} we discuss the formation of early structures;
the effects on the reionization patchiness is presented in 
\S~\ref{sect:large_scale_reion}; the convergence with volume of the 
reionization history, its milestones and some of the resulting observables is 
discussed in \S~\ref{sect:conv_w_volume}; finally, our summary and 
conclusions are in \S~\ref{sect:summary}. 


Results of the detailed, large-volume RT + N-body simulation of reionization
described here have already been applied elsewhere to predict some
additional observational probes of reionization, for which the large
box size had distinct advantages.  \citet{2013ApJ...769...93P} used
these simulations, along with those in boxes of smaller size,
$114\,h^{-1}{\rm Mpc}=163\,$Mpc on a side, to predict the contribution
of patchy reionization to the secondary cosmic
microwave background (``CMB'') temperature fluctuations from the
kinetic Sunyaev-Zel'dovich (``kSZ'') effect. 
We demonstrated there that a correction
based upon linear perturbation theory for the growth of the underlying
matter density and velocity fluctuations is necessary in order to account for
the missing power on large scales when predicting the effect of patchy
reionization on the kSZ using the smaller-box simulations, while
finding that this correction is small for the larger box size. 
\citet{2013PhRvL.110o1301S} and \citet{Jensen2013} applied these big-box
simulation results to model the anisotropy of the 3D power spectrum of 
21-cm brightness
temperature fluctuations from the epoch of reionization caused by the
redshift-space distortion introduced by the Doppler shift from
velocity fluctuations along the line of sight, to determine if
measurements could be used to separate out the effects of cosmology
from astrophysics. 
A large box was required to be able to 
make predictions at the small wavenumbers that will be
probed by upcoming 21-cm radio surveys, with enough modes in the box 
to make the sampling variance small enough.  
Finally, \citet{2012MNRAS.424..762D} used this large-box reionization
simulation to predict the signature of luminous QSOs in the 21-cm
background maps, to show that a matched filter technique,  could detect the
ionized patches surrounding the QSOs as a gap in the signal.  
This required a box large enough to find a rare galactic halo massive
enough to host a bright QSO, while also much larger than the size of
the H II region that might surround such a source.

Here we will describe the underlying, large-box RT + N-body
simulation, itself, in some detail, and address a range of measures of
the effect of box size on the results for patchy reionization and its
observable consequences, with the latter focused on statistical 
predictions of the 21-cm background from the epoch of reionization.  

The simulations presented in this work use the following set of 
cosmological parameters $\Omega_\Lambda=0.73, \Omega_{\rm M}=0.27, 
\Omega_{\rm b}=0.044, h=0.7, \sigma_8=0.8, n_\mathrm{s}=0.96$ where 
$H_0 = 100h$\,km\,s$^{-1}$\,Mpc$^{-1}$, consistent with the {\it WMAP} 
5-year data \citep{2009ApJS..180..330K} and the recent Planck results 
\citep{2013arXiv1303.5076P}.

\section{Simulations}
\label{sect:sims}
We simulate the formation of high-redshift cosmic structures using 
a set of high-resolution N-body simulations, summarised in 
Table~\ref{summary_sim_table}. Our main simulation has $5488^3$ 
(165 billion) particles in a volume of $425\,\rm h^{-1}$Mpc per side. 
For comparison, we also consider two smaller-volume simulations with 
a slightly higher resolution, each following $3072^3$ particles in a 
volume of $114\,\rm h^{-1}$Mpc per side and each with a different 
random realisation of the initial Gaussian-random field to control 
for cosmic variance. The simulations were performed with the P$^3$M 
N-body code CubeP$^3$M \citep{2012arXiv1208.5098H} on the (now retired)
{\it Ranger} computer at the Texas Advanced Computing Center (TACC; 
SunBlade x6420 with AMD x86 64 Opteron Quad Core, 2.3 GHz, 9.2 
GFlops/core “Barcelona” processors and Infiniband networking). The 
smaller-volume, $114~h^{-1}$Mpc simulations were previously presented 
in \citet{2012MNRAS.423.2222I}, while the large-volume simulation was 
part of the suite of simulations used in \citet{will2013} for deriving 
the halo mass function and its evolution with redshift. The large-volume, 
$425\,h^{-1}$Mpc-box, N-body simulation required 2 million core-hours to 
reach redshift $z=6$ and was ran using 2,744 MPI processes. 
Each of MPI node initially contained 4 OpenMP-threaded computing cores and 
8 GB shared memory (for a total of 10,976 cores). During the later stages 
of the simulation, strong clustering meant that the size of an MPI node 
had to be increased to 8 cores and 16~GB memory (for a total of 21,956 
cores) to accommodate the larger memory requirement. The CubeP$^3$M code 
shows an excellent (weak) scaling up to 4,096 cores and very good scaling 
up to 22,000 computing cores. For full details on the CubeP$^3$M code and 
its performance see \citet{2012arXiv1208.5098H}.

\begin{table*}
\caption{Simulation parameters. Background cosmology is based on 
the WMAP 5-year results. 
}
\label{summary_sim_table}
\begin{center}
\begin{tabular}{@{}llllllllll}
\hline
simulation & boxsize & $N_{part}$  & mesh  & force softening & $m_{particle}$ & $M_{halo,min}$ & RT grid & $g_{\gamma,HMACH}$ & $g_{\gamma,LMACH}$
\\\hline
XL2 & 425 $\,h^{-1}$Mpc & $5,488^3$ & $10,976^3$ & 3.88$\,h^{-1}$kpc & $4.97\times10^7\,M_\odot$ & $9.93\times10^{8}\,M_\odot$ & $504^3$& 1.7 & 7.1
\\[2mm]
L2.1 & 114 $\,h^{-1}$Mpc & $3,072^3$ & $6,144^3$ & 1.86$\,h^{-1}$kpc & $5.47\times10^6\,M_\odot$ & $1.09\times10^8\,M_\odot$ & $256^3$& 1.7 & 7.1
\\[2mm]
L2.2 & 114 $\,h^{-1}$Mpc & $3,072^3$ & $6,144^3$ & 1.86$\,h^{-1}$kpc & $5.47\times10^6\,M_\odot$ & $1.09\times10^8\,M_\odot$ & $256^3$& 1.7 & 7.1
\\[2mm]
\hline
\end{tabular}
\end{center}
\end{table*}

All simulations were started at redshift $z=300$ and with initial conditions
generated using the Zel'dovich approximation and power spectrum of the linear
fluctuations given by the CAMB code \citep{Lewis:1999bs}. This starting 
redshift is sufficiently early to guarantee linearity of the density field 
even at such high resolution and thus avoiding numerical artifacts 
\citep{2006MNRAS.373..369C}, see also \citet{will2013} for discussion of this
point regarding the current simulations. The smaller-volume simulations were 
ran until redshift $z=6$, and the larger-volume to $z=2.6$. We have identified 
collapsed haloes at 76 redshifts equally-spaced in time (every 11.5 Myr)
between $z=30$ and $z=6$ using an on-the-fly halo finder based on the 
spherical overdensity method with overdensity of $\Delta=178$ with respect to
the mean density. We also calculated the density and velocity distributions on 
several regular grids (from $252^3$ up to $2016^3$ for the $425\,h^{-1}$Mpc 
box and $256^3$ and $512^3$ for the $114\,h^{-1}$Mpc boxes) using an SPH-like 
smoothing over the particle distributions, which produces smoother 
distributions than other commonly-used interpolation schemes (e.g. CIC, 
NGP) and minimises noise in low-density regions. These gridded data were 
then used as an input to the radiative transfer simulations of reionization 
discussed below, as well as for subsequent analysis of the results from 
those simulations. 

\begin{figure}
    \includegraphics[width=3.4in]{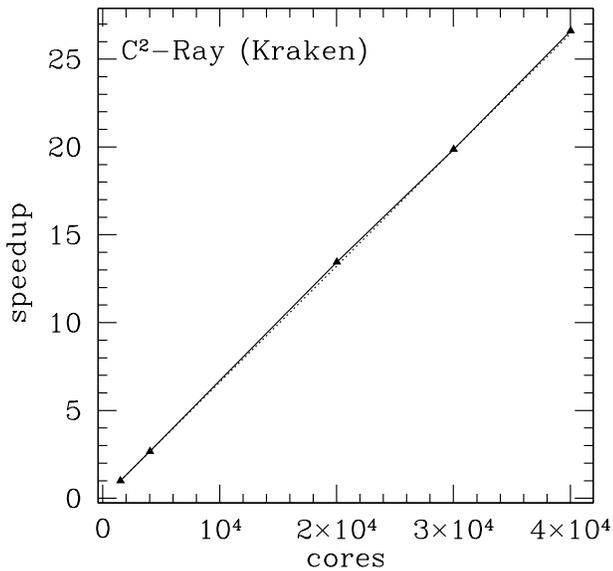}
  \caption{Computational speedup (normalised to the smallest run) vs. number of 
    computing cores (solid line) and the ideal linear weak scaling (dotted line)
    for the C$^2$-Ray code on Kraken (Cray XT5) at University of Tennessee. 
    \label{scaling}}
\end{figure}

The halo catalogues are used to construct the sources of ionizing radiation,
largely using the approach introduced in \citet{2006MNRAS.369.1625I} and 
\citet{2007MNRAS.376..534I}, with some important modifications. The 
smaller-volume, higher-resolution simulations resolve all haloes with mass
of $10^8\,M_\odot$ or more, i.e. roughly all atomically-cooling haloes (ACHs), 
which we split into low-mass atomically-cooling haloes (LMACHs), which are
affected by radiative feedback and high-mass atomically-cooling haloes 
(HMACHs) which are not. On the other hand, the larger-volume simulation only 
resolves haloes of mass $10^9\,M_\odot$ and above, corresponding to the HMACHs 
only. Consequently, for the reionization simulation we model the LMACHs as a 
sub-grid population using a recently-developed extended Press-Schechter type 
model (Ahn et al., in prep.). We do this by calculating a 
cell-by-cell local LMACH population based on a fit to the data from 
higher-resolution simulations. This yields the local LMACH collapsed 
fraction as function of the cell density. Our procedure reproduces the 
mean halo mass function for the whole box and reflects the halo clustering 
down to the grid scale, but at present it does not account for the (modest) 
scatter in the local halo number - overdensity relation observed numerically. 
We then combine the local LMACH collapsed fraction yielded by our model with 
the directly-resolved HMACH population binned onto the same grid cells to 
create the complete ionizing source lists. We do not include minihaloes (MH; 
haloes with masses $10^5\lesssim M/M_\odot \lesssim 10^8$), even though they are 
expected to play an important role in the earliest stage of reionization, as 
previously shown in \citet{2012ApJ...756L..16A}. The reason for this is that 
star formation in MHs is more uniformly distributed on large scales and is 
strongly regulated by the long-range Lyman-Werner feedback, which further
smooths out their distribution and thus we expect their effect to be less 
correlated with large-scale modes in our $425\,h^{-1}$~Mpc volume. We defer
the study of these effects to future work.

Another new element in the current radiative transfer simulations is that
we also include an approximate treatment of the absorber systems which 
limit the mean free path of the ionizing photons during the late stages 
of reionization \citep{2010ApJ...721.1448S}. Numerically this is done by 
limiting the photon propagation to a box of 80~$h^{-1}$Mpc comoving centered 
on each source. This size roughly corresponds to the observed mean free path 
due to Lyman Limit Systems at $z\sim6$. Any photons reaching that boundary
are presumed lost, absorbed by the Lyman-limit systems.  

Once the source lists are constructed, we assign to each an ionizing photon 
production rate per unit time, $\dot{N}_\gamma$, proportional to the total 
mass in haloes within that cell, $M$ as previously discussed in 
\citet{2006MNRAS.369.1625I,2007MNRAS.376..534I,2012MNRAS.423.2222I}:
\be
\dot{N}_\gamma=\frac{g_\gamma M\Omega_b}{\Omega_0 m_p}\left(\frac{10 \;\mathrm{Myr}}{\Delta t}\right)\,, 
\ee
where $m_p$ is the proton mass and $f_\gamma = f_{\rm esc} f_\star N_\star$ is an 
ionizing photon production efficiency parameter which includes the efficiency 
of converting gas into stars, $f_*$, the ionizing photon escape fraction from 
the halo into the IGM, $f_{\rm esc}$ and the number of ionizing photons produced 
per stellar atom, $N_\star$. The latter parameter depends on the assumed IMF 
for the stellar population and could vary significantly, between $\sim4,000$
(Pop II, Salpeter IMF) and $\sim 100,000$ (Pop III, top-heavy IMF). 
HMACH and LMACH haloes are assigned different luminosities, based on the 
reasonable assumption that the latter would typically be less metal-enriched 
and thus their stellar content should be more dominated by massive stars, 
which are efficient producers of ionizing photons. ${\Delta t}=11.46\,$~Myr 
is the time between two snapshots from the N-body simulation. For all 
radiative transfer simulations in this paper we adopt $g_{\rm \gamma,LMACHs}=7.1$ 
and $g_{\rm \gamma,HMACHs}=1.7$, which values yield a reionization history 
consistent with the current observational constraints 
\citep{2012MNRAS.423.2222I}, with electron-scattering optical depth 
$\tau=0.0566$ and overlap reached at redshift $z=6.5$. LMACHs are assumed to 
be suppressed within ionized regions (for ionization fraction higher than 
10\%), through Jeans-mass filtering, as discussed in \citet{2007MNRAS.376..534I}.

\begin{figure*}
  \begin{center}
    \includegraphics[width=3.3in]{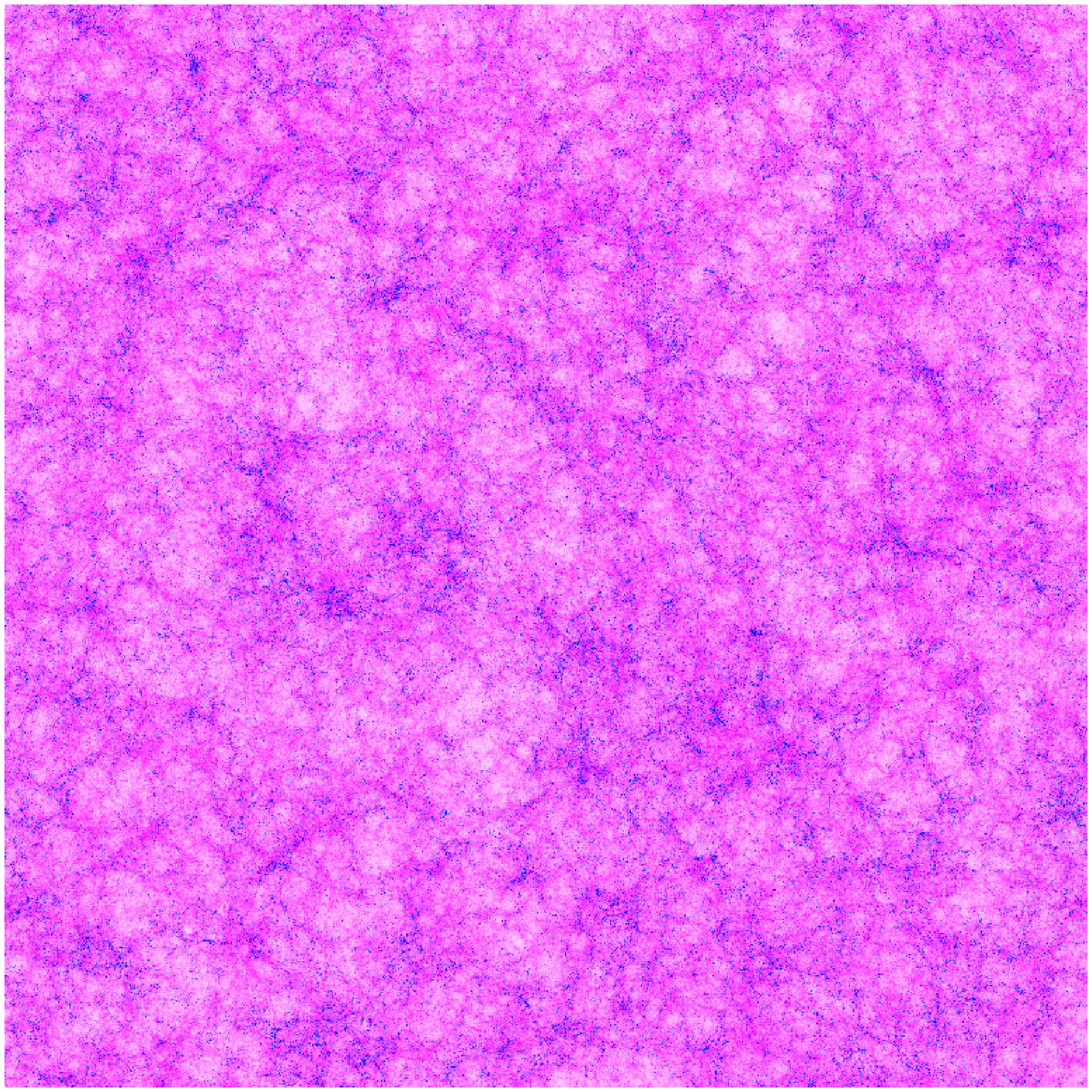}  
    \includegraphics[width=3.3in]{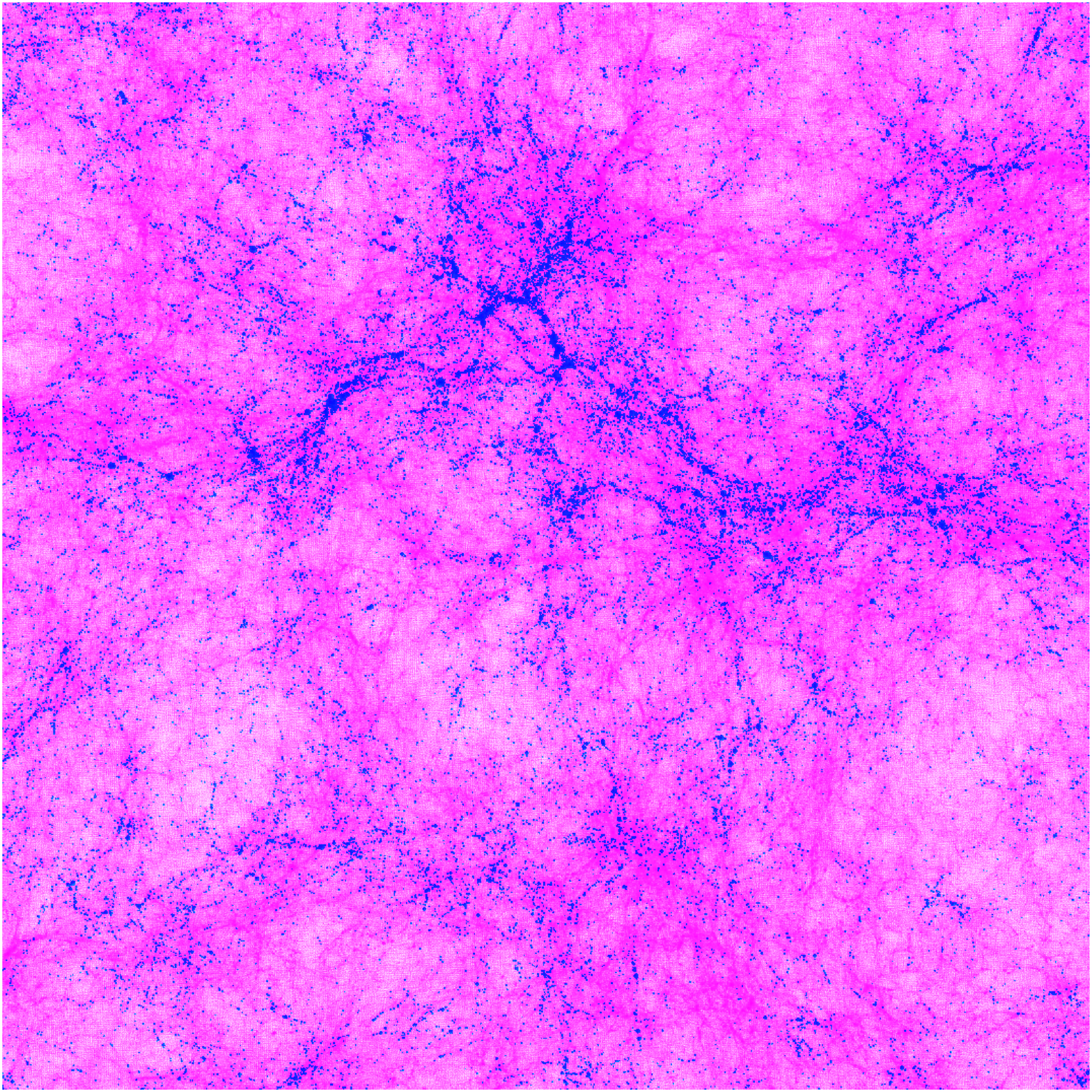}
  \end{center}
  \caption{Large-scale structures and cosmic web at redshift $z=6$ from 
our CubeP$^3$M simulation with $5,488^3$ particles (165 billion) on a 
$10,976^3$ fine grid in a comoving volume of $425/h=607$~Mpc on a side: 
(left) full simulation volume, and (right) zoomed-in $41.5\times41.5$~Mpc 
region of the same image. Shown are the density field (violet) and haloes 
(blue). Slices are $15\,h^{-1}$Mpc thick.  
    \label{image:fig}}
\end{figure*}

Due to the nature of its algorithm, our C$^2$-Ray radiative transfer code
is quite flexible and can run on serial, shared-memory (using OpenMP) or 
distributed parallel systems (using MPI), as well as on hybrid systems 
(OpenMP+MPI), with the latter being the most common mode of usage. The code
can run on any number of computational cores, as long as there is sufficient
memory per processor/MPI process. The $425\,h^{-1}$Mpc-box radiative transfer 
simulation presented here was ran on several different clusters, utilising 
our available allocations. The initial phases were ran on the Apollo cluster 
at The University of Sussex (168 cores at the time) and the Sciama SEPNet 
computer (1,008 cores) at University of Portsmouth. The later stages of the
evolution, when many more sources needed to be handled were ran on larger
clusters, the TACC {\it Ranger} and {\it Lonestar} (22,656 cores, 44.3 TB RAM, 
302 TFlops peak) clusters, the {\it Kraken} Cray XT-5 (112,896 compute cores, 
147 TB RAM, 1.17 PetaFlops) at the National Institute for Computational 
Sciences/University of Tennessee (NICS) and the SciNet computer at University 
of Toronto (30,240 cores, 60 TB RAM, 306 TFlops peak). The simulation was 
performed over 10 months and used up a total of approximately 8 million 
Lonestar-equivalent (3.3 GHz, 13.3 GFlops) core-hours. At different stages 
of the computation we used between 32 and 40,032 cores simultaneously. For 
large enough problems and on machines with fast communication network like 
the NICS Kraken the code scaling is essentially perfect as far as we have 
tested it, here on up to 40,032 computing cores (see Figure~\ref{scaling}). 

\begin{figure}
  \begin{center}
    \includegraphics[width=3.2in]{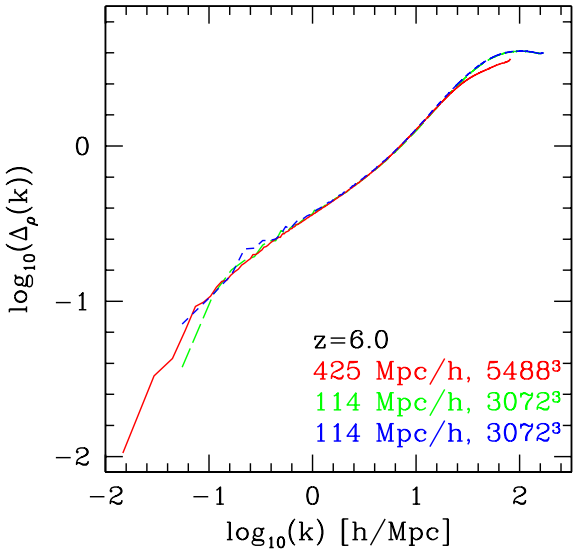}
  \end{center}
  \caption{Power spectrum of the density field at redshift $z=6$ for 
the $425\,h^{-1}$Mpc volume (solid, red) and two realisations of 
$114\,h^{-1}$Mpc volume (dashed, green and blue). 
    \label{ps:fig}}
\end{figure}

\begin{figure}
  \begin{center}
    \includegraphics[width=3.2in]{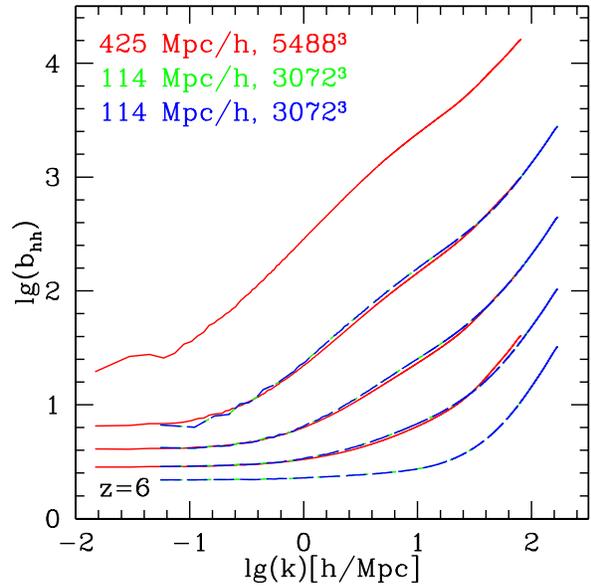}
  \end{center}
  \caption{The halo bias, $b_{\rm hh}=\Delta_{\rm hh}/\Delta_\rho$, at 
redshift $z=6$ for the $425\,h^{-1}$Mpc volume (solid, red) and two 
realisations of $114\,h^{-1}$Mpc volume (dashed, green and blue). 
Lines are for haloes binned by decades of mass (bottom to top curve) 
$10^8M_\odot\leq M_{\rm halo}<10^9M_\odot$ (LMACHs), 
$10^9M_\odot\leq M_{\rm halo}<10^{10}M_\odot$, 
$10^{10}M_\odot\leq M_{\rm halo}<10^{11}M_\odot$,
$10^{11}M_\odot\leq M_{\rm halo}<10^{12}M_\odot$,
and $10^{12}M_\odot\leq M_{\rm halo}$ (HMACHs all). 
    \label{bias:fig}}
\end{figure}

\section{Results}

\subsection{Early cosmic structure formation}
\label{sect:early_struct}
In Figure~\ref{image:fig} we show an illustrative image of the cosmic web 
at redshift $z=6$ extracted from our $425\,h^{-1}$Mpc volume. It shows a 
$15~h^{-1}$Mpc thick slice of the density distribution with the haloes (in 
their actual sizes) overlayed. It is clear that while the universe is
fairly homogeneous at such large scales, we can clearly observe that the 
long-wavelength density modes result in significant density fluctuations 
even at scales as large as tens to 100 Mpc. As we will see below, these 
large-scale variations have important implications for the reionization 
patchiness on large scales. The N-body haloes are strongly clustered around 
the high-density peaks, again at scales of up to tens of Mpc, and along the 
filaments and there are very large, with sizes of up to 50~Mpc or more, 
underdense regions mostly devoid of haloes. 

\subsubsection{Density power spectra and halo bias}

In Figure~\ref{ps:fig} we show the dimensionless power spectra of the 
density field, $\Delta_{\rho}=(k^3P(k)/2\pi^2)^{1/2}$, for all three 
simulations. The density fields were obtained by interpolating the N-body 
particles on a regular grid ($10,976^3$ for the $425\,h^{-1}$Mpc volume and 
$6,144^3$ for the $114\,h^{-1}$Mpc volumes) using cloud-in-cell (CIC) 
interpolation. All three power spectra agree closely over the wavenumber
range which they all resolve well, $-0.5<{\rm log_{10}}(k {\rm h/Mpc})<1.4$. At 
smaller scales (higher $k$) there is less power in the large box due to its 
slightly lower resolution, while large scales, near the respective box sizes
are affected by cosmic variance. Apart from these expected differences, in the
425$\,h^{-1}$Mpc box there is clearly significant additional power at large 
scales ($k\lesssim0.06\,h/$Mpc, corresponding to scales of 
$\gtrsim100\,h^{-1}$Mpc), which fluctuations are completely missing in the 
smaller volumes. 

The halo bias derived from our simulations, defined as 
$b_{\rm hh}=\Delta_{\rm hh}/\Delta_\rho$, where $\Delta_{hh}$ is the halo-halo 
auto-correlation power spectrum calculated on the same grid as the matter 
power spectrum $\Delta_{\rho}$, again using CIC interpolation, is shown in 
Figure~\ref{bias:fig}. We plot the halo bias in five mass bins, one per 
each decade in mass, starting from $10^8M_\odot$. Due to resolution the 
lowest mass bin, $10^8M_\odot\leq M_{\rm halo}<10^9M_\odot$ (LMACHs), is only 
present in the smaller, $114\,h^{-1}$Mpc boxes, while the largest-mass bin, 
$M_{\rm halo}\geq 10^{12}M_\odot$, is absent for the same smaller boxes due to 
lack of statistics, there are only a few such large haloes present in those
volumes. The results from all three boxes show an excellent agreement at 
all scales and masses present in both simulation volumes, with the two 
$114\,h^{-1}$Mpc box simulations yielding essentially identical results. At 
small scales the bias is highly nonlinear and can reach quite high values. 
At large scale all the curves asymptote to a constant value, which for the 
lower mass bins roughly corresponds to the large scale bias derived in 
\citet{2001MNRAS.323....1S}. However, for higher masses (i.e. very rare 
haloes) the numerically-calculated bias is well above the analytical 
prediction, indicating the latter is not valid for such rare haloes.



\subsubsection{Dark matter halo mass function}
In a recent paper \citep{will2013}, we presented detailed results and 
fits of the dark matter halo mass functions based on a large ensemble 
of N-body simulations which included the ones used here. This was done
based on three different methods for halo finding and included precision 
fits to the mass function, including one specifically targeted at 
high-redshift data. The results confirmed the trend observed previously 
by us and others \citep{2006MNRAS.369.1625I,2007MNRAS.374....2R,
2007ApJ...671.1160L,2012MNRAS.423.2222I}, namely that at the high redshifts
of interest here the halo mass functions differ significantly from both the 
Press-Schechter \citep{1974ApJ...187..425P} and Sheth-Tormen 
\citep{2002MNRAS.329...61S} analytical fits. This departure is particularly 
important for very rare haloes (corresponding to the massive haloes at the 
lower end of the redshift range we consider, $z\sim6-7$, and to all ACHs 
at higher redshifts), whose abundance is very strongly under-estimated by 
Press-Schechter and over-estimated by Sheth-Tormen (both by up to an order 
of magnitude at e.g. $z=17$). Therefore, such analytical approximations should 
not be used in analytical and semi-numerical modelling involving high-redshift 
haloes and instead more precise fits, such as the ones we presented in 
\citet{will2013} should be applied.

\subsection{Large-scale cosmic reionization}
\label{sect:large_scale_reion}
\begin{figure}
  \begin{center}
    \includegraphics[width=3.4in]{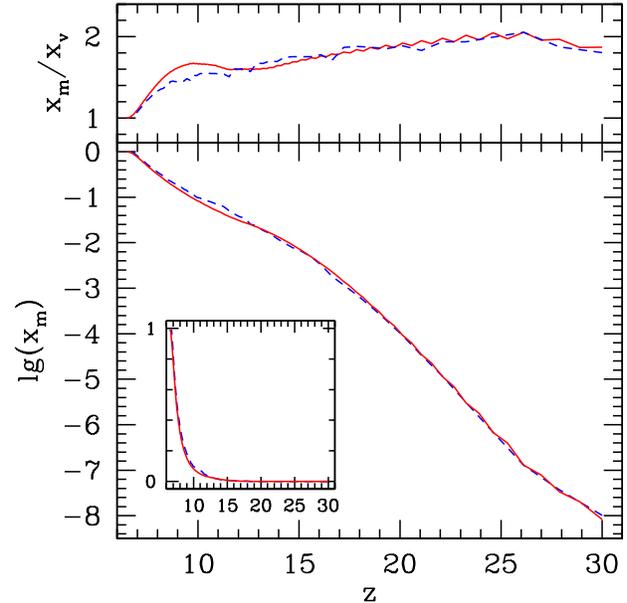}
  \end{center}
  \caption{(bottom) Mass-weighted reionization histories, $x_m$, for 
    XL2 (red, solid) and L2.1 (blue, dashed). (top) Ratio of the 
    mass- and volume-weighted mean ionized fractions ($x_m$ and $x_v$, 
    respectively)
    \label{reion_history:fig}}
\end{figure}

\begin{figure*}
  \begin{center}
    \includegraphics[width=3.in]{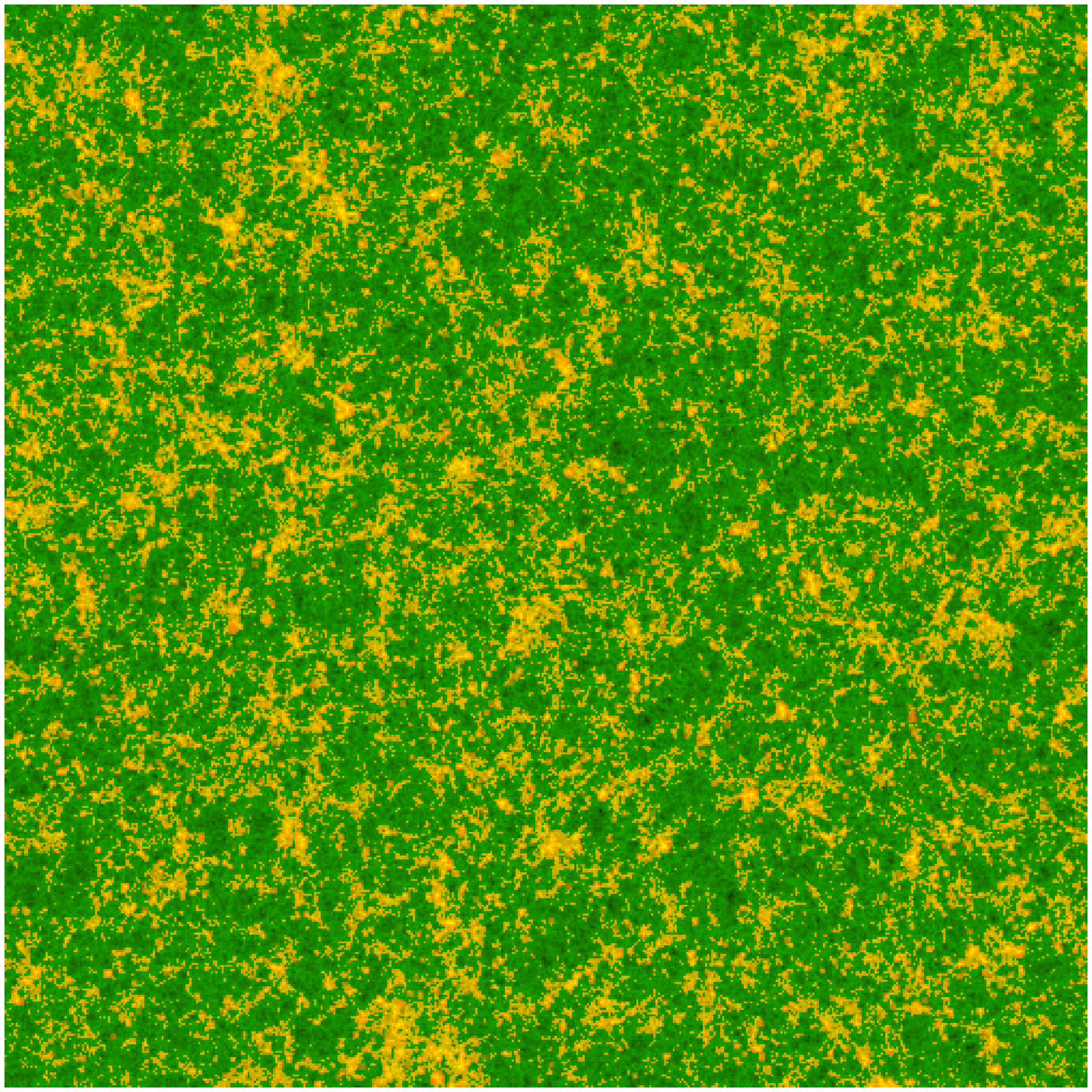}
    \includegraphics[width=3.in]{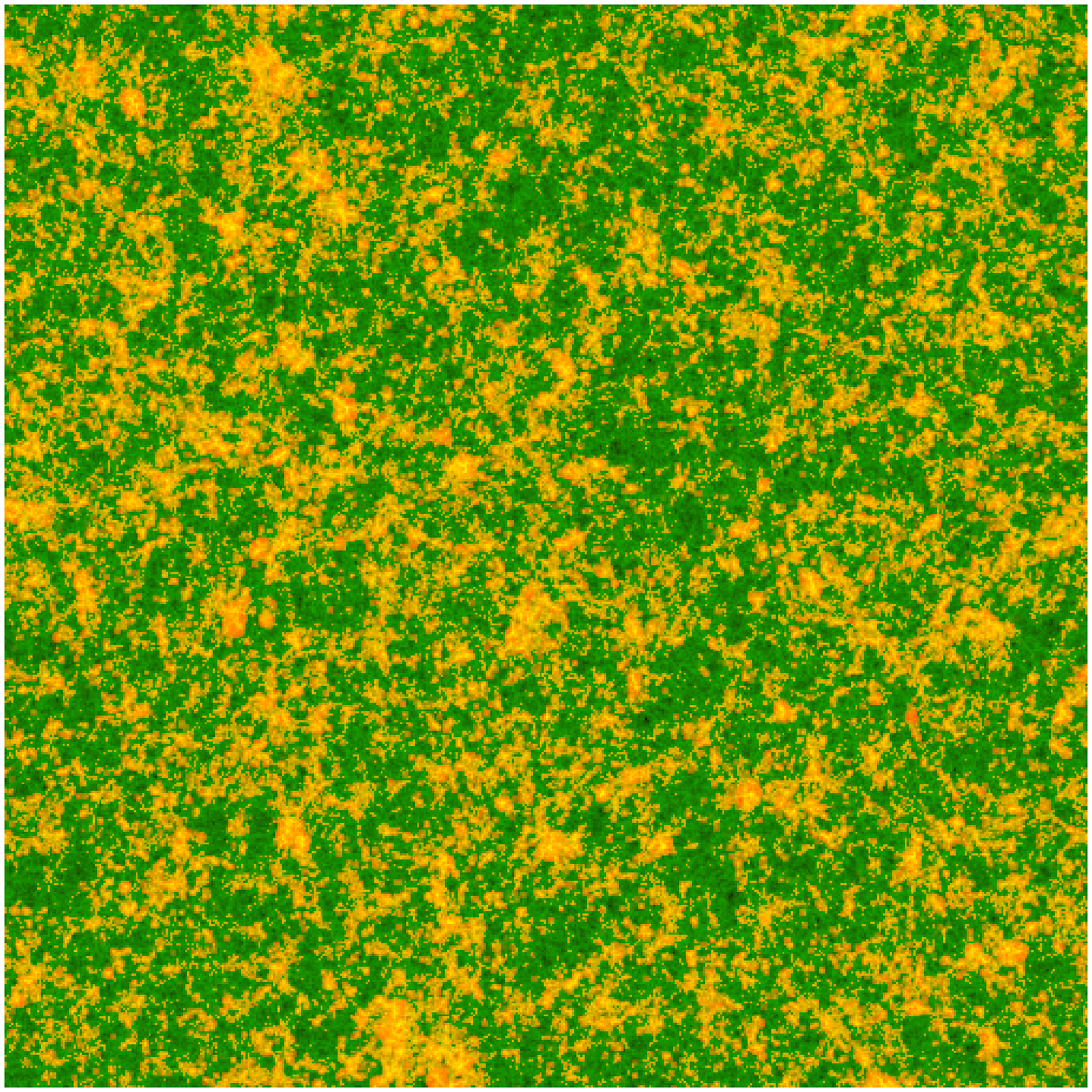}
    \includegraphics[width=3.in]{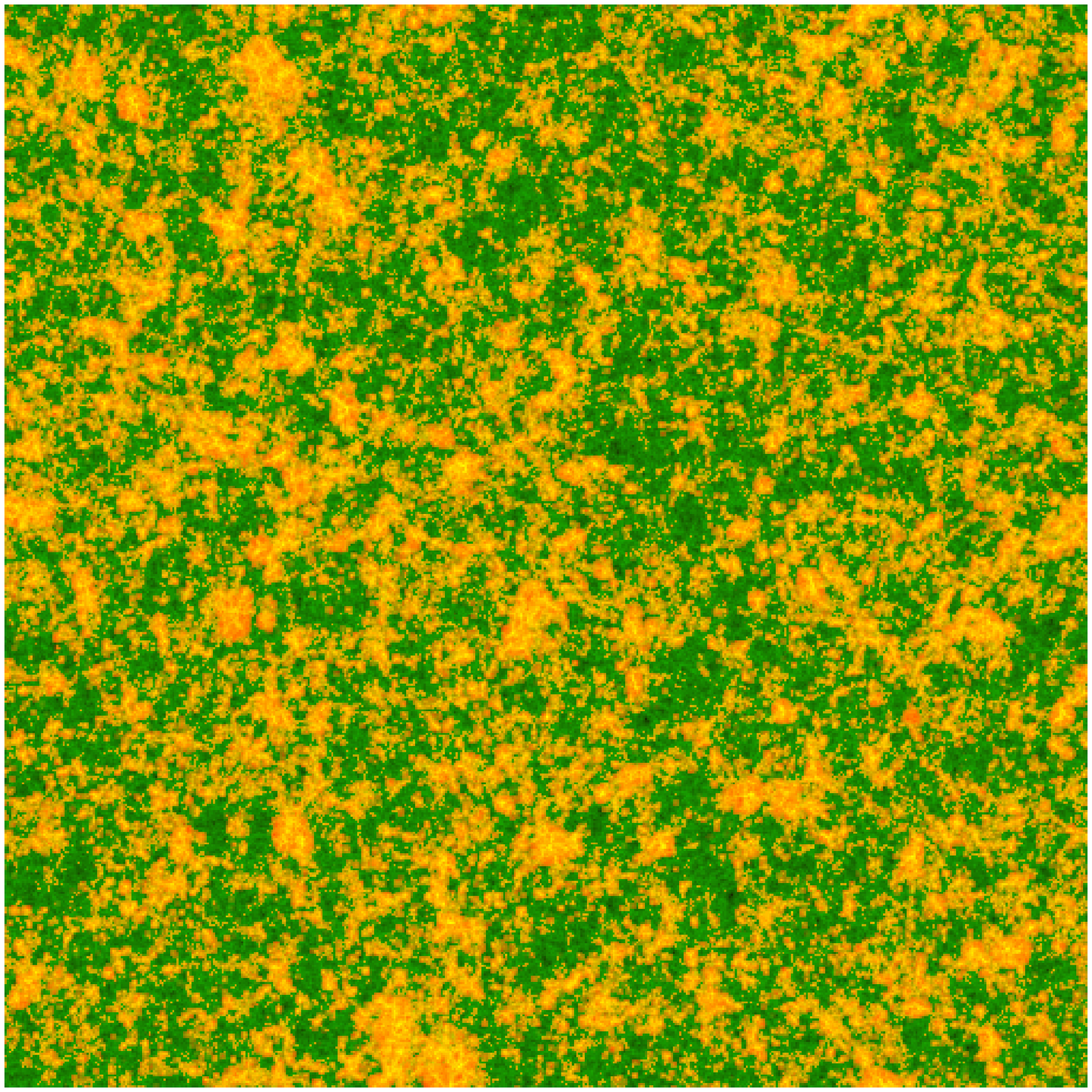}
    \includegraphics[width=3.in]{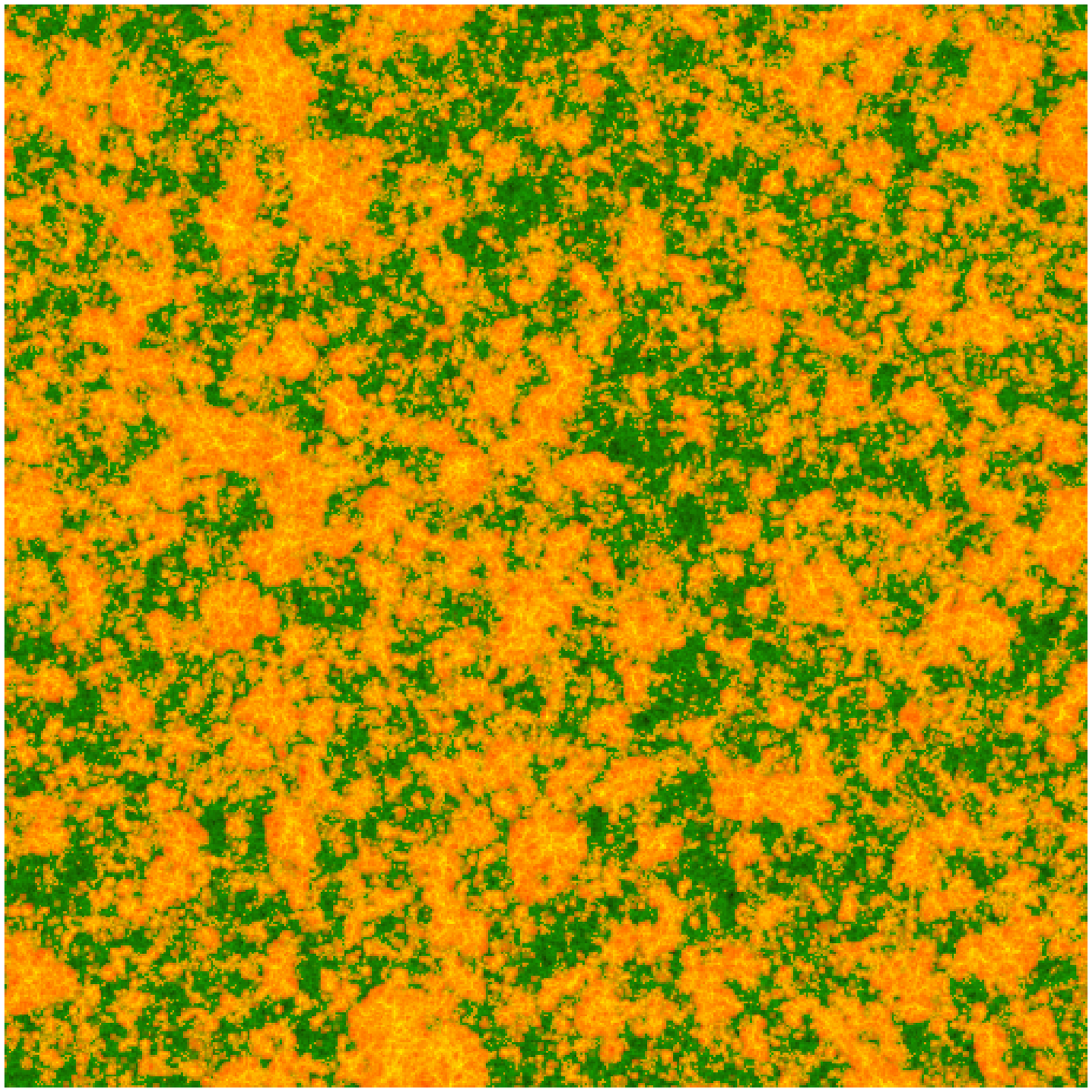}
    \includegraphics[width=3.in]{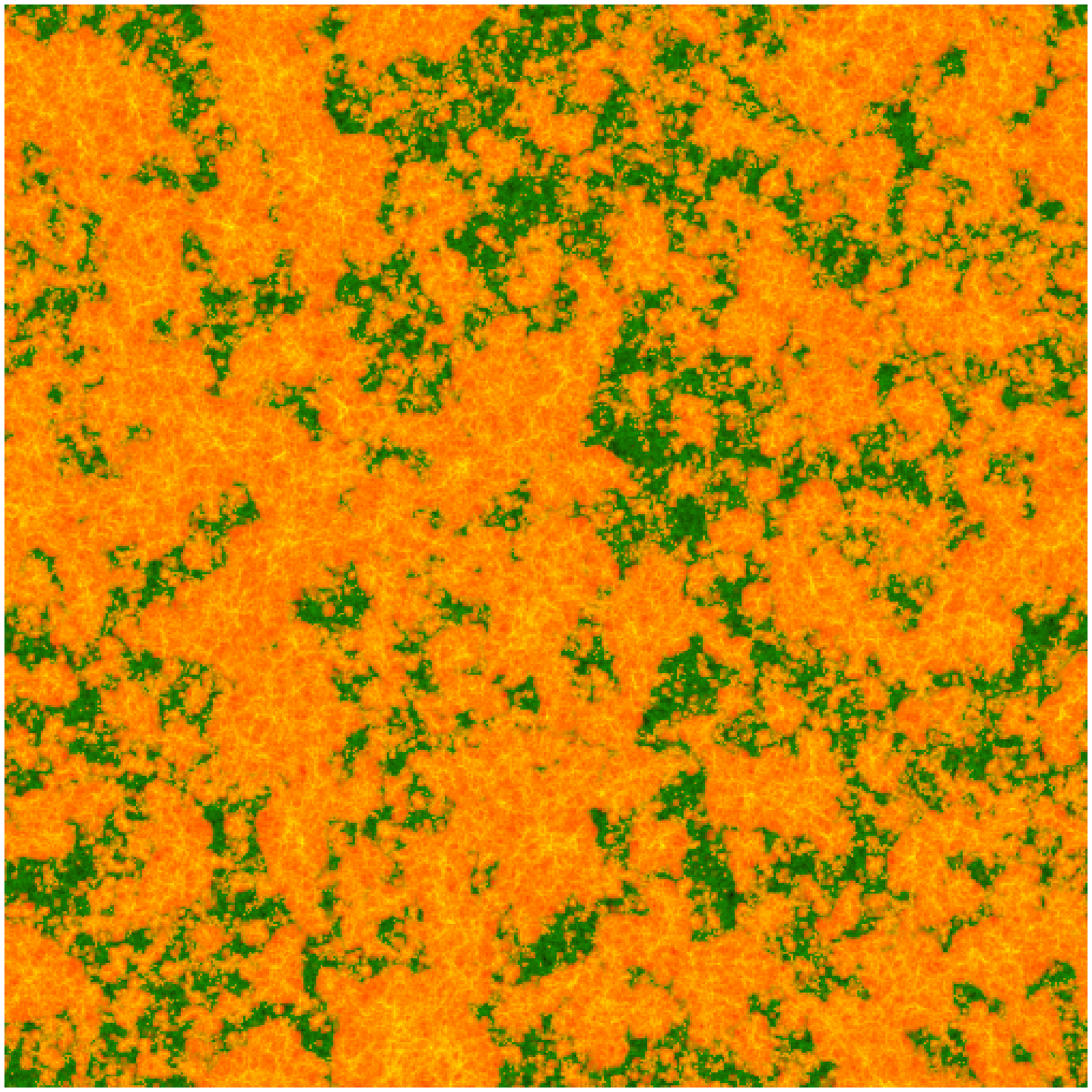}
    \includegraphics[width=3.in]{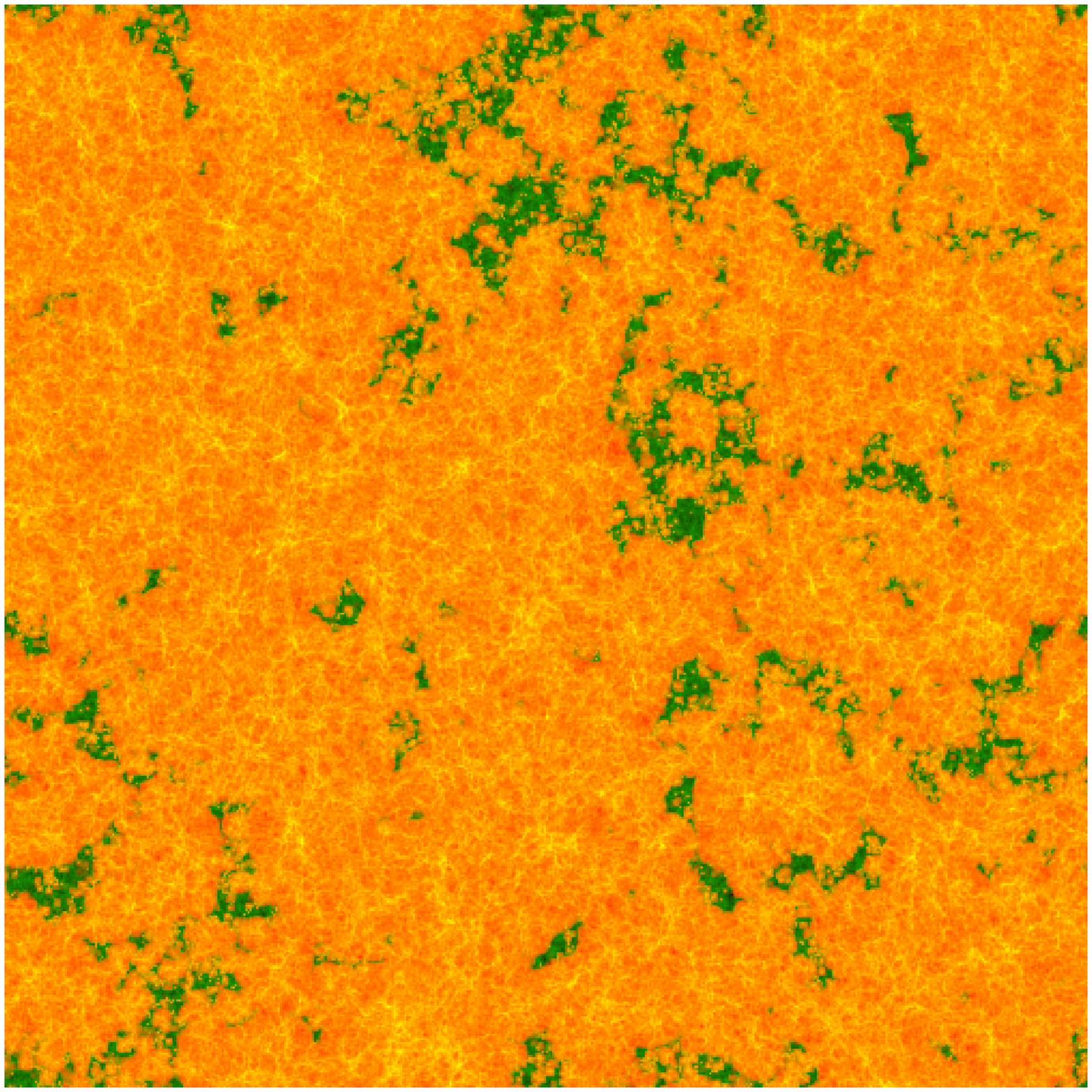}
  \end{center}
  \caption{(left to right and top to bottom) Spatial slices of the 
    ionized and neutral gas density from our XL2 radiative transfer 
    simulation with boxsize $425/h$~Mpc at box-averaged ionized 
    fractions by mass of $x_m=0.1$ ($z=9.6$), $x_m=0.2$ ($z=8.6$),
    $x_m=0.3$ ($z=8.1$), $x_m=0.5$ ($z=7.5$), $x_m=0.7$ ($z=7.1$),
    $x_m=0.9$ ($z=6.8$). Shown are the density field (green) 
    overlayed with the ionized fraction (orange/yellow). 
    \label{reion_image:fig}}
\end{figure*}

\begin{figure*}
  \begin{center}
    \includegraphics[width=3.2in]{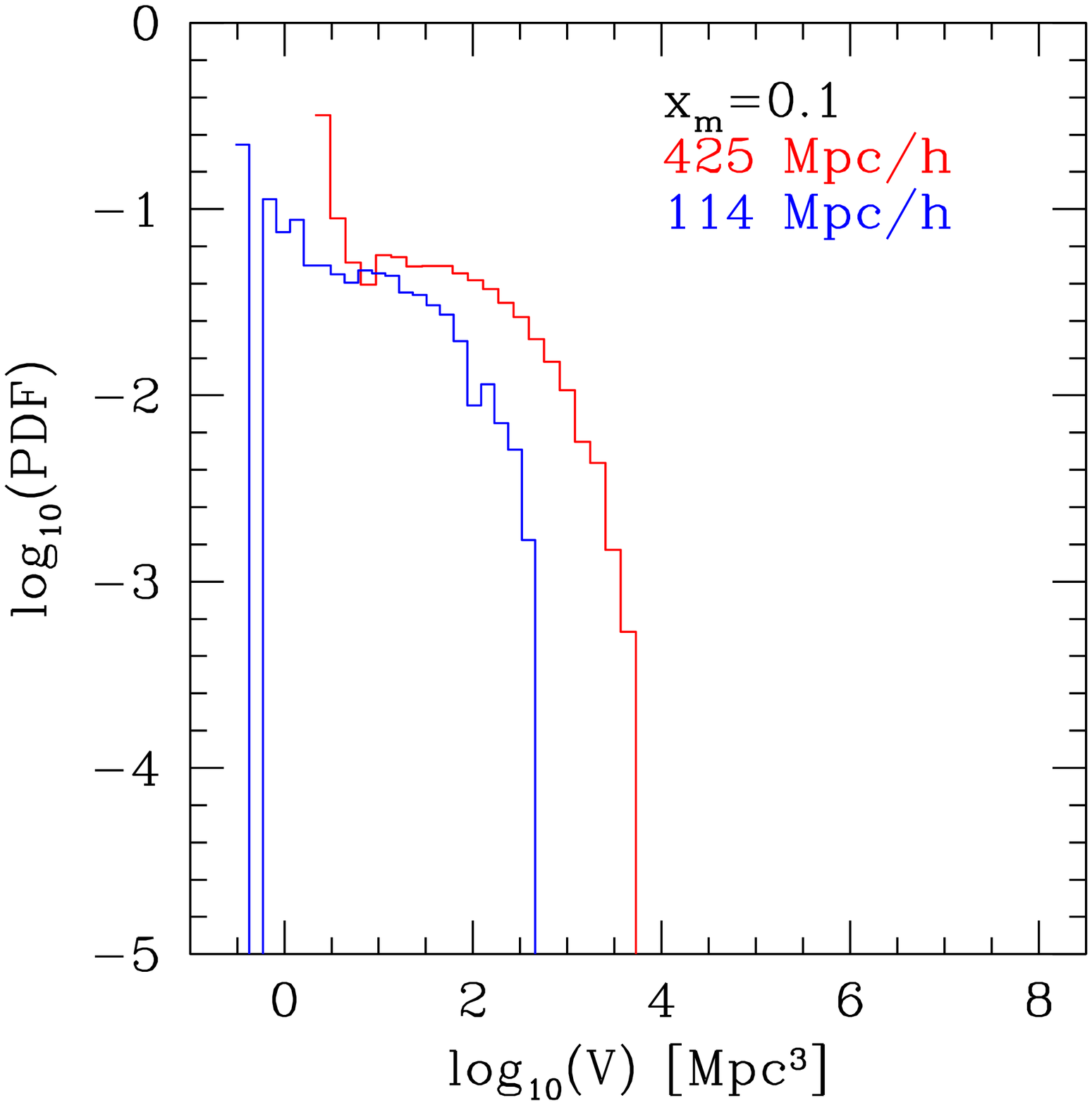}
    \includegraphics[width=3.2in]{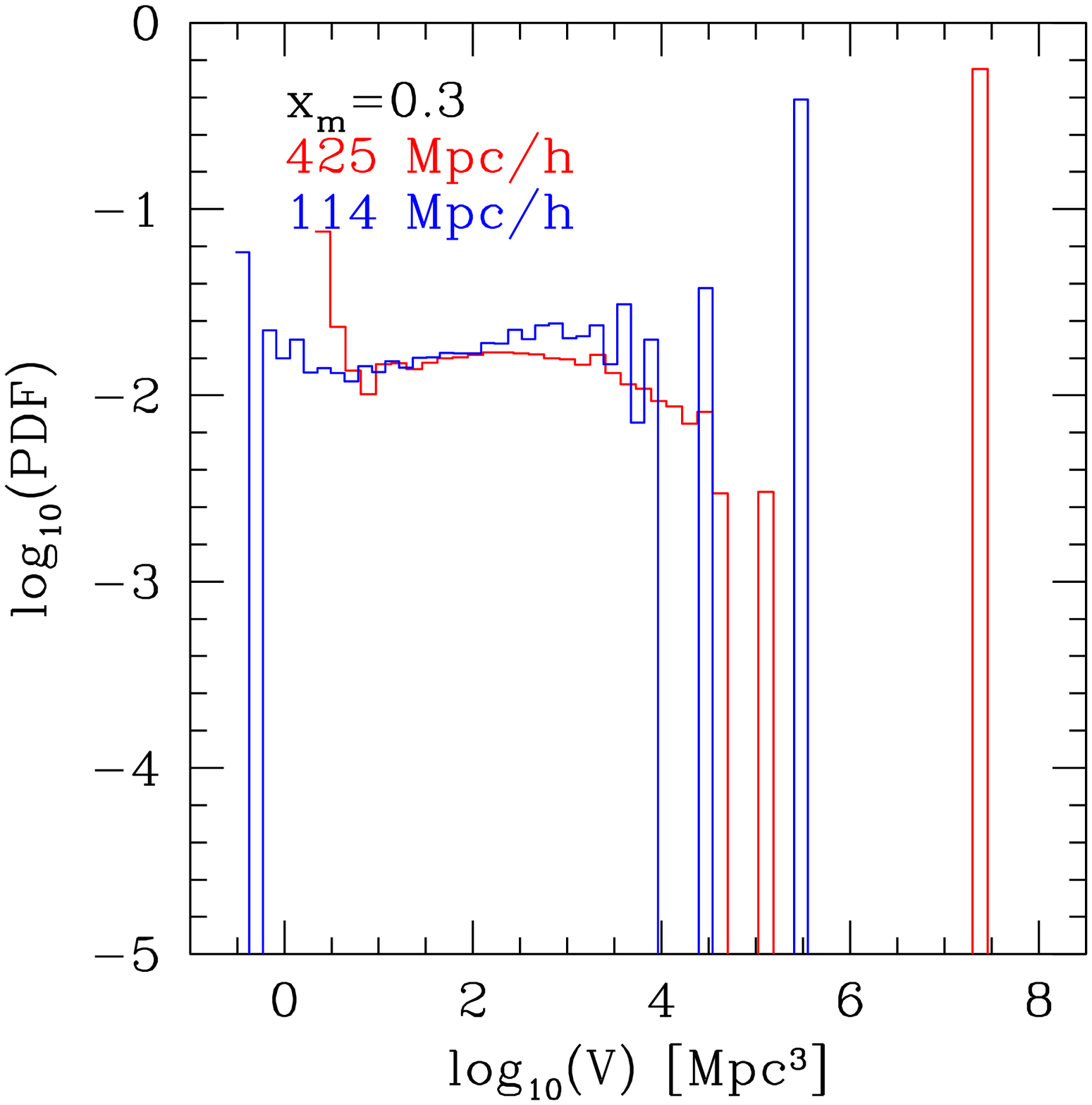}
    \includegraphics[width=3.2in]{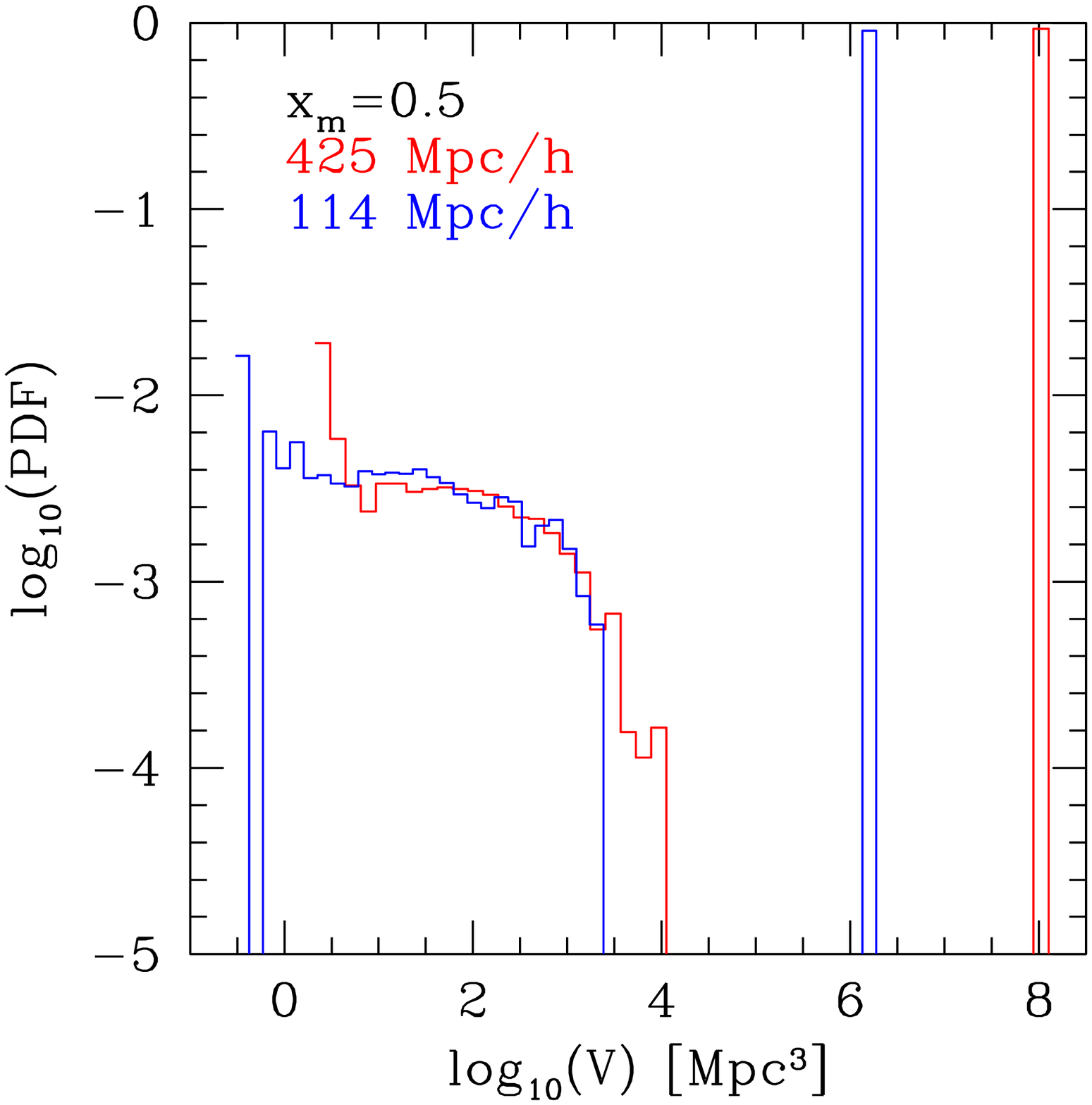}
    \includegraphics[width=3.2in]{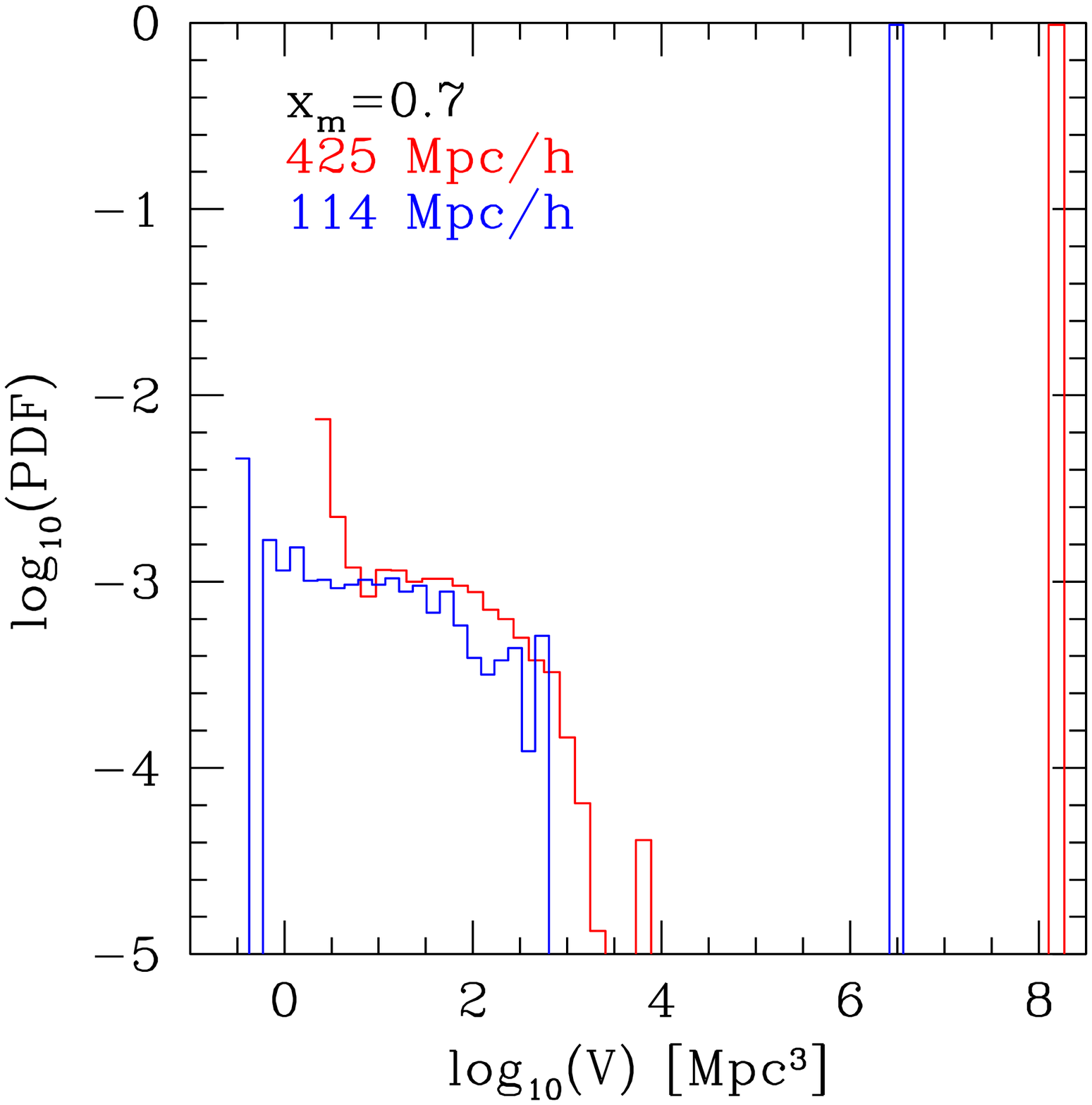}
  \end{center}
  \caption{Volume-weighted probability distribution function, $V(dp/dV)$,
for H~II regions extracted from our simulations at different stages of the
reionization process, as labelled. Shown are the results based on XL2 (red) 
and L2.1 (blue).
    \label{fof:fig}}
\end{figure*}

\subsubsection{Reionization history}
The reionization history resulting from our radiative transfer simulations 
is shown in Figure~\ref{reion_history:fig}. In all simulations the first 
sources appear at $z\sim30$ (in the large-box case those sources are 
sub-grid) and the end-of-reionization (defined by ionized fraction by mass 
of $x_m=0.99$) is reached at similar redshifts of $z=6.5$ and 6.6 for XL2 
and L2.1, respectively. The evolution of the ionized fractions is roughly 
exponential at early times, tracking the exponential rise in the number of 
(mostly LMACH) ionizing sources. The reionization history thereafter flattens 
starting at $\sim15$, due to the LMACHs suppression becoming more wide-spread, 
and finally again becomes exponential starting from at $z\sim8$, at which point 
the unsuppressible HMACH sources come to dominate the evolution. The overall 
reionization history is near-identical for the two volumes due to the same 
assumed ionizing photon efficiencies $g_\gamma$ for both types of sources, 
LMACHs and HMACHs. The only notable differences between the two runs are 
observed for $z\sim10-12$, where the reionization of the larger volume 
slightly lags the smaller one. The reason for this lag is the different 
modelling of the LMACH sources, which is based on the directly-resolved 
haloes (L2.1) or as a sub-grid population (XL2). As we discussed above, the 
latter is solely based on the average relation between collapsed fraction in 
haloes and the local density and at present does not model the scatter around 
that mean relation observed in simulations. This results in slightly higher 
suppression of LMACHs due to their increased local clustering. The inside-out 
character of reionization, evidenced by the ratio of mass-weighted to volume 
ionized fraction, $x_m/x_v$, being always above one \citep{2006MNRAS.369.1625I}
 is largely unaffected by the size of the simulation volume and the difference 
in grid resolution. 

\begin{figure*}
  \begin{center}
    \includegraphics[width=3.2in]{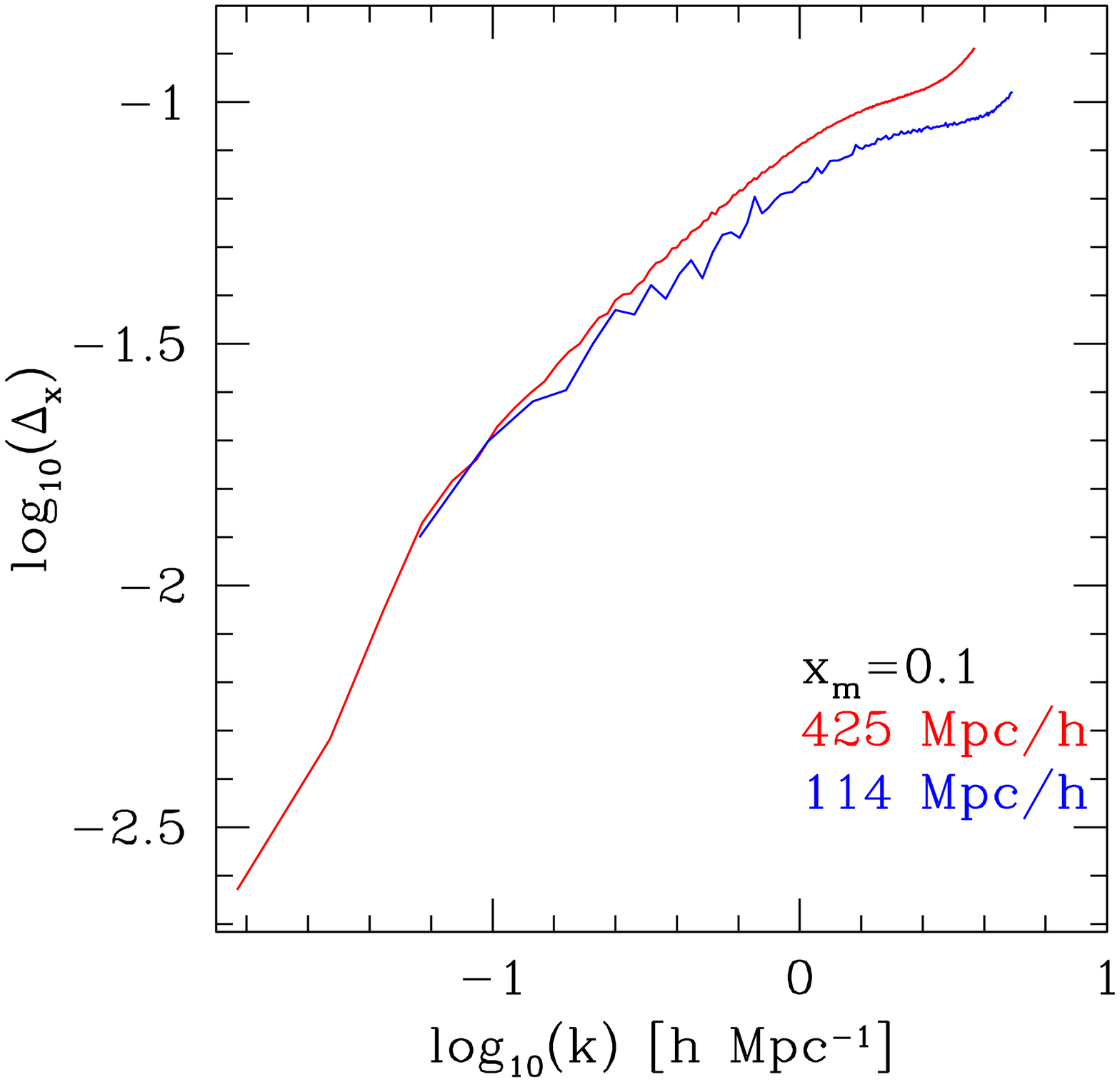}
    \includegraphics[width=3.2in]{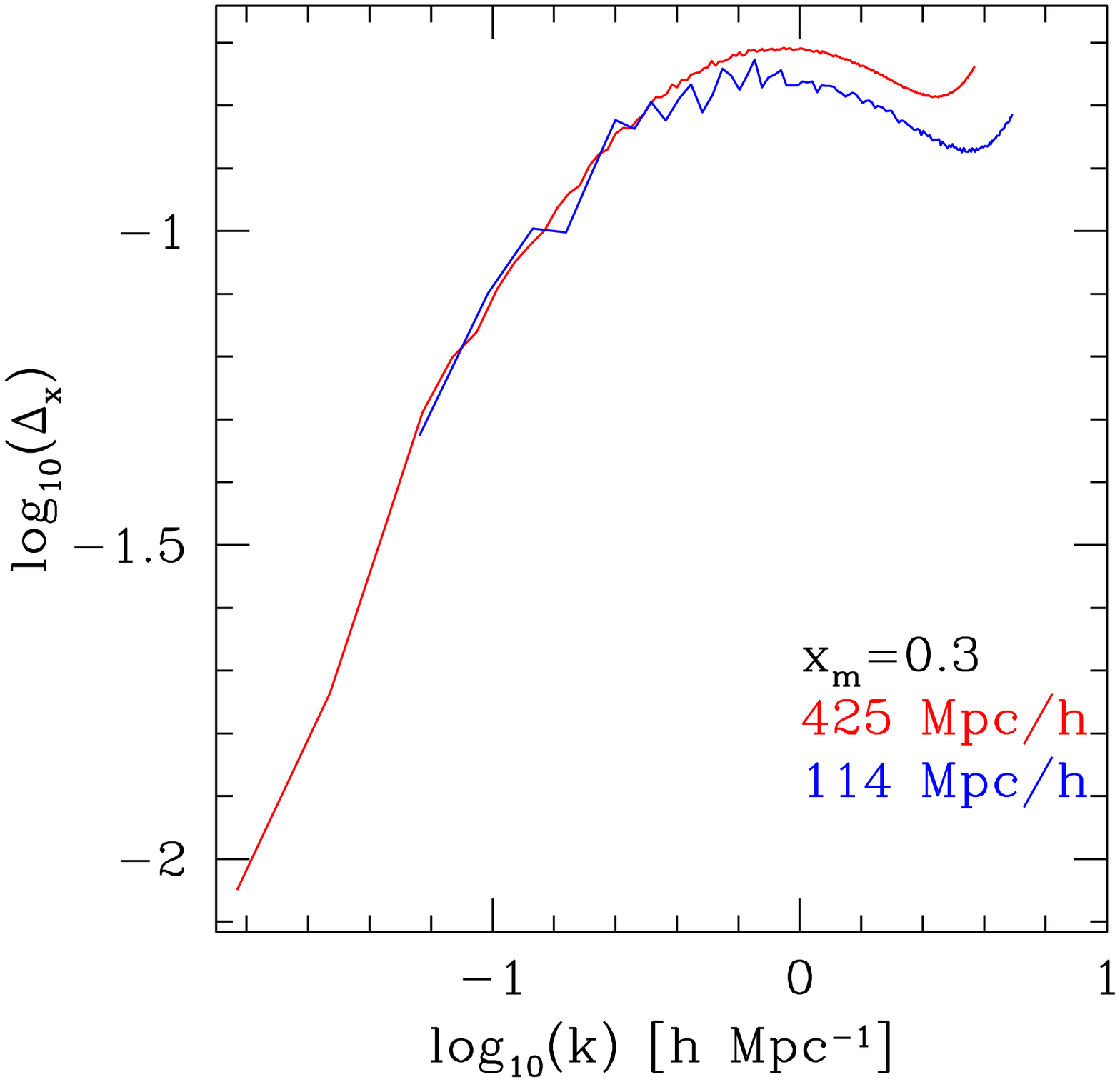}
    \includegraphics[width=3.2in]{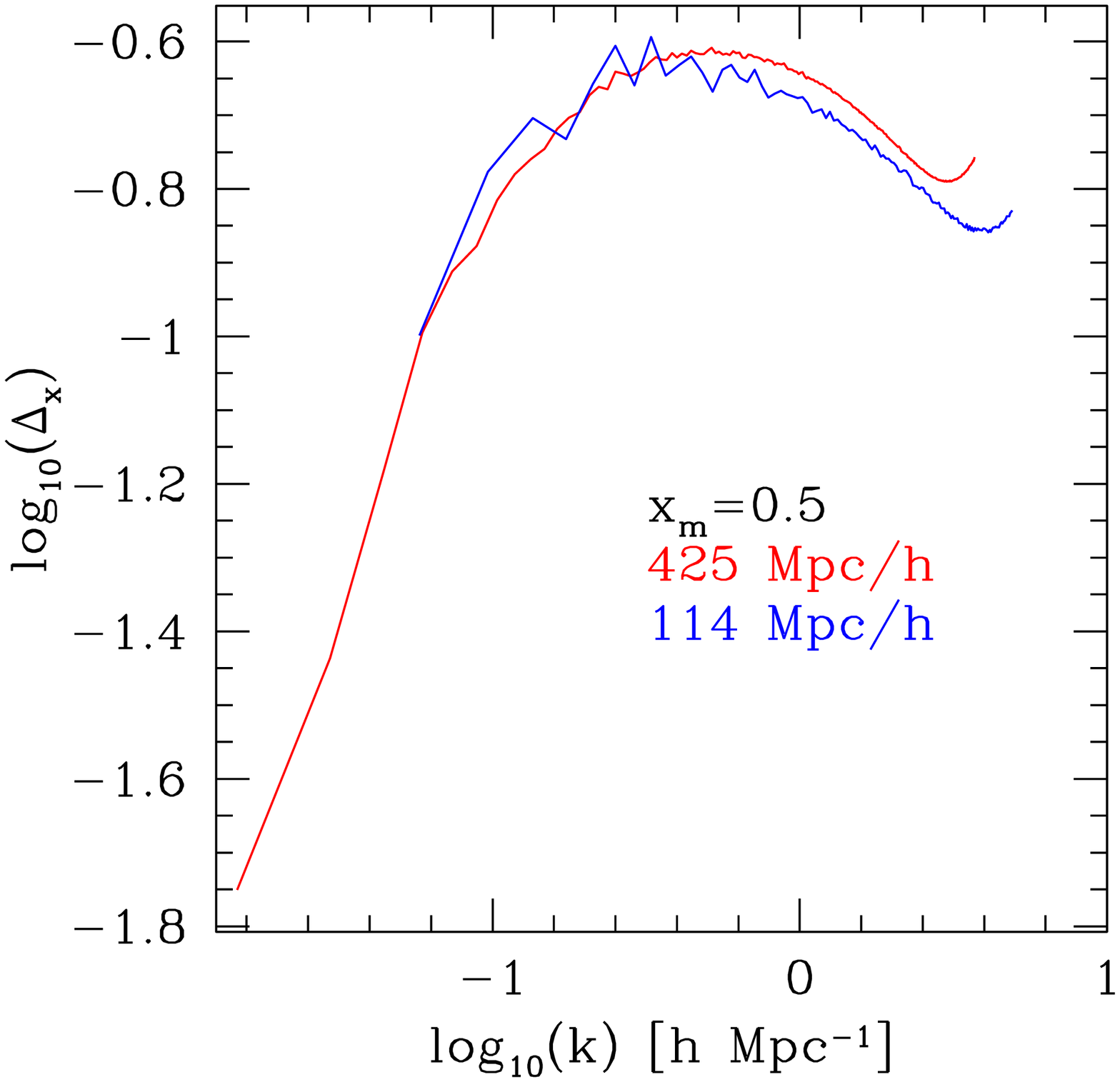}
    \includegraphics[width=3.2in]{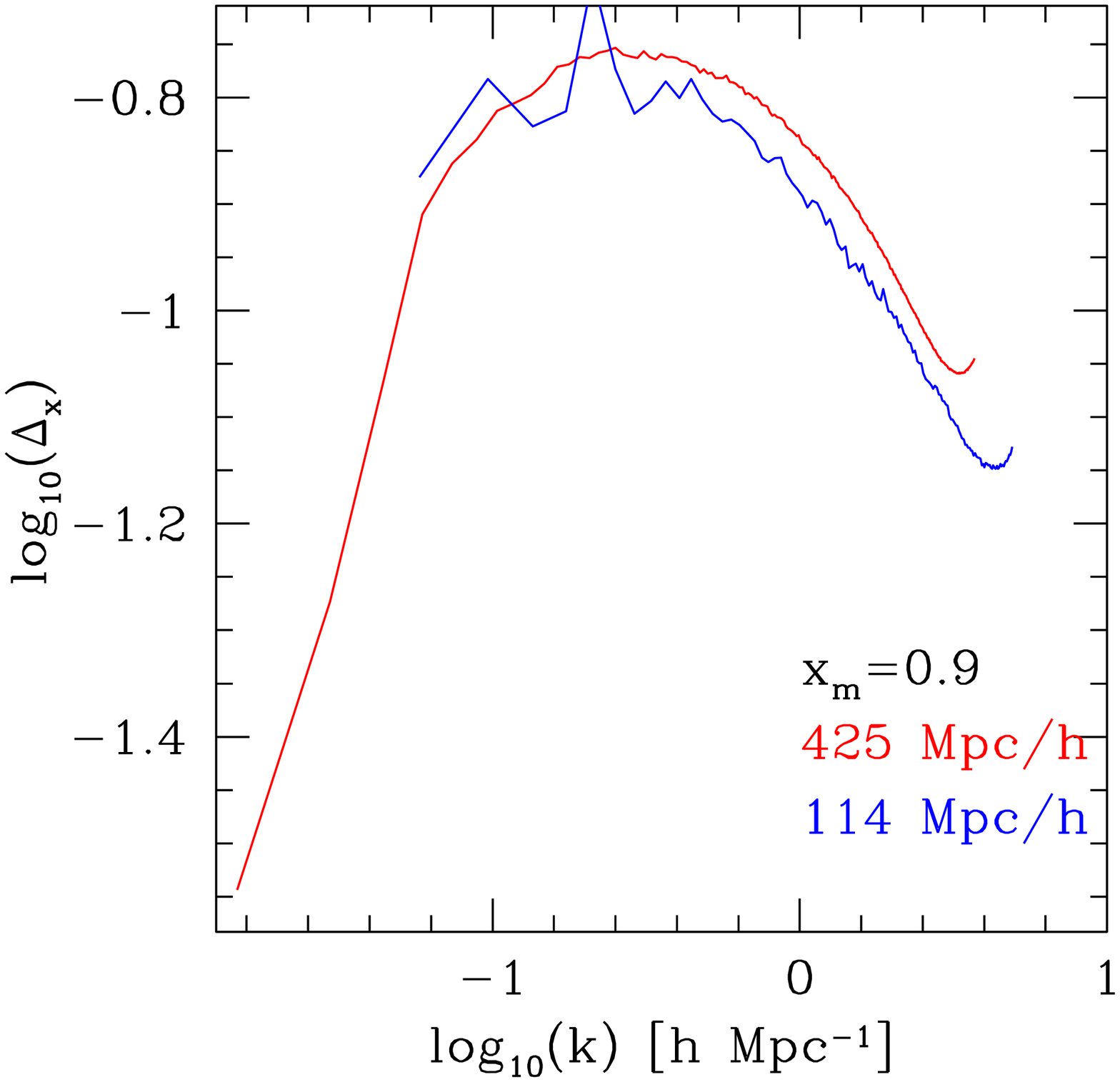}
  \end{center}
  \caption{Dimensionless power spectra, $\Delta_x$, of the volume-weighted 
ionized fraction, $x_v$, for H~II regions extracted from our simulations at 
different stages of the reionization process, as labelled. Shown are the 
results for XL2 (red) and L2.1 (blue). \label{ps_x:fig}}
\end{figure*}

\subsubsection{Reionization geometry and H~II region size distributions}
\label{sizes:sect}

The significant density fluctuations at very large scales, from tens to 
hundreds of comoving Mpc, and the strong clustering of high-$z$ galaxies 
discussed on the previous section could be expected to result in enhanced 
large-scale patchiness of reionization. In Figure~\ref{reion_image:fig} we 
illustrate the reionization geometry at several key stages of the evolution, 
corresponding to mass-weighted ionized fraction of $x_m=0.1, 0.2, 0.3, 0.5, 
0.7$ and 0.9, or a redshift range of $z=9.6$ to $z=6.8$. Initially, a large 
number of fairly small, Mpc-size H~II regions form. They are strongly
clustered on small scales, following the clustering of the sources, but are
relatively uniformly distributed on larger scales, where the long-wavelength 
modes modulating the density fluctuations are still very low. Locally these 
small H~II regions quickly start merging into larger ones, with sizes between 
few and $\sim10~$Mpc across. We note that of course these are 2D cuts of the
ionization field and H~II regions can, and do, have different sizes depending
on the direction, as quantified e.g. in \citet{2008MNRAS.391...63I}. 
Significant large-scale percolation of the H~II regions only occurs when the 
universe reaches $\sim50\%$ ionization by mass, 
at which point many ionized regions reach sizes of tens of Mpc and become 
connected by bridges to other nearby large ionized regions of similar sizes. 
At the same time there are also similarly-sized regions which remain largely 
neutral. The H~II regions continue percolating up to still larger scales and 
by $x_m=0.7$ some reach hundreds of Mpc across, with large, while at the same
time large, tens of Mpc across, neutral regions remain between them, both 
reflecting the large-scale fluctuations of the underlying density field. 
Finally, when the volume is 90\% ionized all H~II regions have percolated 
into one, although fairly large neutral regions still exist even in this late 
phase. 

\begin{figure*}
  \begin{center}
    \includegraphics[width=3.2in]{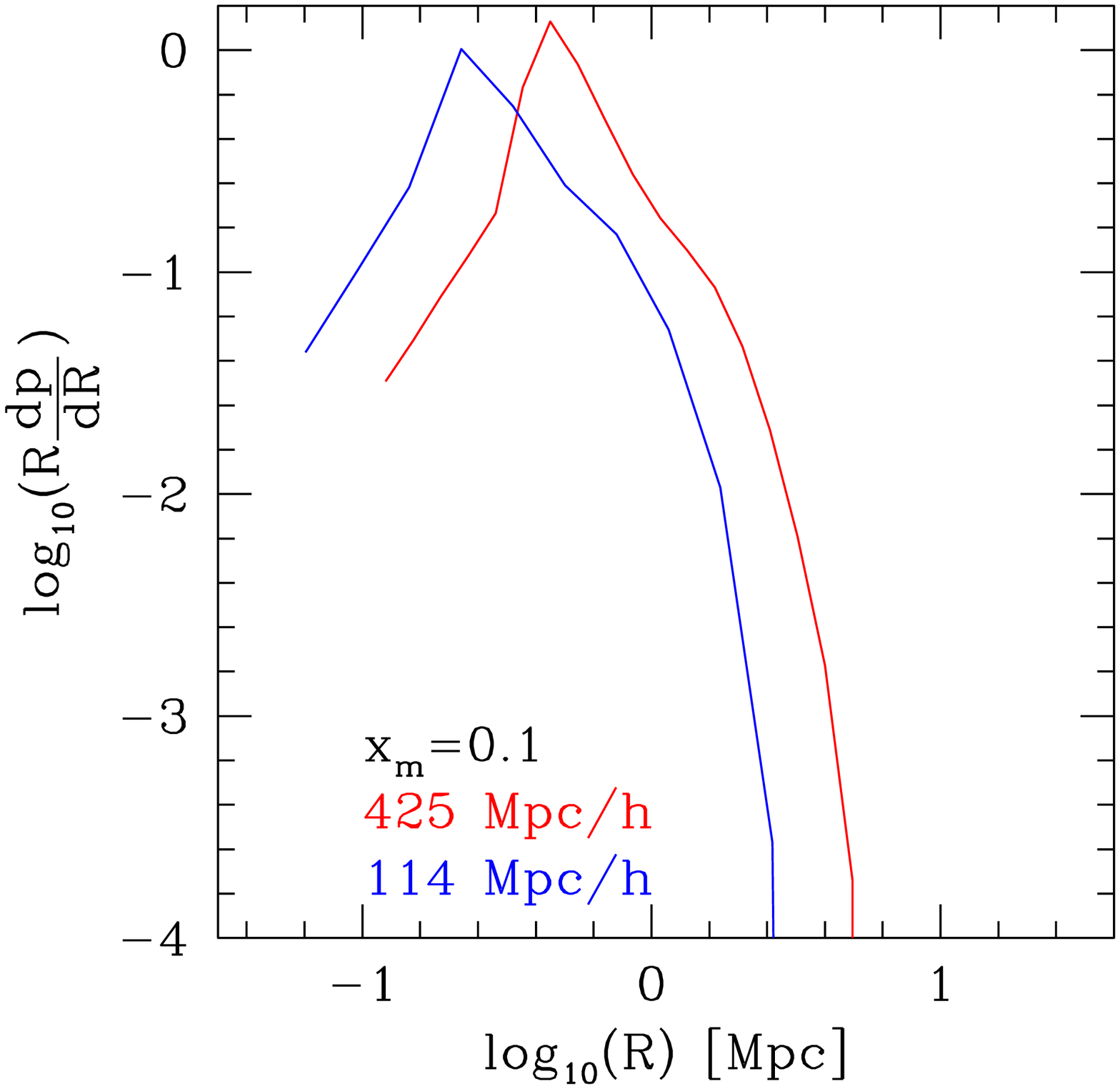}
    \includegraphics[width=3.2in]{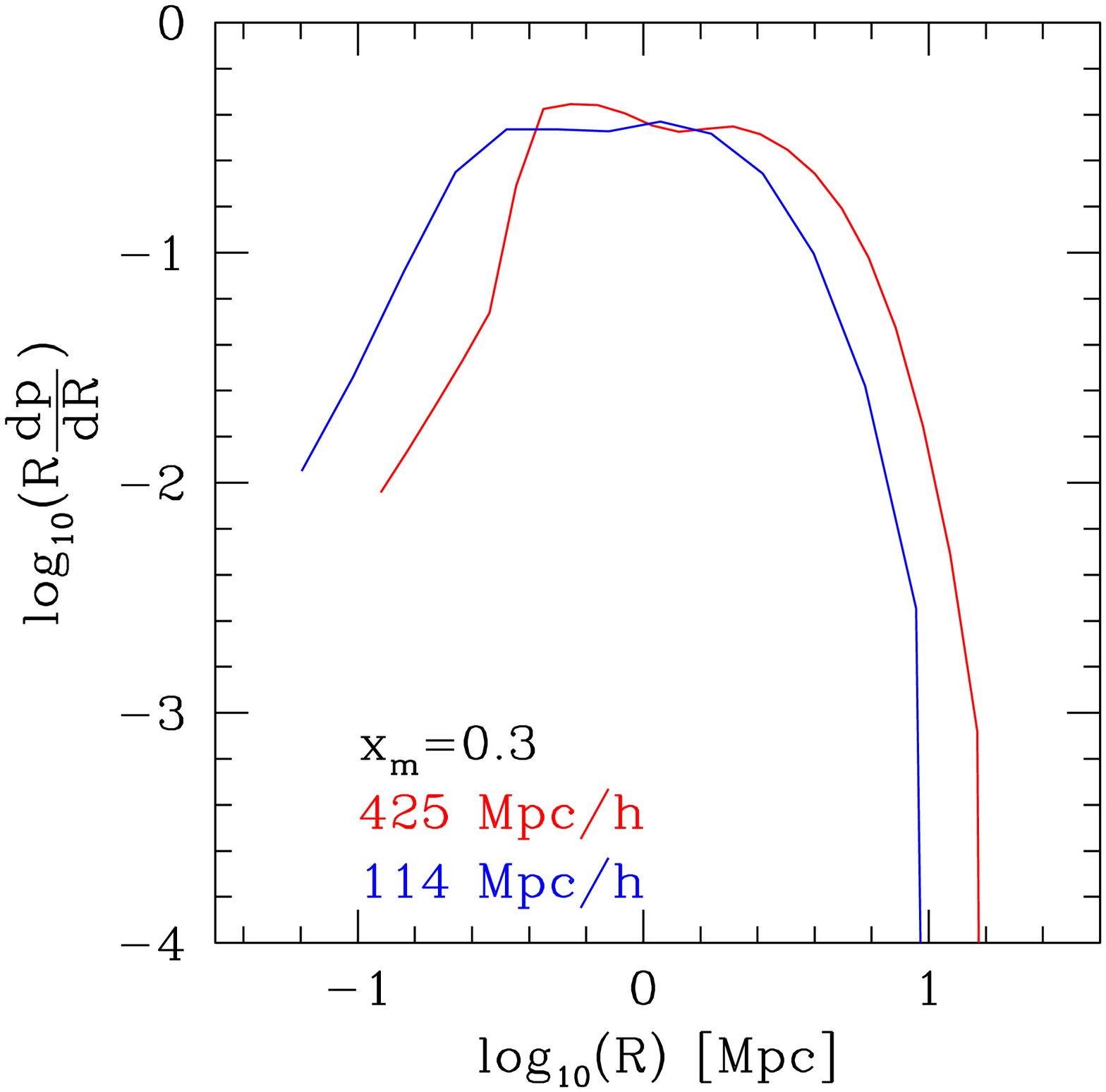}
    \includegraphics[width=3.2in]{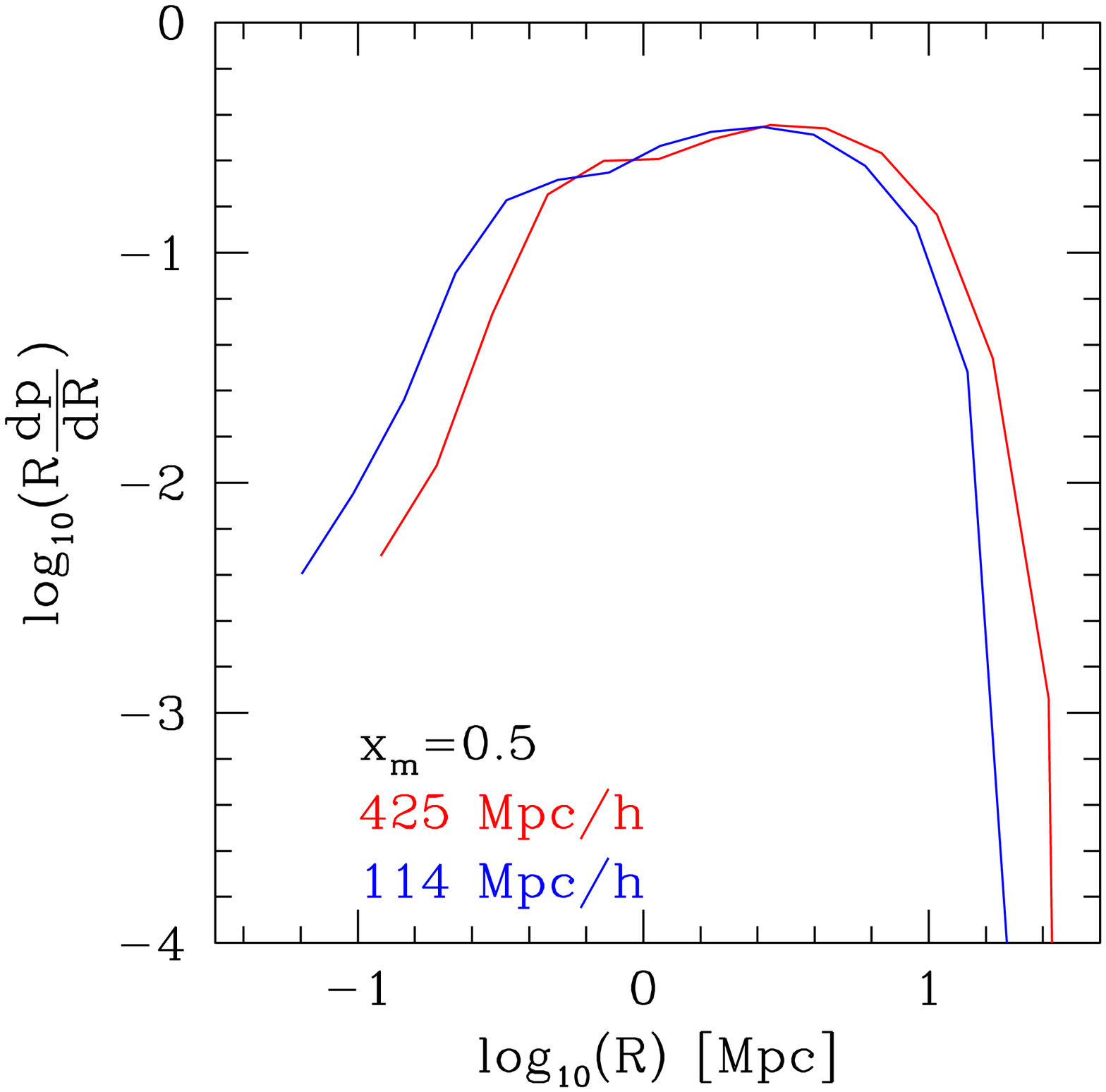}
    \includegraphics[width=3.2in]{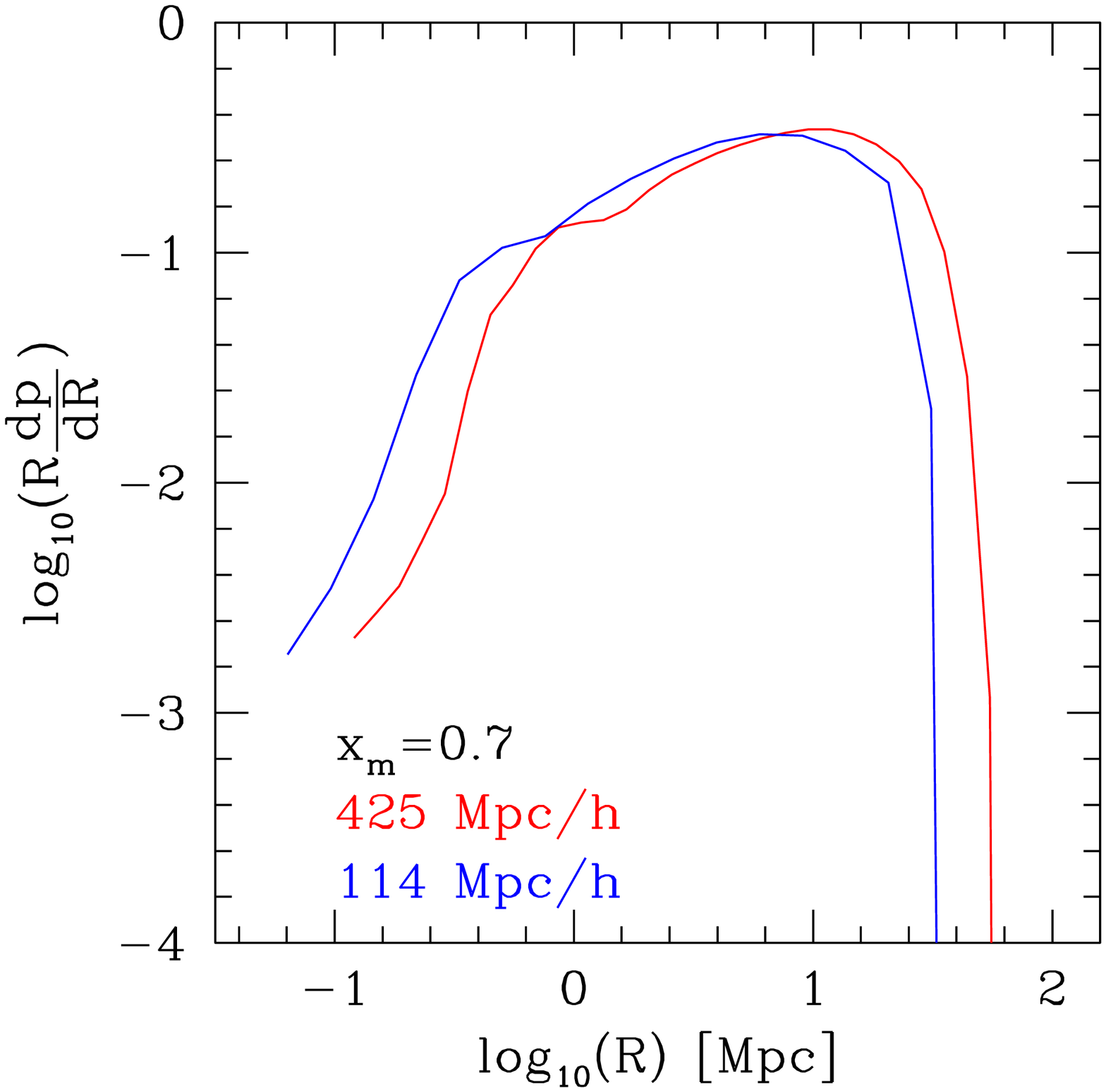}
  \end{center}
  \caption{Probability distribution function per logarithmic radial bin, 
$R\frac{dp}{dR}$, spherical regions with radius $R$ as given by the SPA
method, based on the ionized distribution given by our simulations. Shown
are different stages of the reionization process, for ionized fraction by 
mass $x_m=0.1, 0.3, 0.5$ and 0.7, as labelled. Shown are the results based 
on XL2 (red) and L2.1 (blue).
    \label{spa:fig}}
\end{figure*}

These qualitative observations are further supported by more quantitative
measures of the size distributions of the ionized regions obtained based 
on different approaches: the friends-of-friends (FOF) method 
\citep{2006MNRAS.369.1625I,2011MNRAS.413.1353F}, the spherical average 
method \citep[SPA,][]{2007ApJ...654...12Z,2007MNRAS.377.1043M} and the 
three-dimensional power spectra of the ionization fraction field 
\citep{2006MNRAS.369.1625I,2011MNRAS.413.1353F}. The results from the FOF 
method, which links together topologically-connected ionized regions, at 
several representative stages of reionization and our two different 
simulation volumes are shown in Figure~\ref{fof:fig}. During the early 
stages of reionization (ionized fraction by mass of $x_m=0.1$) the two 
distributions have similar shape, with the majority of the ionized volume 
found in small, but numerous H~II regions with typical sizes ranging 
from less than a Mpc up to a few Mpc. The fraction of the simulation volume 
occupied by the smallest regions which are resolved in both simulations 
($V\sim10\,\rm Mpc^3$) is very similar. However, in XL2 there are 
considerably more larger ionized regions, ($V>30\,\rm Mpc^3$). In L2.1 
there are no H~II regions with volume larger than about $500 \,\rm Mpc^3$, 
while the largest ones in XL2 reach volumes in excess of $10^3 \,\rm Mpc^3$ 
even at this early stage, indicating a longer characteristic scale for local 
percolation in the larger volume.
\begin{figure}
  \begin{center}
    \includegraphics[width=3.2in]{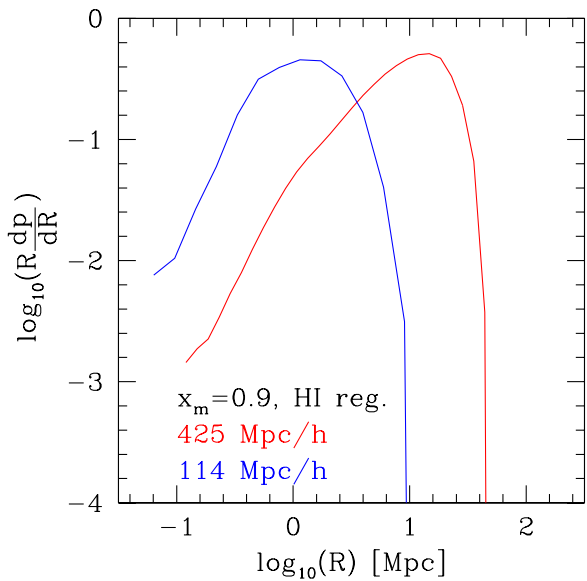}
  \end{center}
  \caption{Same as in Figure~\ref{spa:fig}, but for the {\it neutral}
H~I patches at late times, for ionized fraction by mass $x_m=0.9$. 
    \label{spa2:fig}}
\end{figure}

The ionized regions continue to grow over time and at $x_m=0.3$ we already 
observe considerable local percolation, whereby multiple H~II regions merge 
into a single very large one, with a volume filling a significant fraction 
of the simulation box. Similar behaviour was previously observed in 
\citet{2006MNRAS.369.1625I} and \citet{2011MNRAS.413.1353F}. For both XL2 
and L2.1 this percolated region fills roughly 10\% of the total volume at this 
stage of the evolution. Again there are significant differences between the 
two boxes in terms of the abundances of the smaller H~II regions. The resolved 
regions in the two simulations are in excellent agreement up to 
$V\sim10^4\,\rm Mpc^3$. However, 
in the smaller volume there are no ionized regions with volume above 
$V\sim3\times10^4\,\rm Mpc^3$, while such regions still take a significant 
fraction of the large volume. Similar trends are observed during the later
stages of the evolution ($x_m>0.5$). The percolation of the ionized regions 
proceeds much further, with one large, connected region filling most of the 
volume in either case (with its volume limited only by the total simulation 
volume), while the smaller regions decrease in both abundance and size. There 
is agreement in the abundances of smallest resolved H~II regions 
($V<10^3\,\rm Mpc^3$ at $x_m=0.5$, $V<10^2\,\rm Mpc^3$ at $x_m=0.7$), but in 
the larger simulation volume there are still many more mid-sized regions, 
some reaching up to $V\sim10^4\,\rm Mpc^3$ in volume.

\begin{figure*}
  \begin{center}
    \includegraphics[width=3.2in]{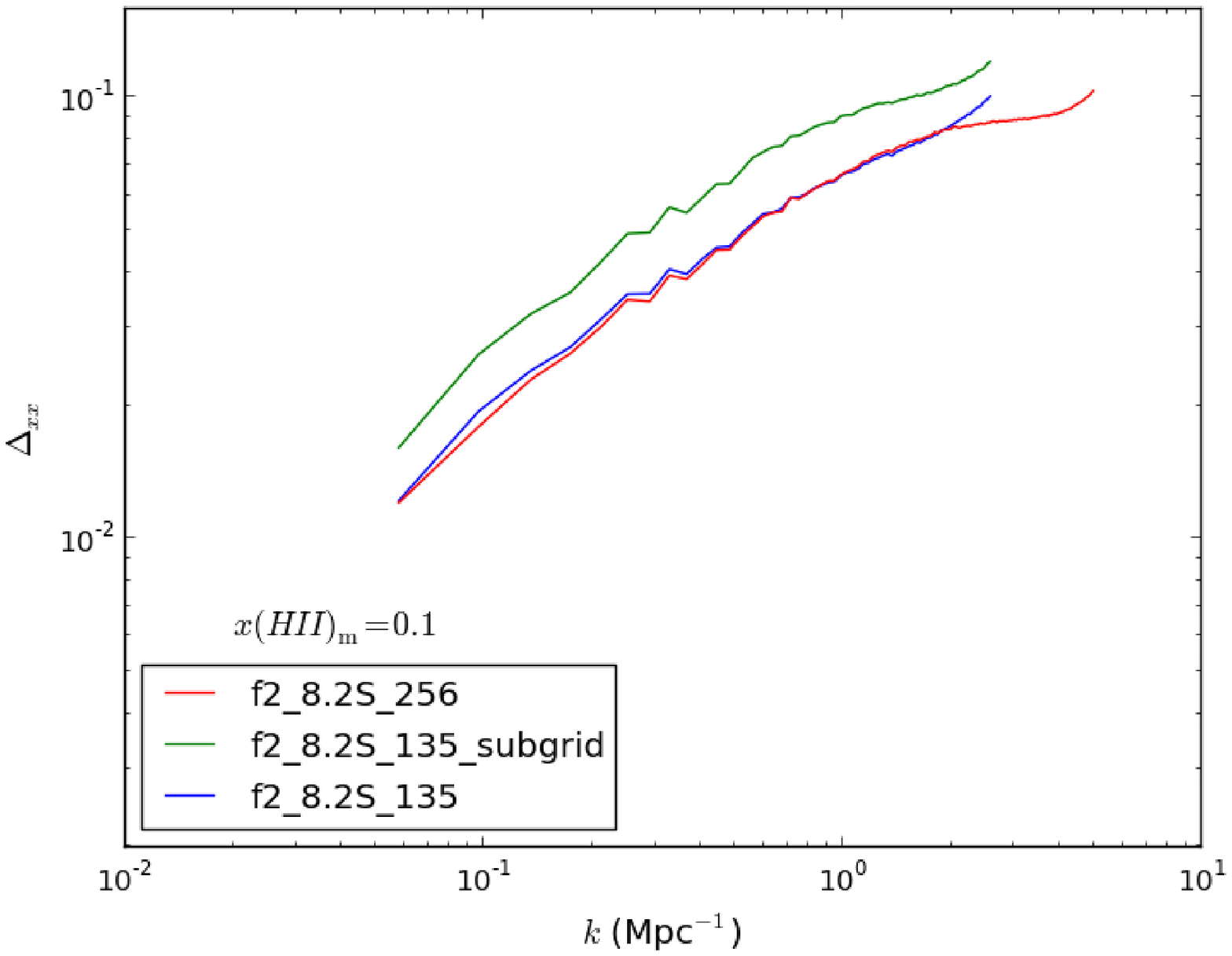}
    \includegraphics[width=3.2in]{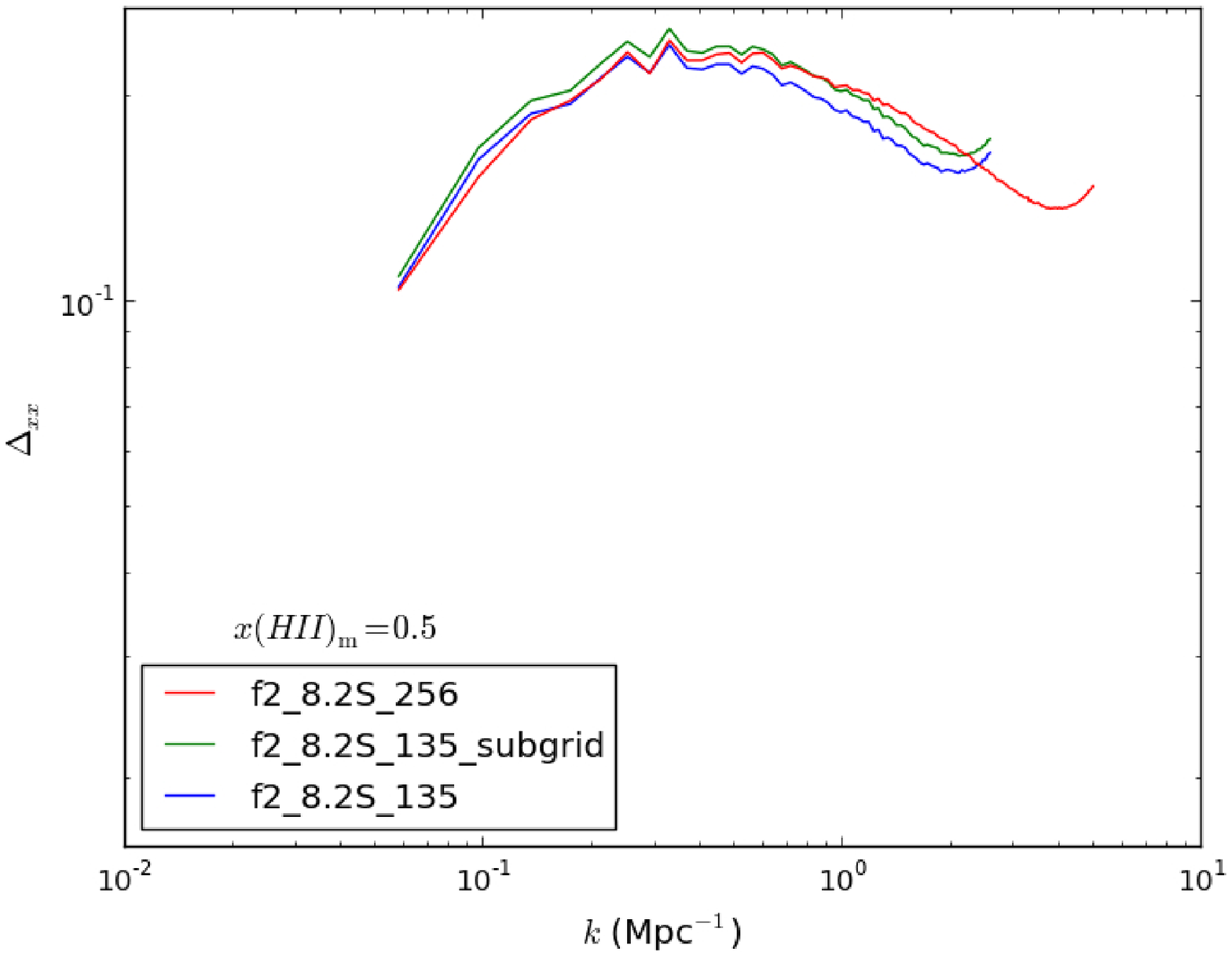}
    \includegraphics[width=3.2in]{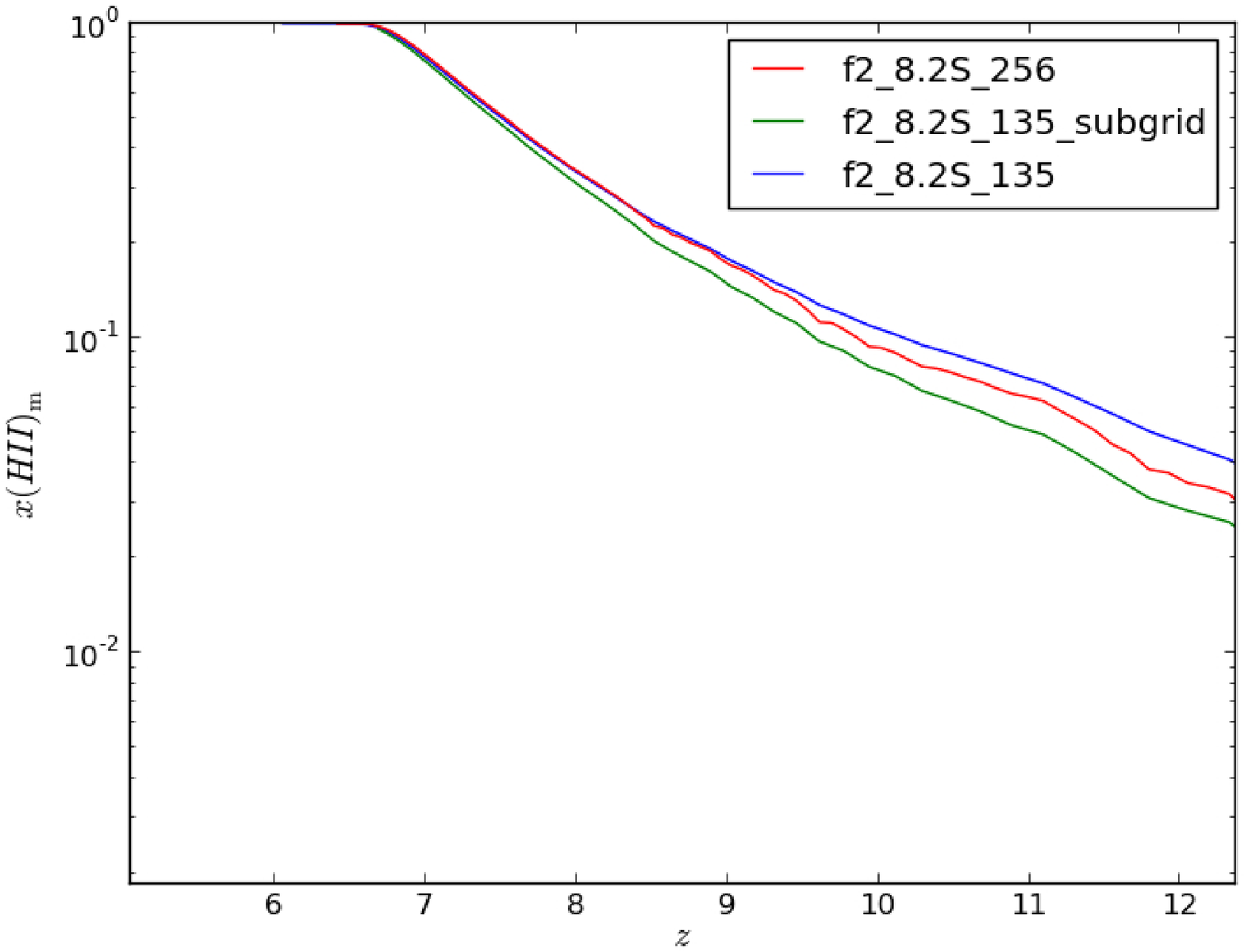}
    \includegraphics[width=3.2in]{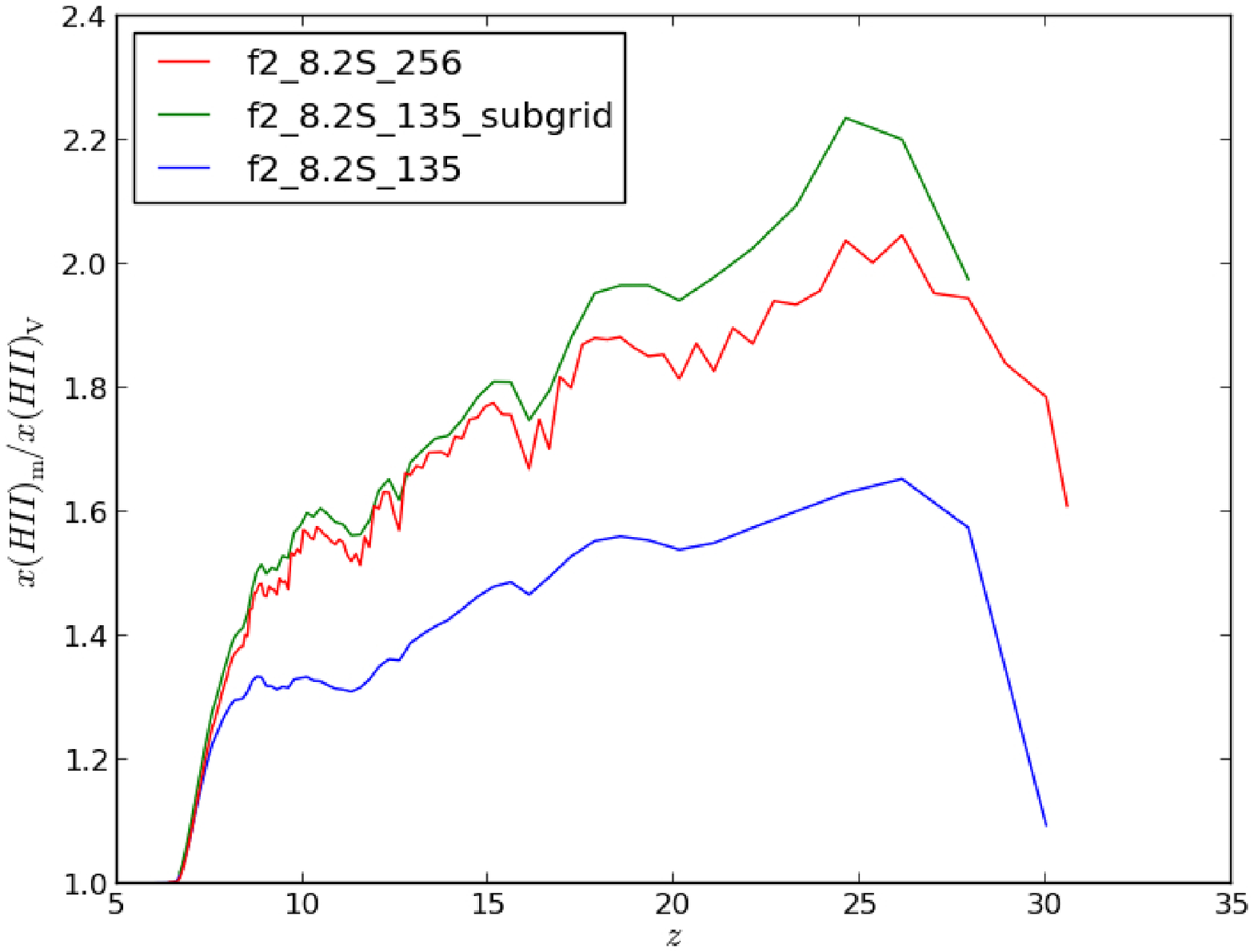}
  \end{center}
  \caption{Effect of resolution and sub-grid source modelling: 
power spectra of the ionized field, $\Delta_{xx}$, at mass-weighted ionized 
fraction of $x_m=0.1$ (top left) and $x_m=0.5$ (top right); late-time 
($z<12$) mass-weighted reionization history, $x_m(z)$ (bottom left) and 
ratio of the mass- and volume-weighted ionized fractions vs, redshift 
(bottom right) for the simulations L2 (f2\_8.2S\_256), L2lr1 (f2\_8.2S\_135),
and L2lr2 (f2\_8.2S\_135\_subgrid), as labelled.  
    \label{res:fig}}
\end{figure*}
We therefore find that, in agreement with our previous results, a bimodal 
H~II region distribution develops once the local percolation starts becoming
wide-spread, around $x_m\sim 0.3$. Ionized patches cannot surpass certain 
size before they merge into one very large H~II region which encompasses 
significant fraction of the total volume. While this quantitative behaviour 
is always observed, independent of the simulation volume, the maximum volume 
which isolated ionized patches can reach before percolation is dependent on 
the box size. They can reach up to $V\sim10^4-10^5\,\rm Mpc^3$ in the larger 
volume, but rarely surpass $V\sim10^3\,\rm Mpc^3$ in the smaller one (except 
around $x_m\sim0.3$, where some reach $V\sim3\times10^4\,\rm Mpc^3$, while in 
the larger volume they reach up to $V\gtrsim10^5\,\rm Mpc^3$ at the same 
epoch). The reason for this is that in the larger volume the voids (which 
due to the inside-out nature of reionization remain largely neutral for 
longer than do high-density regions) are considerably larger, thus allowing 
the H~II regions to grow further before merging. This has important 
implications for the redshifted 21-cm line fluctuations, which we will 
discuss below. We note that in neither case the value of this maximum volume 
of isolated regions reaches more than a fraction of a percent of the total 
volume and is thus unlikely to be affected by straightforward finite box 
effects.

We next turn our attention to the other, complementary measures of the H~II 
region size distribution we previously mentioned, namely the 3D power spectra 
of the ionization fraction field, shown in Figure~\ref{ps_x:fig}, and the SPA 
size distribution, shown in Figures~\ref{spa:fig} and \ref{spa2:fig}. During 
the initial phases of reionization ($x_m=0.1$) the 3D ionization field power 
spectra have similar shapes, uniformly-rising towards smaller scales without 
a clearly-identified characteristic scale. However, the larger volume 
yields up to 50\% more power than the smaller one at the (relatively small)
scales corresponding to the HII regions at that time, as well as some 
additional power at very large scales which are not present in the smaller 
box. Some of differences at the smallest scales is due to the different grid 
resolutions, however these scales fall below the resolution of most current 
observational experiments. As reionization proceeds, the local percolation 
of H~II regions starts and a characteristic peak scale for the power spectra 
develops in both volumes, albeit at slightly different scales and 
overall magnitude. By the time ionized fraction reaches 50\% the power 
spectra become essentially the same in shape over the common scales of the two
 simulations. However, the power spectrum derived from XL2 contains noticeably 
more power at all scales (and appears smoother) compared to L2.1. In both cases
the peak scale gradually moves towards larger scales, tracking the growth of
the ionized regions.

The SPA method yields similar results. At very small scales there is a 
sharp cutoff in the size distribution which is related to the grid 
resolution in each case and is thus a numerical effect, rather than a 
physical one. At all times there are more large H~II regions in XL2, 
which reach tens of Mpc in radius at late times. On the other hand, the 
number of intermediate-size ionized regions (radii from few Mpc to below 
one Mpc) is quite similar in the two boxes, except during the early 
evolution, before a characteristic scale develops. This again indicates 
consistency in the results at scales that are well below the size of the 
computational box and thus not affected by it. In contrast, the larger 
ionized patches are affected by the volume size and large-scale density 
fluctuations, and show to be considerably larger in the larger box. 

As we noted above, at late times all H~II regions percolate into one large,
connected region, thus at that epoch from point of view of the SPA method 
it makes better sense to study the {\it neutral} patch distribution instead, 
as those are the patchy regions which still keep their clear identity. Results 
for the epoch when the ionized fraction by mass is $x_m=0.9$ are shown in 
Figure~\ref{spa2:fig}. Once again, there are many more large neutral patches 
in the larger volume. Those also reach much larger sizes, up to $R\sim45\,$Mpc, 
while in the smaller volume they do not surpass $R=10\,$Mpc. In both cases 
there is a clear peak of the distribution, which is at $R\sim1-2\,$Mpc for 
L2.1, and at $R\sim15\,$Mpc for XL2. 
In the latter case the peak is less broad and thus the peak scale is 
better-defined. The grid resolution-related cutoff at small scales is still 
present, but is now at scales well below the peak scale and is therefore not 
the reason for it, i.e. the peak scale we observe is physical, not numerical. 
The presence of such large neutral patches at late times, not seen in 
smaller-volume simulations, is important because it means better detectability 
at the redshifted 21-cm and other probes. Most such probes rely on fluctuations 
between sky patches, which is clearly enhanced here. Furthermore, radio 
sensitivity is significantly better at lower redshifts/higher frequencies, and 
the low-frequency foregrounds are much weaker, making the late stages of 
reionization an important target for the first, likely statistical, detection, 
though some limited imaging might also be possible \citep{2012MNRAS.425.2964Z}.

\subsubsection{Effect of resolution and sub-grid modelling}

As discussed in \S~\ref{sect:sims}, there are differences in 
the numerical modelling in the two cases discussed above, which 
could potentially affect our results. First, the large-volume,
$425\,h^{-1}$Mpc radiative transfer simulation has lower grid 
resolution than the $114\,h^{-1}$Mpc simulation, by almost a 
factor of 2 per dimension. Furthermore, the low-mass, suppressible 
LMACH sources in the larger volume are modelled using sub-grid 
prescription, as their host haloes are not directly resolved in 
the corresponding N-body simulation. In order to evaluate the 
effect of these differences, we performed two additional radiative 
transfer simulations, both using the $114\,h^{-1}$Mpc volume and grid 
resolution reduced to $135^3$, corresponding to the spatial resolution 
of the large radiative transfer volume. The first of these simulations 
(which will be referred to as L2lr1) used the LMACHs directly based on 
the resolved haloes, as done in the higher-resolution production 
simulation, while the second one (L2lr2) used the sub-grid prescription 
to model the LMACH host haloes. Therefore, by directly comparing them to 
simulation L2 (f2\_8.2S\_256 in our previous, more expanded notation) and 
each other we can evaluate the effect of resolution (L2lr1, same as 
f2\_8.2S\_135 in our previous notation) and the sub-grid model (L2lr2, 
same as f2\_8.2S\_135\_subgrid). 

\begin{figure*}
  \begin{center}
    \includegraphics[width=2.2in]{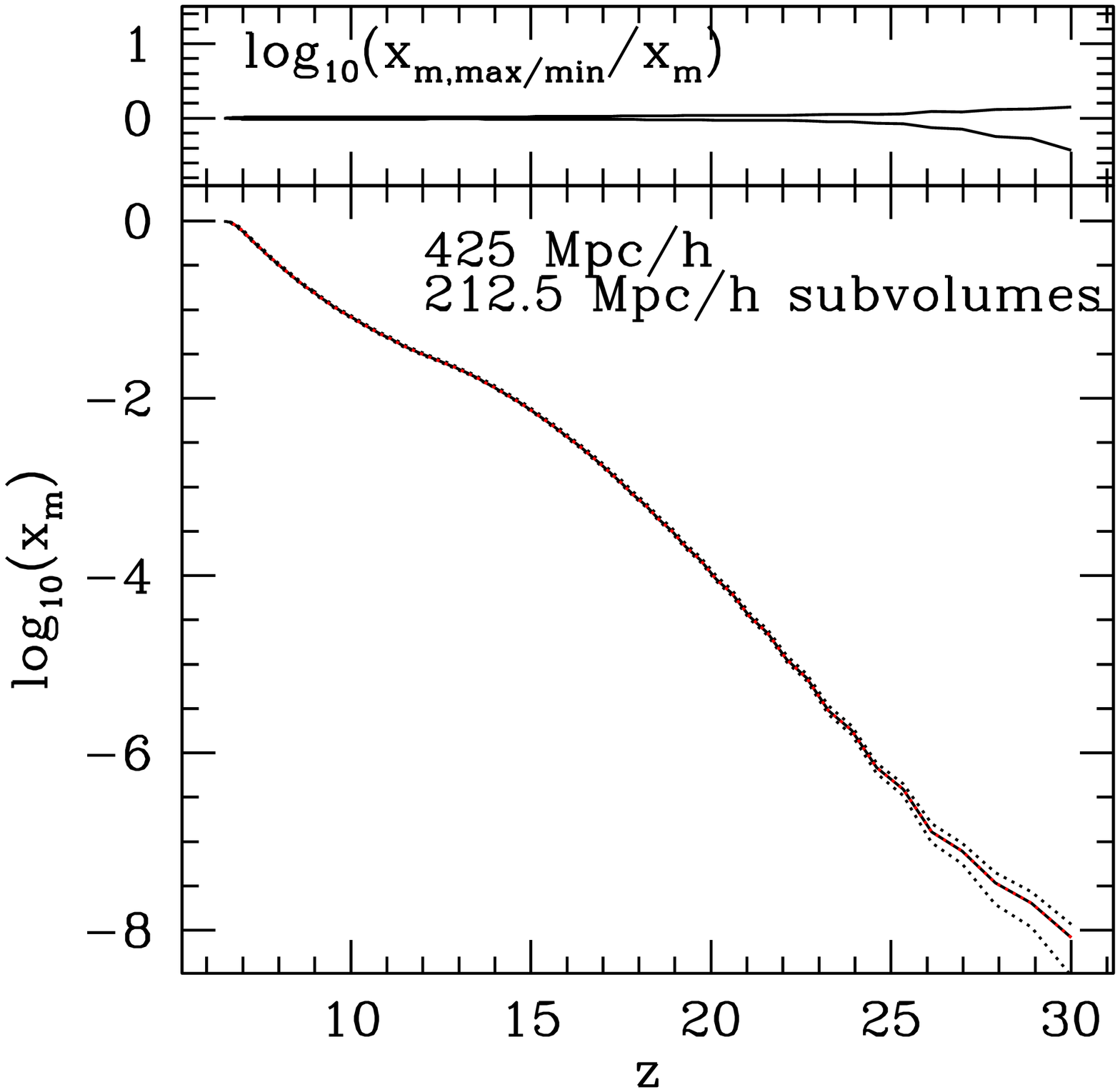}
    \includegraphics[width=2.2in]{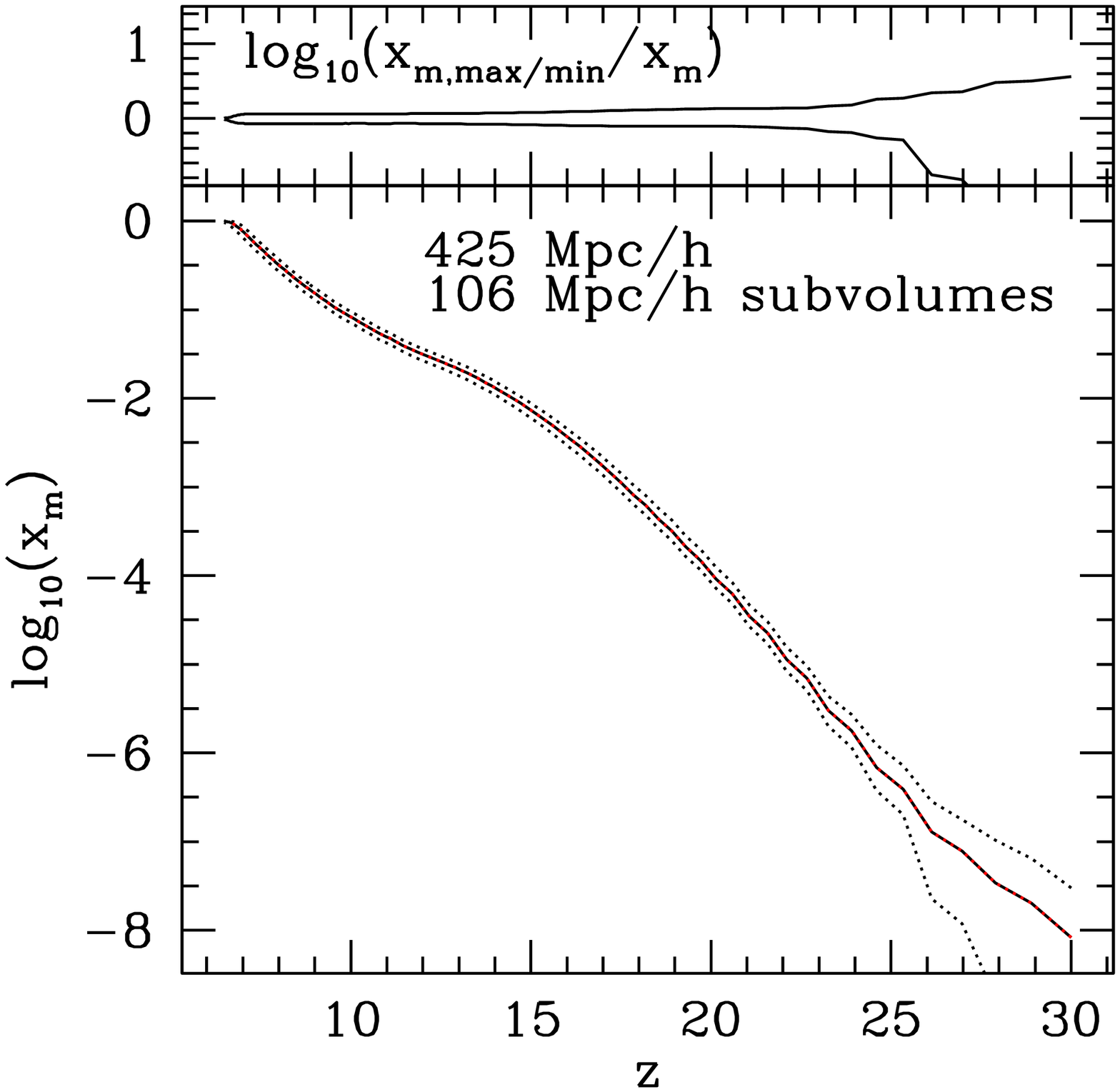}
    \includegraphics[width=2.2in]{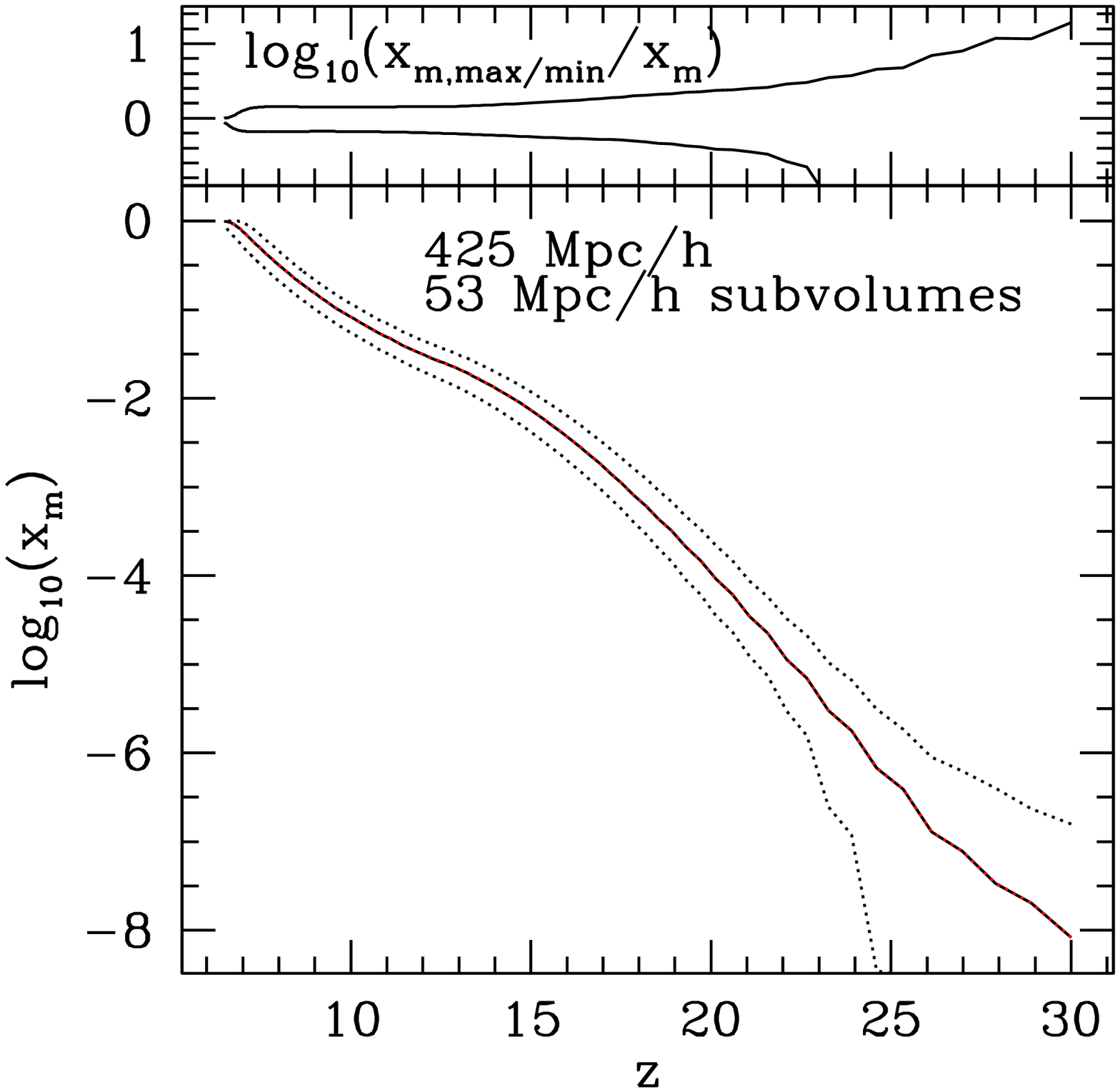}
  \end{center}
  \caption{(bottom panels) Reionization history variation in (non-overlapping) 
    subvolumes of a different size: (left) $212.5\,h^{-1}$Mpc;
    (middle) $106.25\,h^{-1}$Mpc; and (right) $53.125\,h^{-1}$Mpc.
    Shown are the mean, minimum and maximum of all sub-regions 
    (black) and the mean mass-weighted reionization history of 
    the full volume (red). (top panels) Ratios of the largest 
    and smallest sub-region mass-weighted ionized fraction and
    the mean one over the full volume. 
    \label{reion_hist:fig}}
\end{figure*}
When the resolution is decreased (L2 vs. L2lr1), one obvious effect is the
under-resolving of the smaller H~II regions, resulting in smoothing of the 
corresponding fluctuations and in less power on small scales. However, since 
the resolution length in both cases is sub-Mpc, this only affects patches 
smaller than a few Mpc, or roughly arcminute scales (Figure~\ref{res:fig},
top panels), which will not be 
resolved in the current generation of reionization experiments due to 
insufficient sensitivity. An additional, more subtle resolution effect is 
related to the way the LMACHs suppression is implemented -- based on an 
ionization threshold. Averaging the ionized fraction over larger cells means 
that a cell could remain under the threshold even if sub-volumes of it would 
be above it if resolved. This yields a lower level of suppression, particularly 
at the edges of ionized regions in the more coarsely resolved simulation. The 
result is a somewhat faster-proceeding reionization (Figure~\ref{res:fig},
bottom left panel). It also becomes less inside-out in character, with the 
mean mass-weighted ionized fraction closer to the volume one 
(Figure~\ref{res:fig}, bottom right panel). However, this has no significant 
effect on the reionization patchiness and the observable signatures. This is 
demonstrated by all available measures - FOF, SPA and power spectra, which at 
large scales are essentially the same in the two cases throughout the 
reionization history. The only exception is 
the FOF PDF distribution at late times ($x_m=0.7$), where the coarser-grained 
simulation yields a small fraction ($\sim10^{-3}$ of the total volume) of larger 
H~II regions which become topologically connected together due to the blurred 
boundaries between them due to the lower resolution. 

The LMACHs sub-grid modelling (L2lr2) is currently done deterministically -- 
following the mean density-halo number relation derived from higher-resolution 
simulations -- and does not take into account the scatter in that relation. 
This has the effect of concentrating the model LMACHs towards the higher density 
regions more than is the case for the directly-resolved LMACHs. Due to the 
inside-out nature of reionization, where the higher-density regions are 
preferentially ionized earlier, the model LMACHs are more strongly suppressed, 
which delays reionization slightly (Figure~\ref{res:fig}, bottom left panel). 
The result is that any given mean ionized fraction is reached at a later 
time and therefore the H~II regions manage to become slightly larger. This is 
especially noticeable during the early ($x_m=0.1-0.3$) and late ($x_m=0.7$) stages 
of the reionization process in the power spectra and SPA patchiness measures, 
while during the middle stages ($x_m=0.5$) this effect largely disappears
(Figure~\ref{res:fig}, top panels). In contrast, the FOF PDF distribution is 
only noticeably affected early on ($x_m=0.1$) and very little after that time.
Interestingly, the separate effects of lower resolution and sub-grid model on
the mass-weighted to volume-weighted ionized fraction evolution largely cancel
out (Figure~\ref{res:fig}, bottom right panel), indicating that when both effects
are present the inside-out ionization properties are very close to the fully
resolved simulation.

Overall, compared to the large-volume effects discussed in \S~\ref{sizes:sect}, 
the consequences of the lower resolution and sub-grid model on the reionization 
patchiness are much smaller and therefore do not invalidate any of our results.

\subsection{Convergence of the reionization history and observables
with simulation volume}
\label{sect:conv_w_volume}

\subsubsection{Reionization history}

As we showed above, the mean reionization history is already well converged
for $\sim100$~Mpc volumes \citep[or even for ones of few tens of Mpc across, 
see ][]{2006MNRAS.369.1625I}. However, locally for smaller patches it does 
vary significantly from patch to patch. In order to better quantify the 
convergence of the reionization history with the sample volume size, we 
split our XL2 simulation volume into a number of (non-overlapping) volumes 
of different sizes and then calculate their individual reionization histories, 
i.e. mean mass-weighted ionized fraction, $x_m$ vs. redshift. In 
Figure~\ref{reion_hist:fig} we show the reionization history averaged over 
all sub-volumes, as well as the maximum and minimum subvolume ionized fraction 
at each redshift. Neither the maximum nor the minimum value necessarily belong 
to the same sub-volume at different times, thus the minimum and maximum lines 
are in in fact enveloping the full variety of reionization histories which 
occur in the sub-volumes. For reference and easier direct comparison we also 
show the mean reionization history of our full volume discussed in the 
previous section.

\begin{figure*}
  \begin{center}
    \includegraphics[width=2.2in]{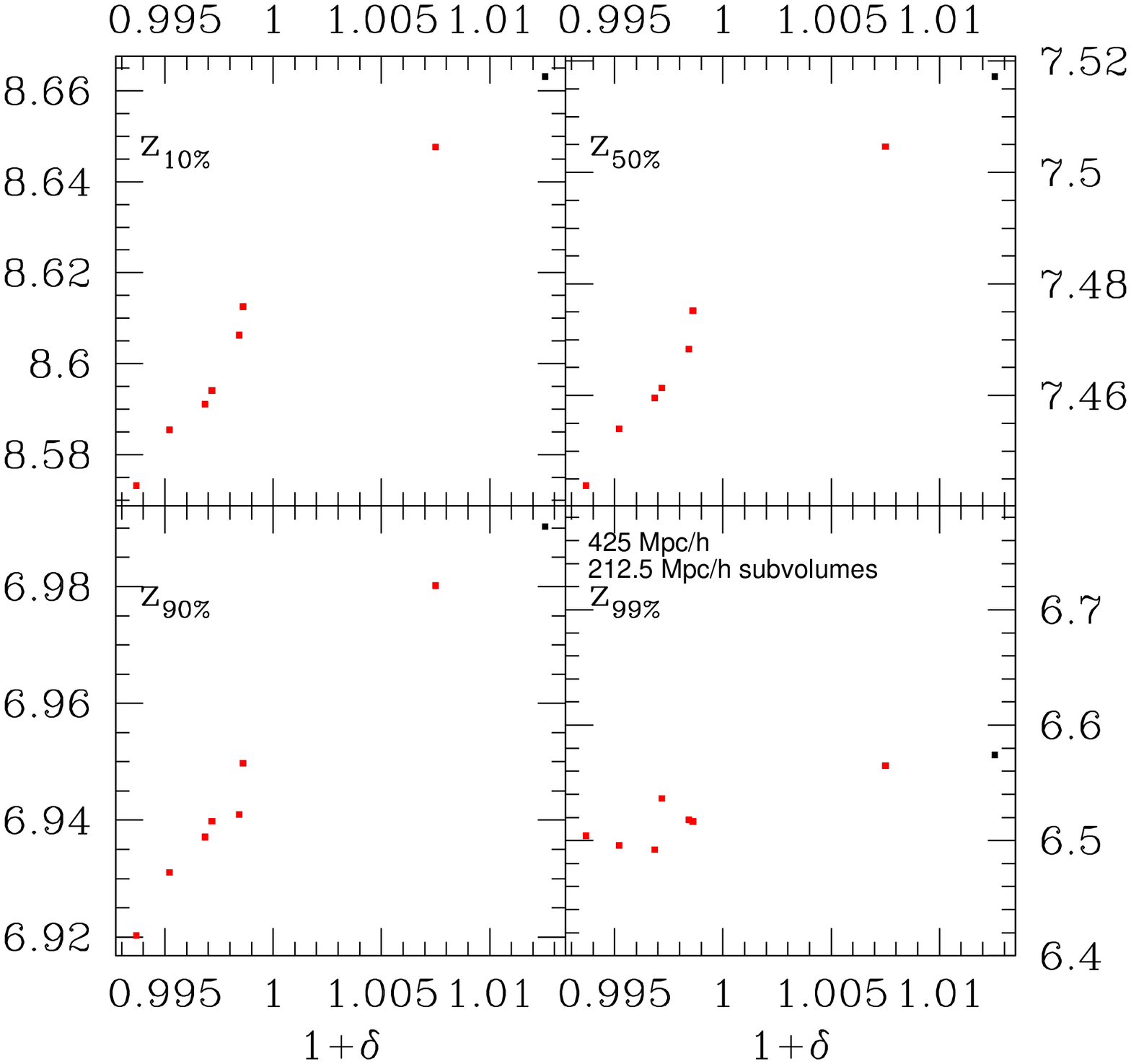}
    \includegraphics[width=2.2in]{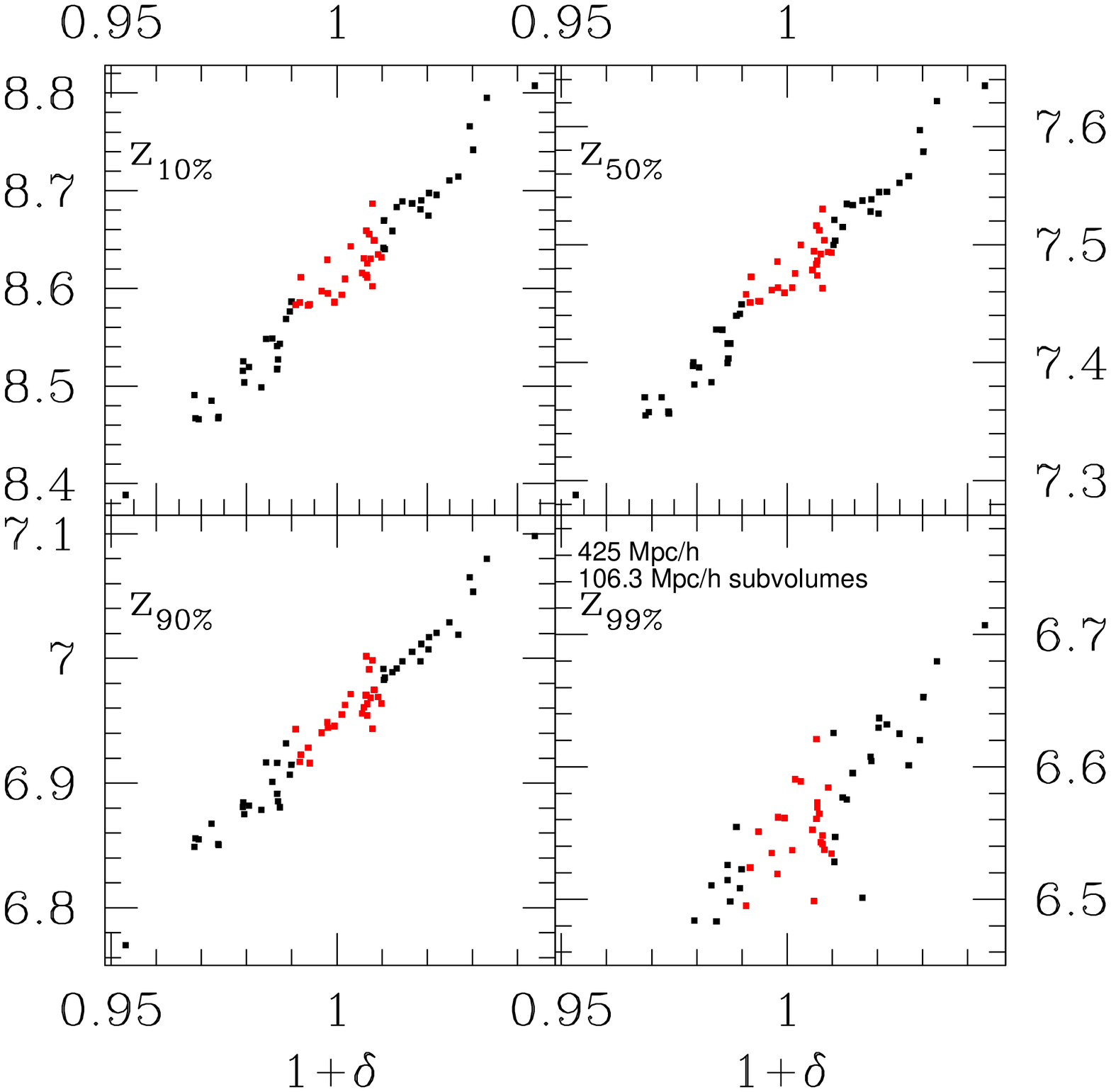}
    \includegraphics[width=2.2in]{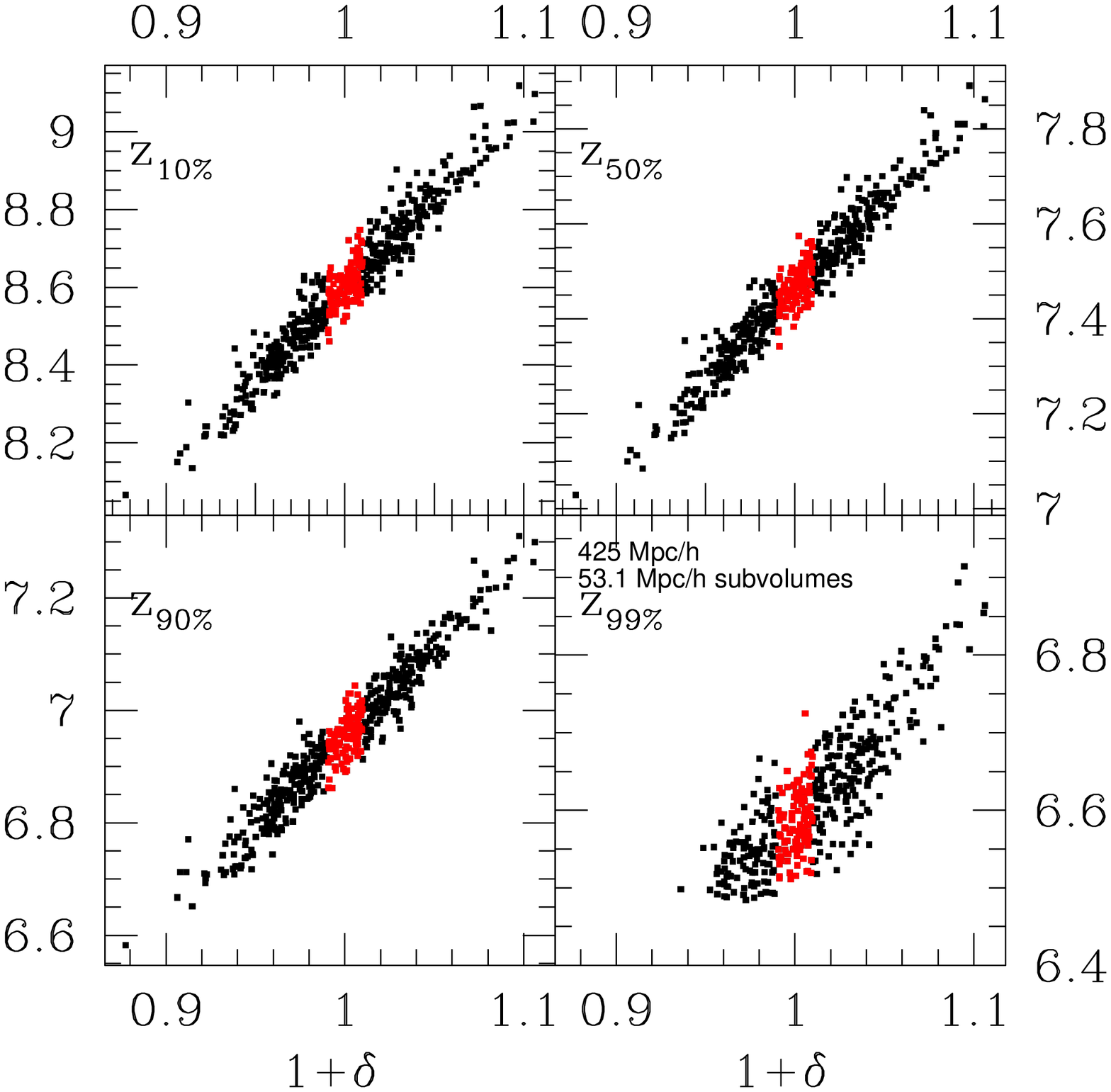}
  \end{center}
  \caption{Redshift at which certain reionization stages are reached: 
    $x_m=0.1$ ($z_{10\%}$),
    $x_m=0.5$ ($z_{50\%}$), $x_m=0.9$ ($z_{90\%}$), and $x_m=0.99$ 
    ($z_{99\%}=z_{ov}$) vs. $1+\delta=\rho_{\rm sub-region}/\rho_{\rm global\, mean}$,
    the average density of that sub-region in units of the global mean 
    for different size sub-regions: (left) $212.5\,h^{-1}$Mpc; (middle) 
    $106.25\,h^{-1}$Mpc; and (right) $53.125\,h^{-1}$Mpc. The regions with 
    roughly mean density (within 1\% of the mean) are shown in red.
    \label{reion_milestones:fig}}
\end{figure*}

We first note that for all subvolume sizes considered here the mean reionization 
history of the sub-volumes is essentially identical to the average one for 
the full volume. This statement would of course be identically true for the 
volume-based reionization histories, however it is easy to show that in general 
this is not the case for the averaged mass-weighted histories, particularly when 
the mean densities of the sub-volumes vary significantly.  

On the other hand, there is a clear variation in the reionization histories 
of different sub-volumes. Naturally, the spread is strongest at early times 
and smaller sub-volumes. For sub-volumes of $212.5\,h^{-1}$Mpc on a side we 
observe clear convergence to the mean evolution, apart from modest variation 
at the earliest times, $z>25$. The convergence worsens somewhat, but still 
remains reasonable for sub-region size of $106.25\,h^{-1}$Mpc, which size is 
close to the usually-assumed convergence value of $\sim100\,h^{-1}$Mpc 
\citep[cf.][]{2004ApJ...609..474B}, however there are now clear, albeit 
relatively small deviations from the mean reionization history. The full 
variation range in ionized fraction among the sub-regions is larger than 10 
at the earliest times, $z>27$, when the cosmic variance due to Poisson noise 
is strongest. As the reionization progresses this ratio drops to a value of 
order 2 and finally slowly approaches the mean reionization history at late 
times, although the variations remain present at low level all the way to 
the end of reionization. 

For sub-region size of $53.125\,h^{-1}$Mpc (which is a volume similar to or 
even larger than the one adopted by many reionization simulations) the 
reionization history variations become very considerable. This is a consequence 
of the significant differences in the mean densities of the sub-regions, which 
surpasses 20\% at late times, which is then reflected exponentially in the local 
abundance of ionizing sources, and to a lesser extend in the local recombinations. 
At the earliest epochs, $z>25$, some regions have ionized fractions as much as 10 
times larger than the mean, while others still remain fully neutral. These 
variations decrease over time, but nonetheless stay significant, with a ratio 
between the extreme cases of order a few at all redshifts all the way up to the 
end of reionization. The completion of reionization in each sub-region also 
varies accordingly. 

The integrated electron scattering optical depth corresponding to each 
subvolume's reionization history is tightly correlated to the mean density
of that volume. For the $53\,h^{-1}$Mpc it varies significantly around the 
mean value corresponding to the full simulation volume, ranging between 
$\tau=0.0518$ and 0.0611. For the $106\,h^{-1}$Mpc sub-volumes this variation
decreases and is between $\tau=0.0544$ and 0.0583. For the $212\,h^{-1}$Mpc 
sub-regions the optical depth essentially converges to the mean, ranging 
between $\tau=0.0562$ and 0.0570. The optical depth being an integrated, 
line-of-sight quantity, these values are mostly of theoretical interest. 
Of more practical importance are the variations of the optical depth for 
mean-density sub-volumes, as these are analogous to small-box reionization 
simulations which typically take their average density to be the mean one for 
the universe. For mean-density sub-volumes the optical depth variations are 
small, with a range of $\pm0.001$ for the $53\,h^{-1}$Mpc sub-volumes and a 
negligibly small one for the larger sub-volumes. This indicates that even
relatively small-box simulations could be reliably used for calculating the 
optical depth for a given reionization scenario. We however note that at 
still smaller smaller scales, below 10 Mpc (few arcminutes) the local 
electron scattering optical depth still varies significantly. It is highly 
correlated with the ionization field and anticorrelated with the neutral one 
\citep{pol21} and can contribute strongly to the CMB polarization at arcminute 
scales \citep{cmbpol}.


\subsubsection{Ionization fraction - density correlation}

In order to further quantify the variations of the reionization history 
between the sub-regions and its dependence on the local density, in 
Figure~\ref{reion_milestones:fig} we show a scatter plot (each point 
represents a non-overlapping sub-region) of the the correlation between
the redshift at which certain reionization milestone is reached vs. the
average density (with respect to the global mean) of that sub-region at
the end of reionization. We have chosen as milestones the epochs when ionized 
fractions $x_m=0.1,0.5, 0.9$ and 0.99 are reached, corresponding to early, 
middle, late and final reionization stages, respectively, and the same 
three sub-volume sizes as in the previous section. 
  
The $212.5\,h^{-1}$Mpc sub-regions (left panels) correspond to a small 
range of densities, all within 1.5\% of the mean and the maximum density 
variation is a little above 1\%. This yields a similarly small variation 
of $\sim 1\%$ in the redshifts at which each reionization stage is reached. 
The correlation between redshift and $1+\delta$ is roughly linear. Only one 
of the eight sub-regions deviates from the mean by more than 1\% and just 
two of them are overdense. Accordingly, those two regions reach each of the
reionization stages noticeably earlier than the rest, by $\Delta z\sim0.04$. 

Similar qualitative trends are observed for the smaller-size sub-regions.
The correlation is still roughly linear and fairly tight in all cases, i.e. 
higher-density regions generally ionize earlier than low-density ones, as
expected. However, the smaller the sub-regions are, the larger the variations 
among them. The sub-region densities vary between $1+\delta=0.95$ and 1.05 for 
$106.25\,h^{-1}$Mpc sub-regions and $1+\delta=0.9-1.1$ for $53.125\,h^{-1}$Mpc
ones. Consequently, the redshifts at which each reionization stage is reached
vary considerably, by as much as $\Delta z=\pm0.2$ and $\Delta z=\pm0.5$, 
respectively. Interestingly, there is a significant scatter even among
sub-regions with the same or similar densities. For sub-regions with density
close to the mean one and size of $106.5$~Mpc we have a maximum variation 
of $\Delta z\sim\pm0.1$, increasing to $\Delta z\sim\pm0.2$ for $53.125$~Mpc 
regions. Therefore, the redshift of end of reionization (or any of the other 
EoR milestones) 
cannot be determined with a good precision based on a small-box simulation even 
for fixed photon production efficiencies. This uncertainty becomes much worse 
for very small volumes, which are therefore quite inappropriate for modelling 
even of the mean ionization history, let alone the EoR patchiness.   

\subsubsection{21-cm emission: power spectra, rms and non-Gaussianity}

\begin{figure}
  \begin{center}
    \includegraphics[width=3.2in]{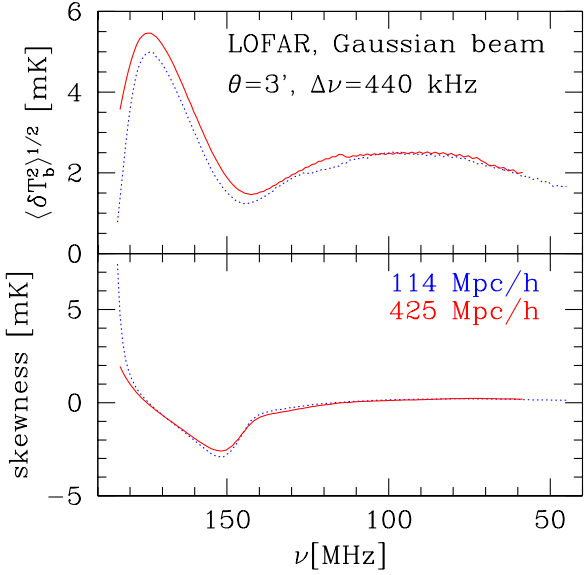}
  \end{center}
  \caption{(top) The evolution of the rms of the 21-cm fluctuations
    smoothed with beam and bandwidth roughly as expected for the LOFAR
    EoR experiment; (bottom) evolution of the skewness of the 21-cm PDFs 
    using the same smoothing. Both quantities are shown for XL2 (red) and 
    L2.1 (blue).
    \label{21cm-rms:fig}}
\end{figure}

We now turn our attention to some of the observable signatures of cosmic 
reionization based on the redshifted 21-cm emission from neutral hydrogen. 
The differential brightness temperature of the redshifted 21-cm emission 
with respect to the CMB is determined by the density of neutral hydrogen, 
$\rho_{\rm HI}$, and its spin temperature, $T_{\rm S}$ and is given by 
\ba
 \delta T_b&=&\frac{T_{\rm S} - T_{\rm CMB}}{1+z}(1-e^{-\tau})\nonumber\\
&\approx&
\frac{T_{\rm S} - T_{\rm CMB}}{1+z}
\frac{3\lambda_0^3A_{10}T_*n_{HI}(z)}{32\pi T_S H(z)}\label{temp21cm}
\\
&=&{28.5\,\rm mK}\left(\frac{1+z}{10}\right)^{1/2}(1+\delta)x_{HI}
\left(\frac{\Omega_b}{0.042}\frac{h}{0.73}\right)
\left(\frac{0.24}{\Omega_m}\right)^{1/2}
\nonumber
\ea
\citep{1959ApJ...129..536F}, where $z$ is the redshift, $T_{\rm CMB}$ 
is the temperature of the CMB radiation at that redshift, $\tau$ is 
the corresponding 21-cm optical depth, assumed to be small when 
writing equation~\ref{temp21cm}, $\lambda_0=21.16$~cm is the 
rest-frame wavelength of the line, $A_{10}=2.85\times10^{-15}\,\rm s^{-1}$
 is the Einstein A-coefficient, $T_*=0.068$~K corresponds to the 
energy difference between the two hyperfine levels, 
$x_{HI}(1+\delta)={n_{\rm HI}}/{ \langle n_H \rangle}$ is the mean number 
density of neutral hydrogen in units of the mean number density 
of hydrogen at redshift $z$, 
\ba
\langle n_H \rangle(z)&=&
\frac{\Omega_b\rho_{\rm crit,0}}{\mu_Hm_p}(1+z)^3\nonumber\\
&=&1.909\times10^{-7}\rm cm^{-3}\left(\frac{\Omega_b}{0.042}\right)
(1+z)^3,
\ea
with $\mu_H=1.32$ the corresponding mean molecular weight for neutral
gas of primordial composition (assuming 24\% He abundance), and $H(z)$ 
is the redshift-dependent Hubble constant,
\ba
  H(z)&=&
H_0[\Omega_{\rm m}(1+z)^3+\Omega_{\rm k}(1+z)^2+\Omega_\Lambda]^{1/2}
                      \nonumber\\  
&=&H_0E(z)\approx H_0\Omega_{\rm m}^{1/2}(1+z)^{3/2},
\ea
where $H_0$ is its value at present, and the last approximation 
is valid for $z\gg 1$. Throughout this work we assume that 
$T_{\rm S} \gg T_{\rm CMB}$ i.e. that there is sufficient Ly-$\alpha$ 
background to completely decouple the 21-cm transition from the CMB 
and that the neutral gas is heated well above the CMB temperature 
(due to e.g. a small amount of X-ray heating). Under these conditions 
the 21-cm line is seen in emission. These assumptions are generally 
well-justified, except possibly during the very early times
\citep[see e.g.][and references therein]{2006PhR...433..181F}. 

\begin{figure*}
  \begin{center}
    \includegraphics[width=2.2in]{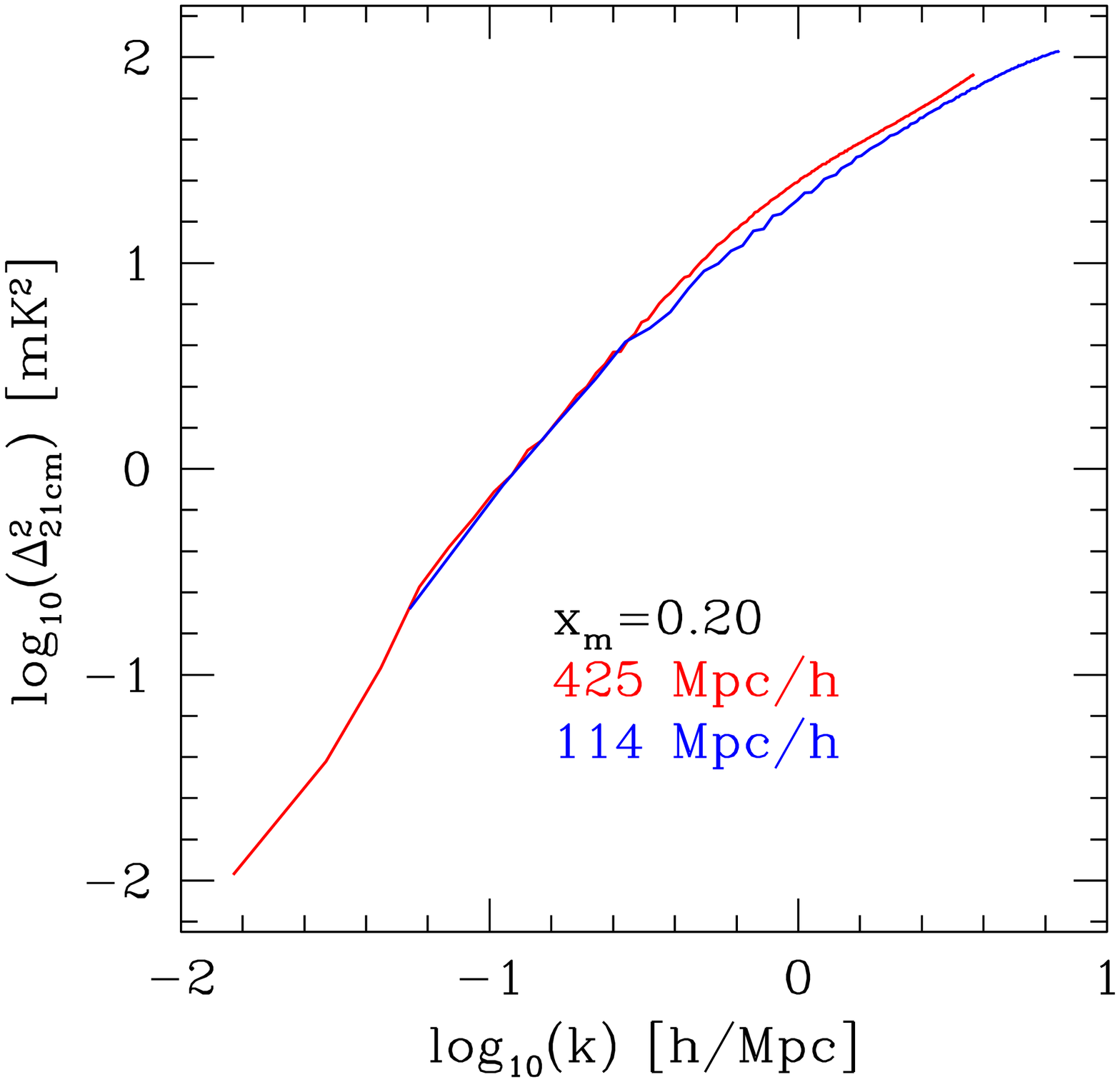}
    \includegraphics[width=2.15in]{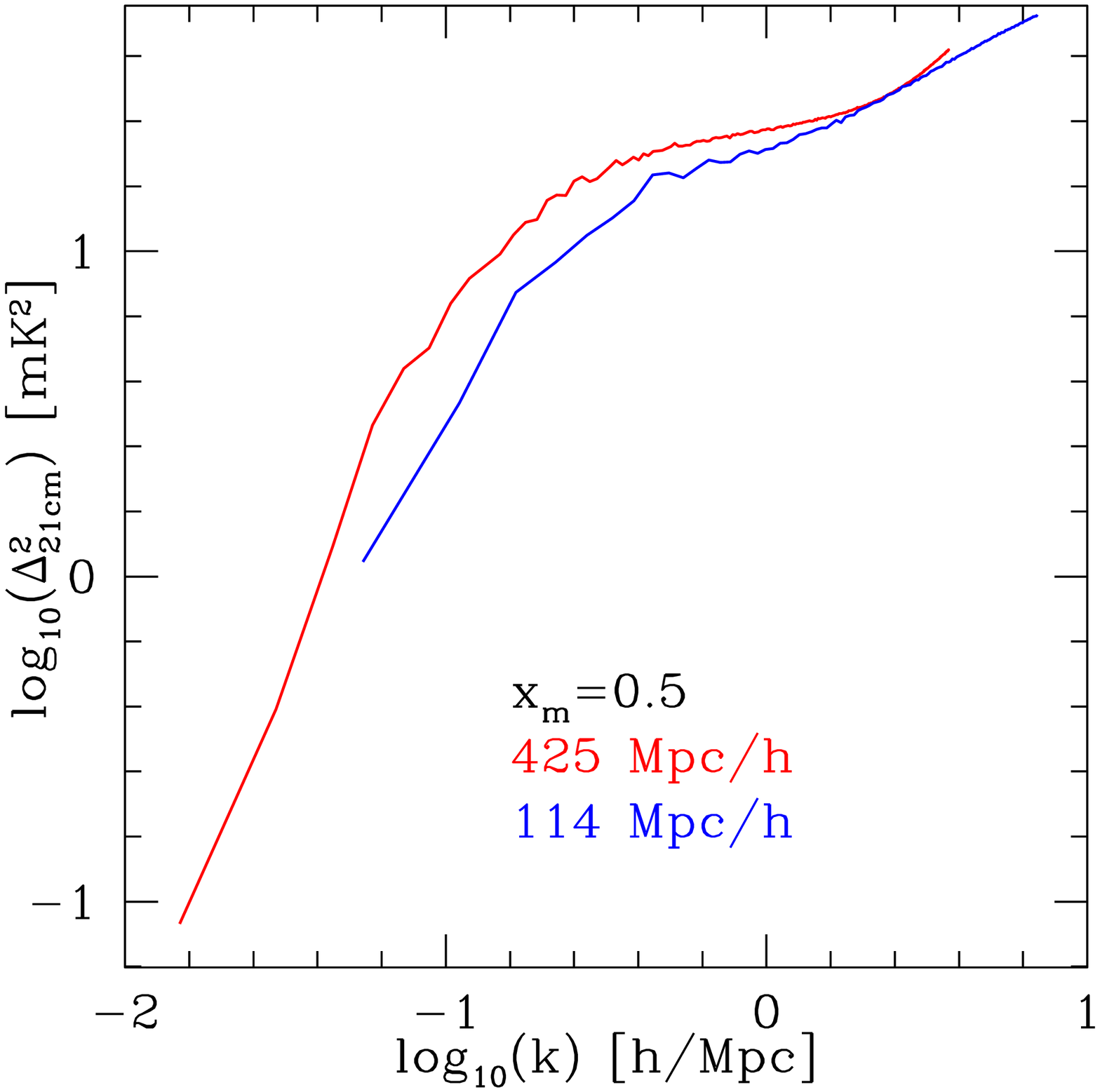}
    \includegraphics[width=2.2in]{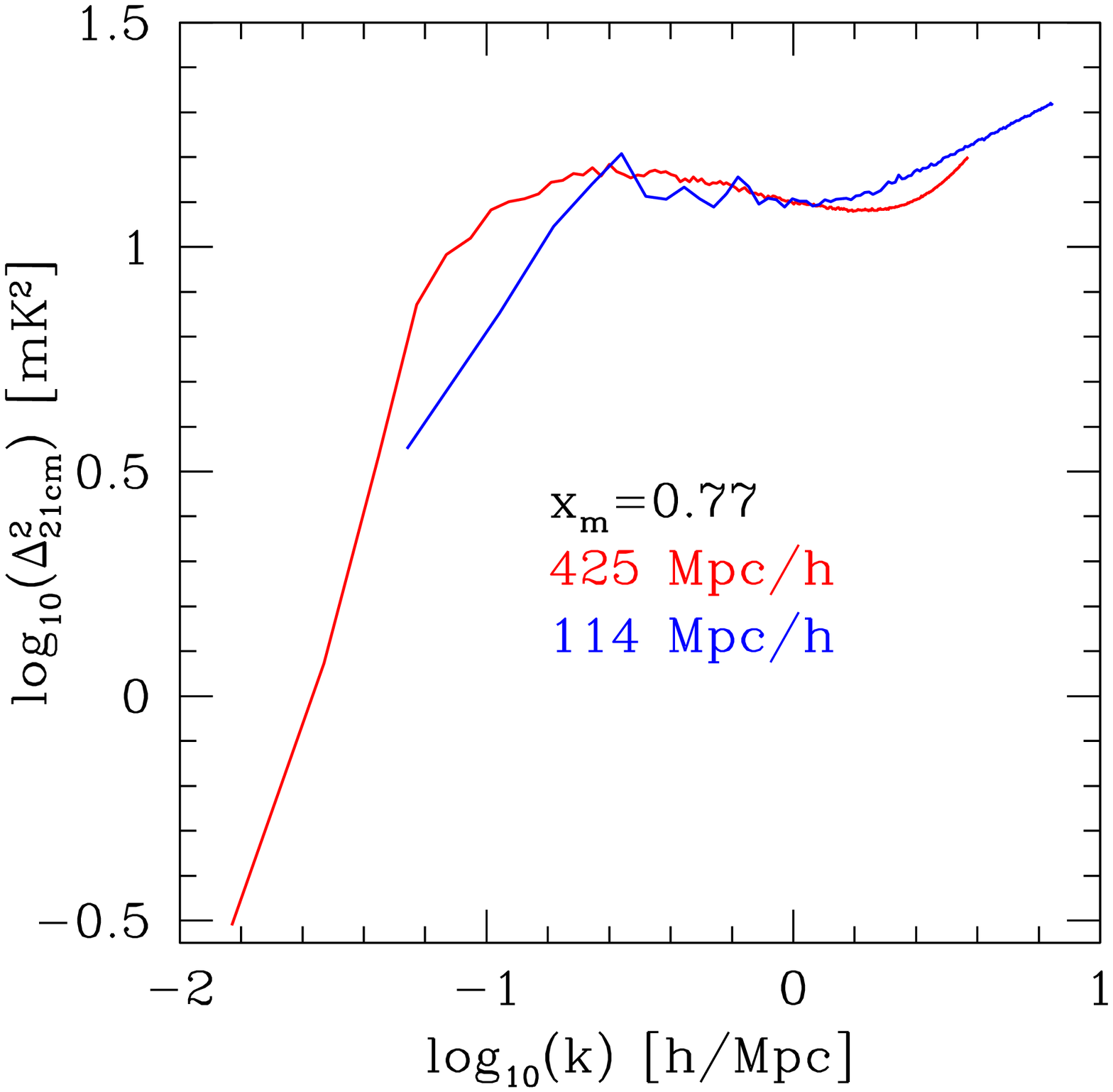}
  \end{center}
  \caption{21-cm differential brightness temperature fluctuation power 
    spectra for the XL2 (red) and L2.1 (blue) at different stages of 
    reionization: (left) $x_m=0.2$, (middle) $x_m=0.5$, and (right) $x_m=0.77$.
    \label{21cm-power:fig}}
\end{figure*}

\begin{figure*}
  \begin{center}
    \includegraphics[width=2.2in]{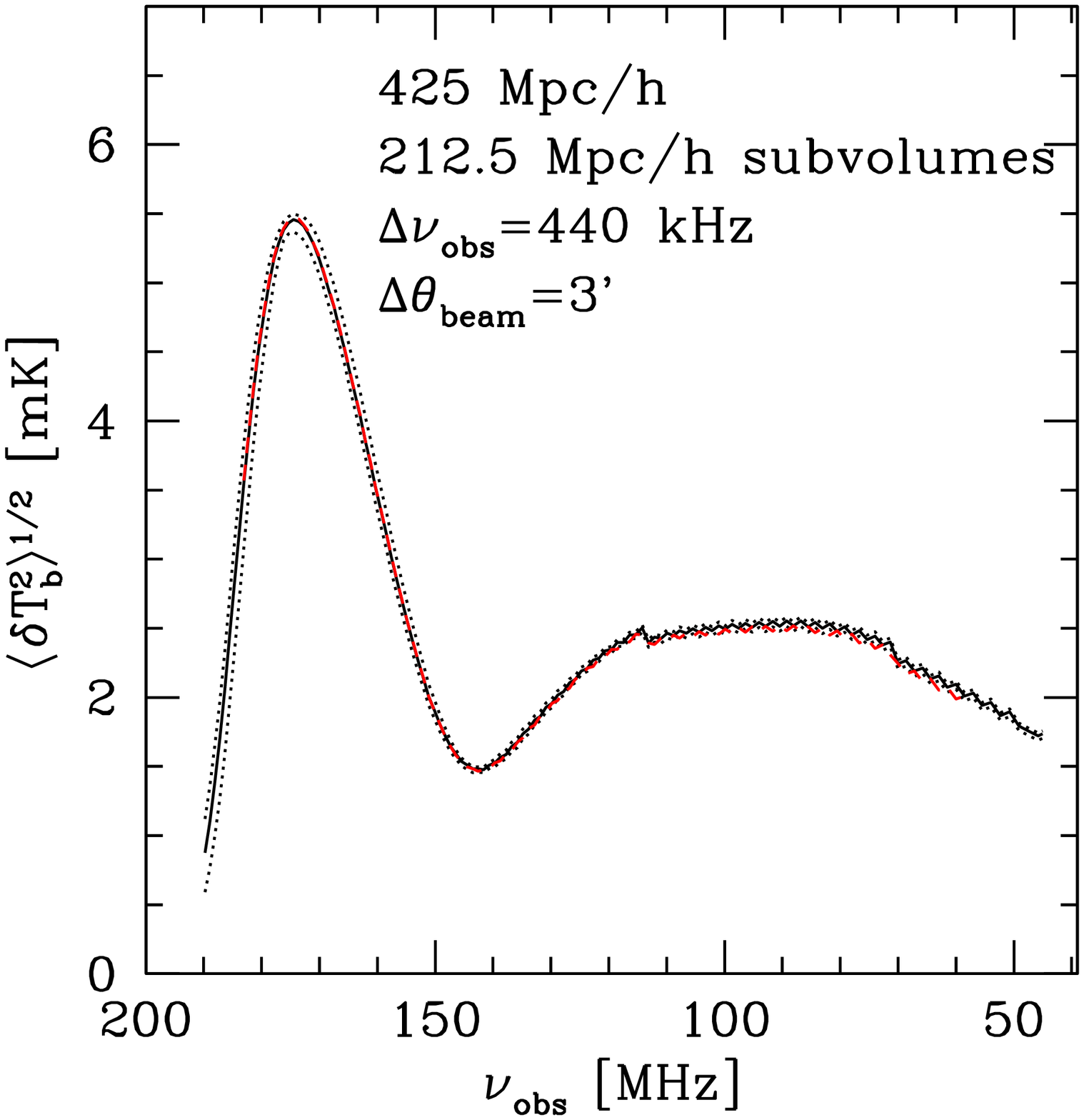}
    \includegraphics[width=2.2in]{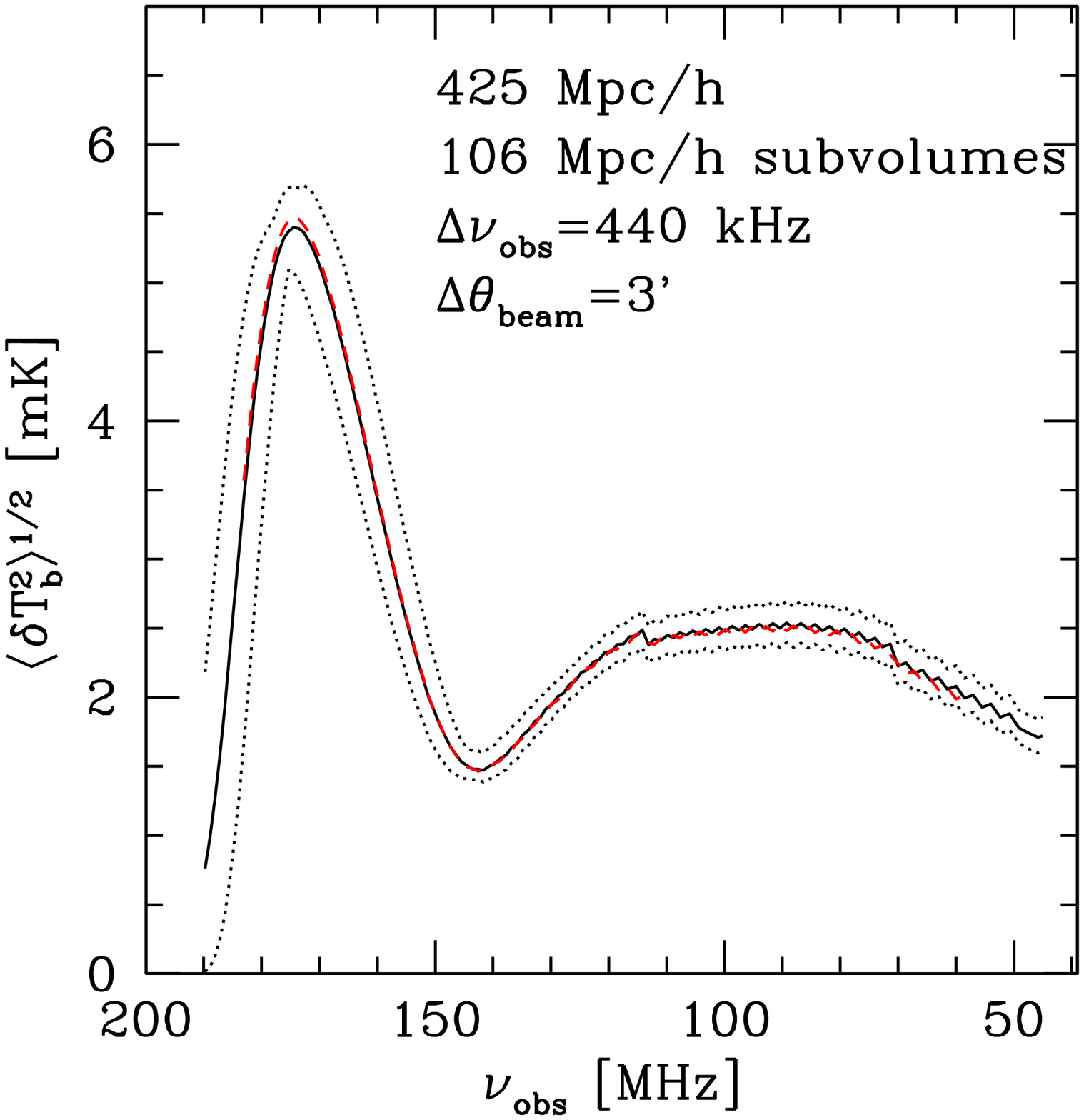}
    \includegraphics[width=2.2in]{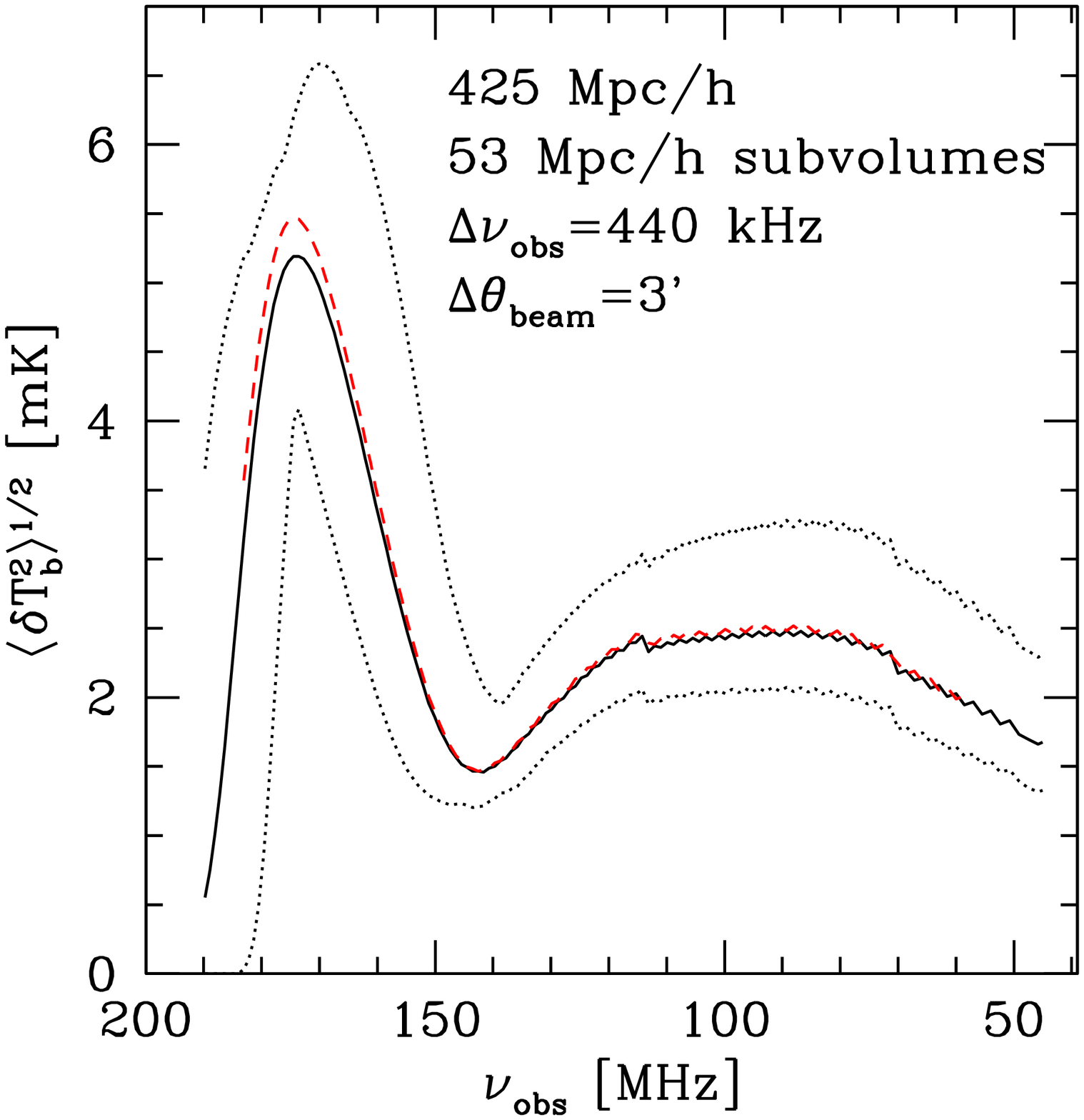}
  \end{center}
  \caption{21-cm differential brightness temperature rms variations, 
smoothed with LOFAR-like beam and bandwidth, between sub-regions of 
XL2 with a given size, as labelled. Shown are the mean of all sub-volumes 
(solid, black) and the minimum and maximum lines enveloping the variations 
between all sub-regions of that size. The average over the full volume is 
also indicated (red).
    \label{dtrms_scatter:fig}}
\end{figure*}

In Figure~\ref{21cm-rms:fig} we show the 21-cm differential brightness 
temperature fluctuations rms and skewness of the PDF distribution smoothed
with a Gaussian beam of size 3' and bandwidth of $\Delta\nu=440$~kHz. 
Comparing the results from our two simulation volumes, XL2 and L2.1, 
we see that during the early evolution ($\nu_{\rm obs}<120\,$MHz) the 
ionized regions are still few and small and thus the 21-cm fluctuations 
are mostly due to the underlying density fluctuations and track each other 
closely. However, once the patchiness becomes significant due to the 
growth of the H~II regions ($\nu_{\rm obs}>120\,$MHz), the XL2 simulation 
yields noticeably higher fluctuations. The largest rms fluctuations 
occur at the same epoch in the two cases, at $\sim127\,$MHz, but with a 
peak amplitude larger by about 10\% in the XL2 case. At late times the
difference grows larger, reaching a factor of $\sim2.5$ at overlap. This
is important since such stronger signal at late times/higher frequencies 
significantly improves the chances of detection, given the lower foregrounds 
at these higher frequencies and the better sensitivity of the radio 
interferometers there.

On the other hand, the skewness -- a measure of the departure of the 21-cm 
beam- and bandwidth-smoothed PDF distribution from a Gaussian one -- is 
largely identical in the two cases, indicating the insensitivity of this
measure to the simulation volume (Figure~\ref{21cm-rms:fig}, bottom panel). 
In both cases the skewness starts close to zero during the initial, density 
fluctuation-dominated phase, and then exhibit a notable dip in the skewness 
to negative values reaching $-3\,\rm mK$ and coinciding with the initial 
stages of the 21-cm rms steep rise due to EoR patchiness. The numerical grid 
resolution does not matter in either of the two cases since resolution elements 
are much smaller than the beam/bandwidth smoothing scale.  

The power spectra of the 21-cm fluctuation fields \citep[including the 
redshift-space distortions, for the calculation methodology see][]{Mao12} 
are shown in Figure~\ref{21cm-power:fig}. Initially, (shown is the epoch 
at which the ionized fraction by mass is $x_m=0.2$) they closely track 
each other within their scales of overlap. Interestingly, this is 
in contrast to the significant boost to the dimensionless power spectra 
of the ionization field in Figure~\ref{ps_x:fig}, which is seen at small 
scales even during the early evolution. This apparent discrepancy is 
explained by the strong dominance of the early 21-cm fluctuations by the 
density auto-correlation and the density-ionized fraction cross-correlation 
terms as compared to the ionized fraction auto-correlation term $\Delta_x$. 
Because these dominant terms are similar in the two simulations, the
resulting 21-cm power spectra are also similar. However, again in 
agreement with the 21-cm rms results above, during the middle ($x_m=0.5$) 
and late ($x_m=0.77$) stages of reionization in simulation XL2 there is 
considerably more power at large scales ($k<1\,\rm h/$Mpc for $x_m=0.5$, 
with the scale of departure between the two increasing to $k<0.3\,\rm h/$Mpc 
for $x_m=0.77$ as the H~II regions grow in size). Therefore, although the 
ionized regions are larger in XL2 at all times, as was shown in 
\ref{sizes:sect}, this is noticeable in the 21-cm rms and power spectra only 
when the patchiness term starts to dominate the total power.

\begin{figure*}
  \begin{center}
    \includegraphics[width=2.2in]{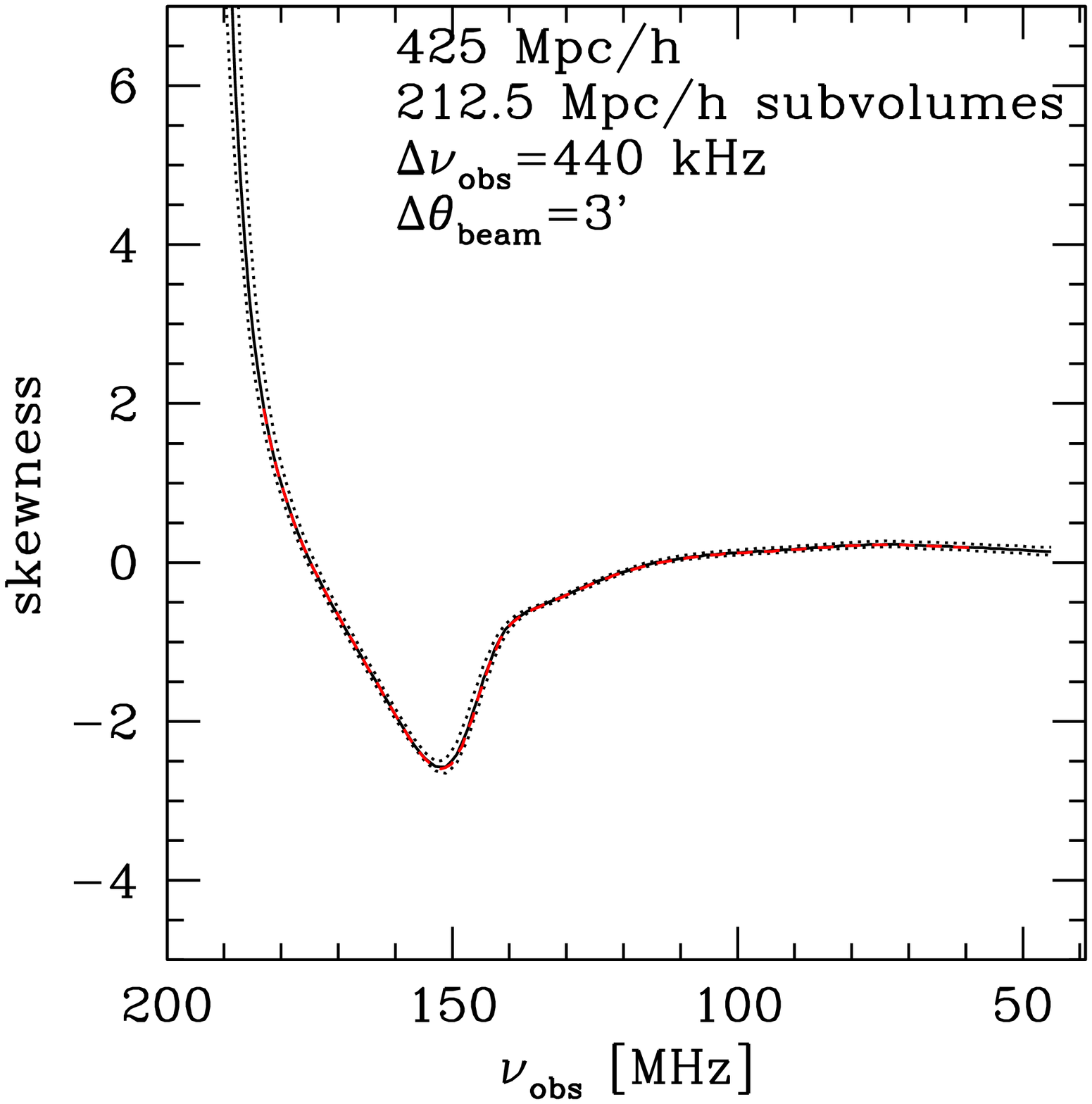}
    \includegraphics[width=2.2in]{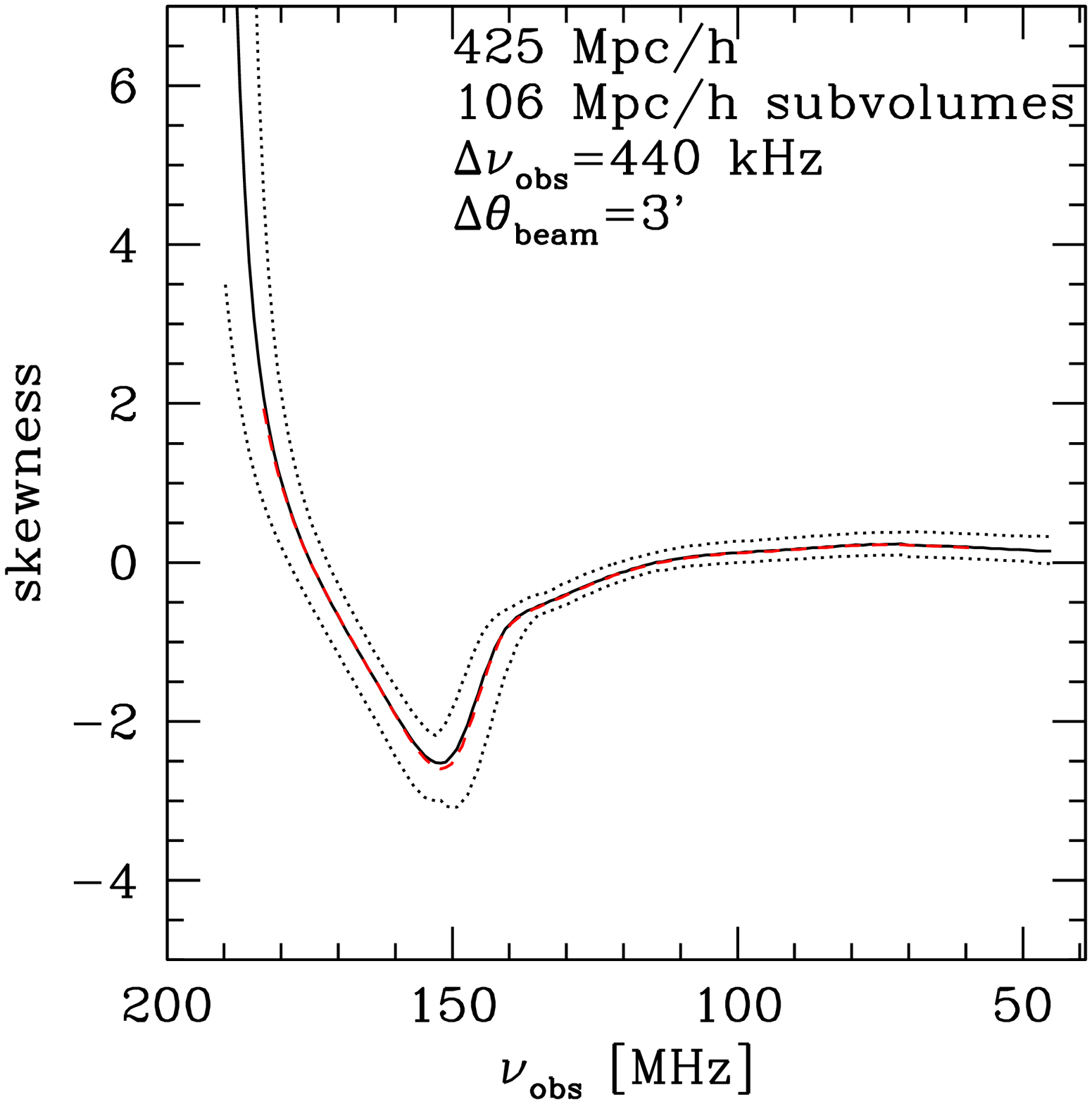}
    \includegraphics[width=2.2in]{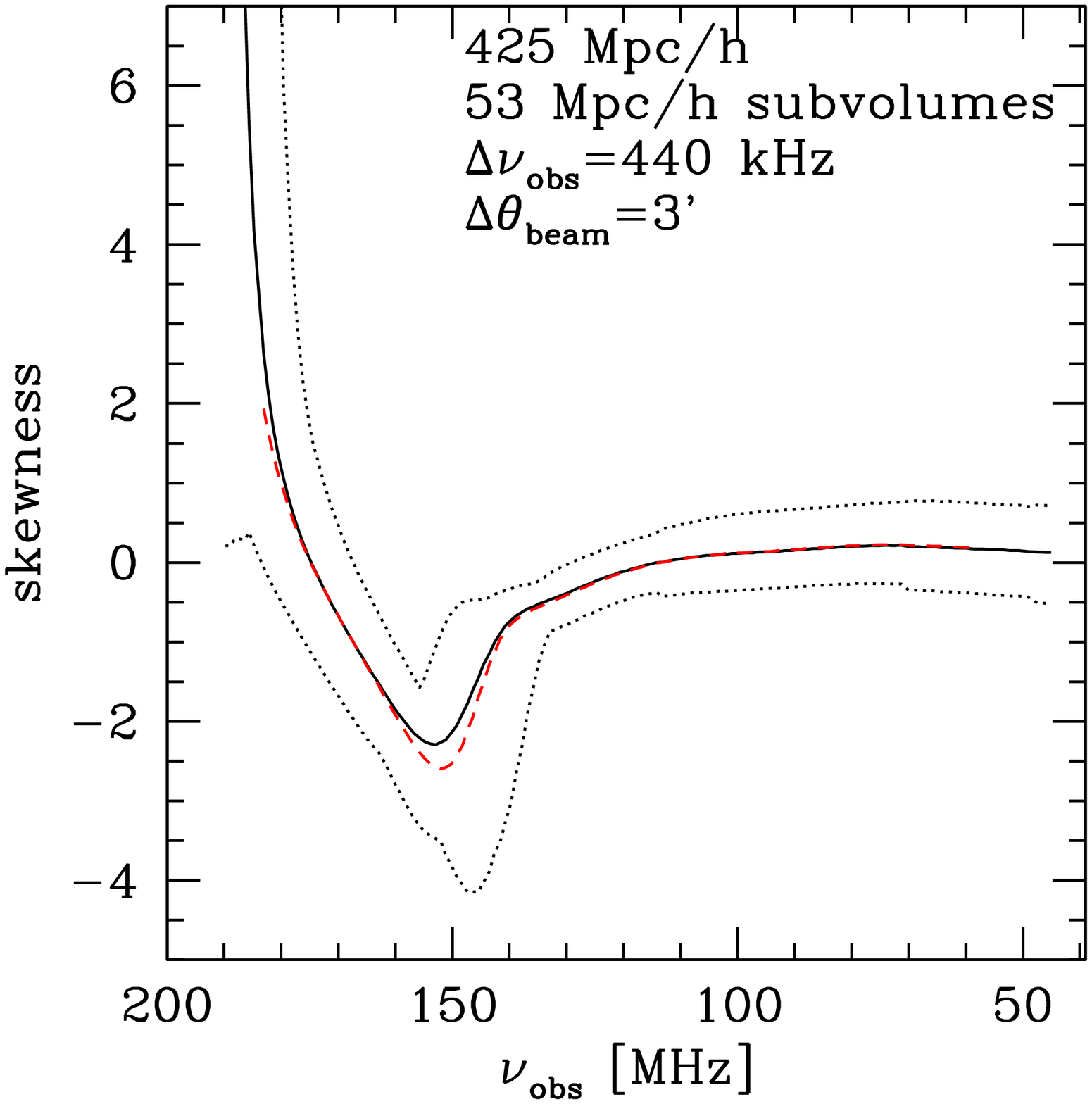}
  \end{center}
  \caption{21-cm differential brightness temperature field skewness 
variations in mK, smoothed with LOFAR-like beam and bandwidth, between 
sub-regions of XL2 with a given size, as labelled. Shown are the mean of 
all sub-volumes (solid, black) and the minimum and maximum lines enveloping
the variations between all sub-regions of that size. The average over 
the full volume is also indicated (red). 
    \label{dt_skew_scatter:fig}}
\end{figure*}

\begin{figure*}
  \begin{center}
    \includegraphics[width=2.2in]{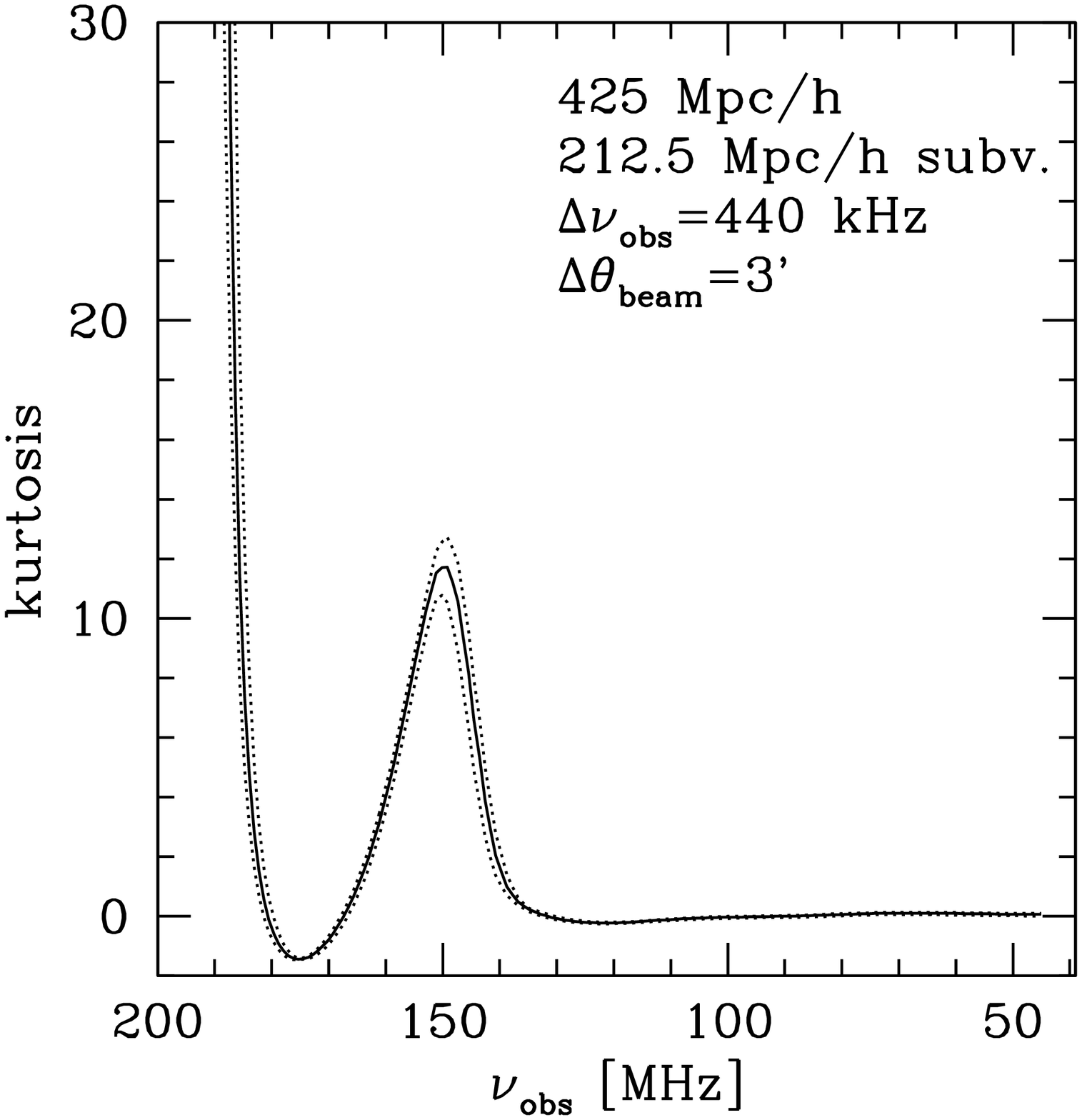}
    \includegraphics[width=2.2in]{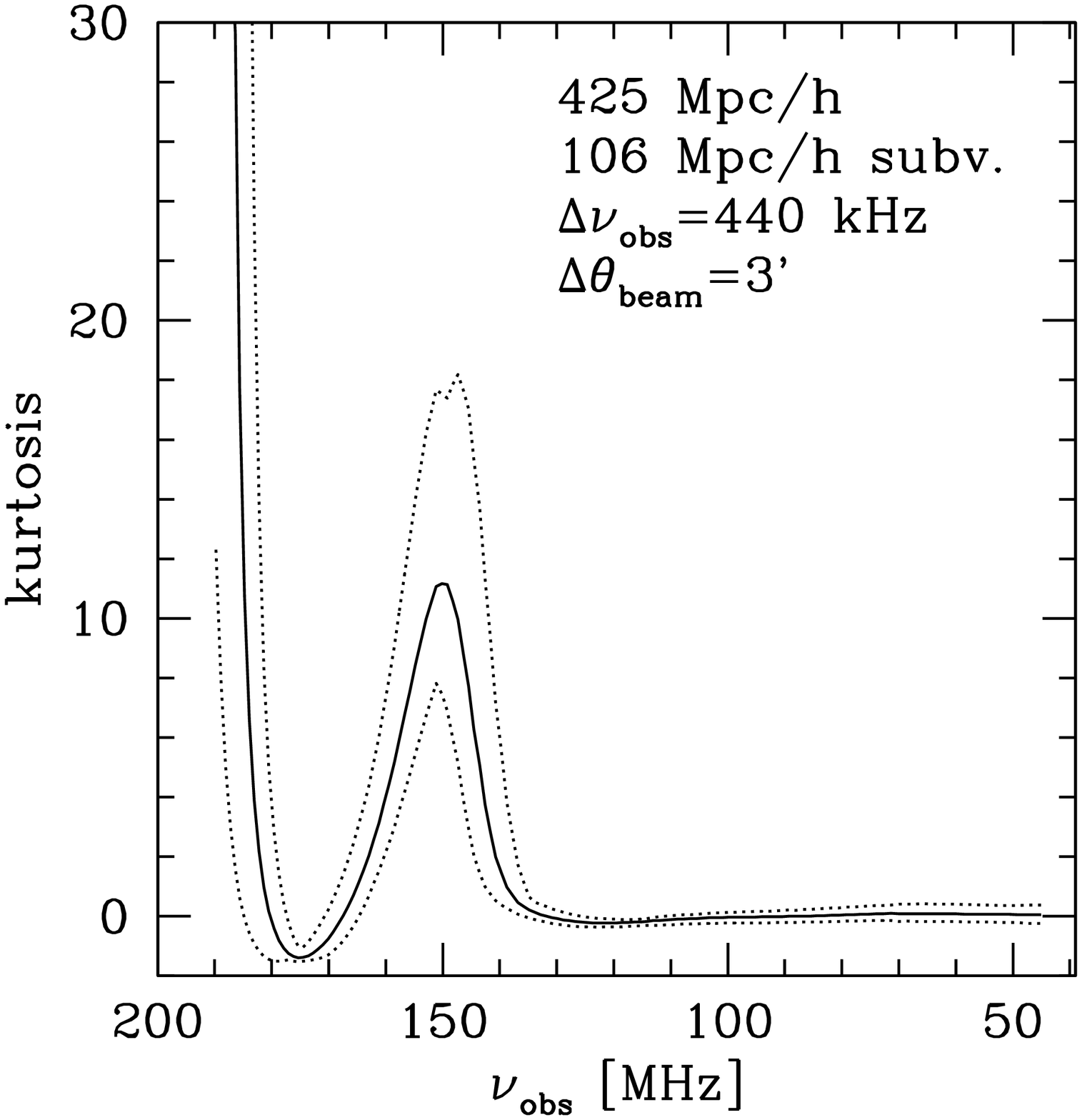}
    \includegraphics[width=2.2in]{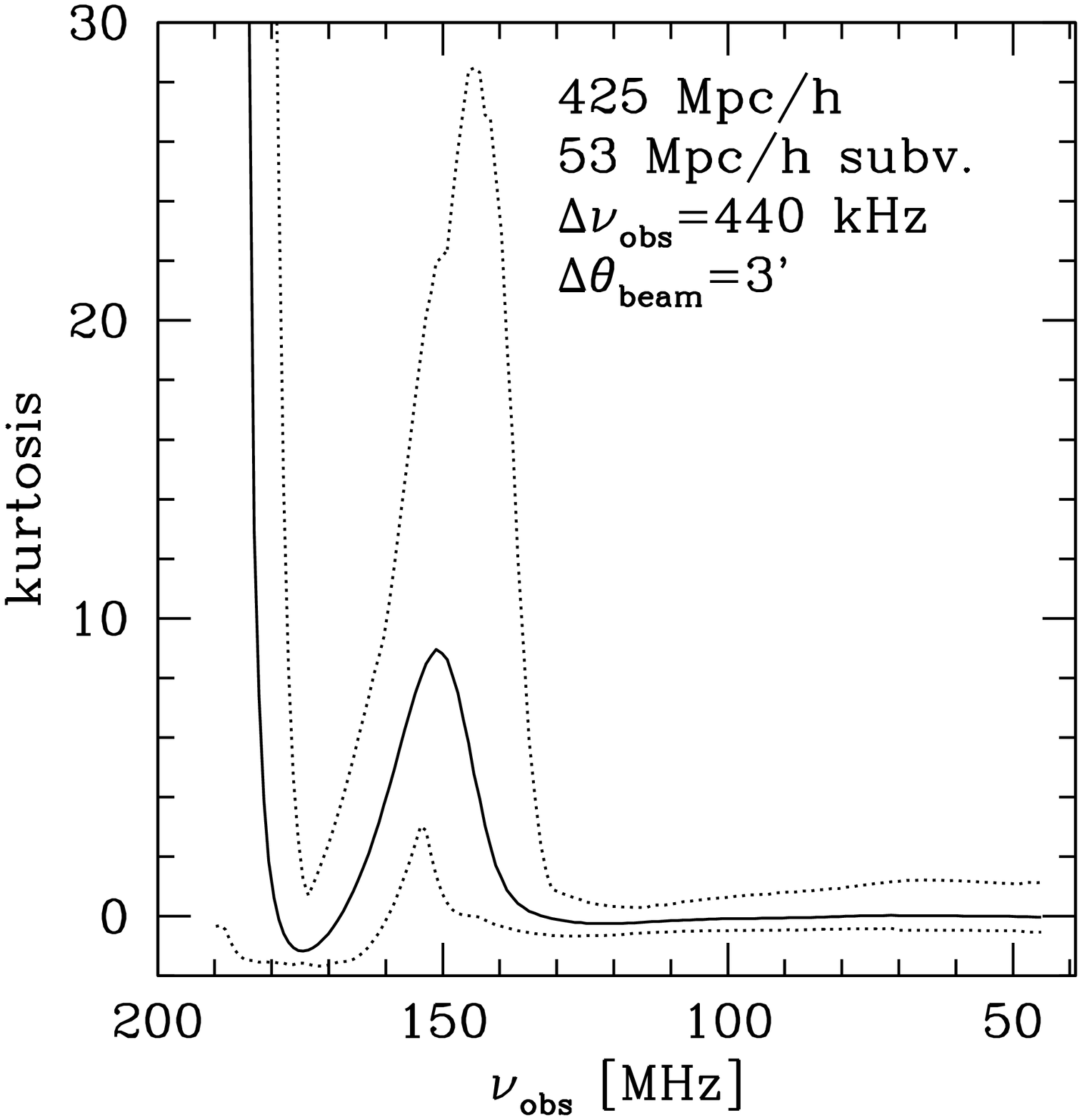}
  \end{center}
  \caption{21-cm differential brightness temperature field kurtosis
variations, smoothed with LOFAR-like beam and bandwidth, between 
sub-regions of XL2 with a given size, as labelled. Shown are the mean of all 
sub-volumes (solid, black) and the minimum and maximum lines enveloping
the variations between all sub-regions of that size. 
    \label{dt_kurt_scatter:fig}}
\end{figure*}

\begin{figure*}
  \begin{center}
    \includegraphics[width=2.2in]{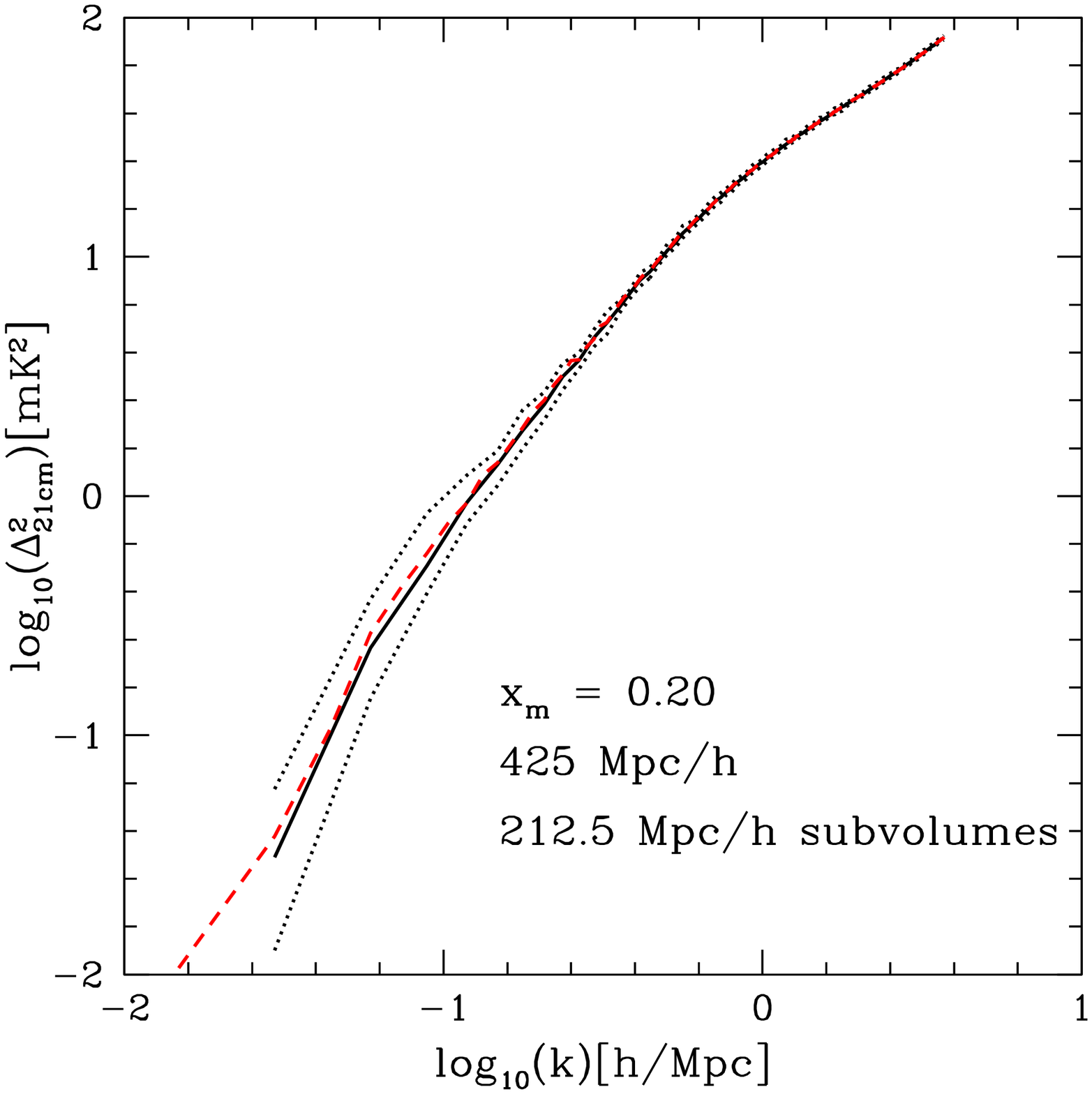}
    \includegraphics[width=2.2in]{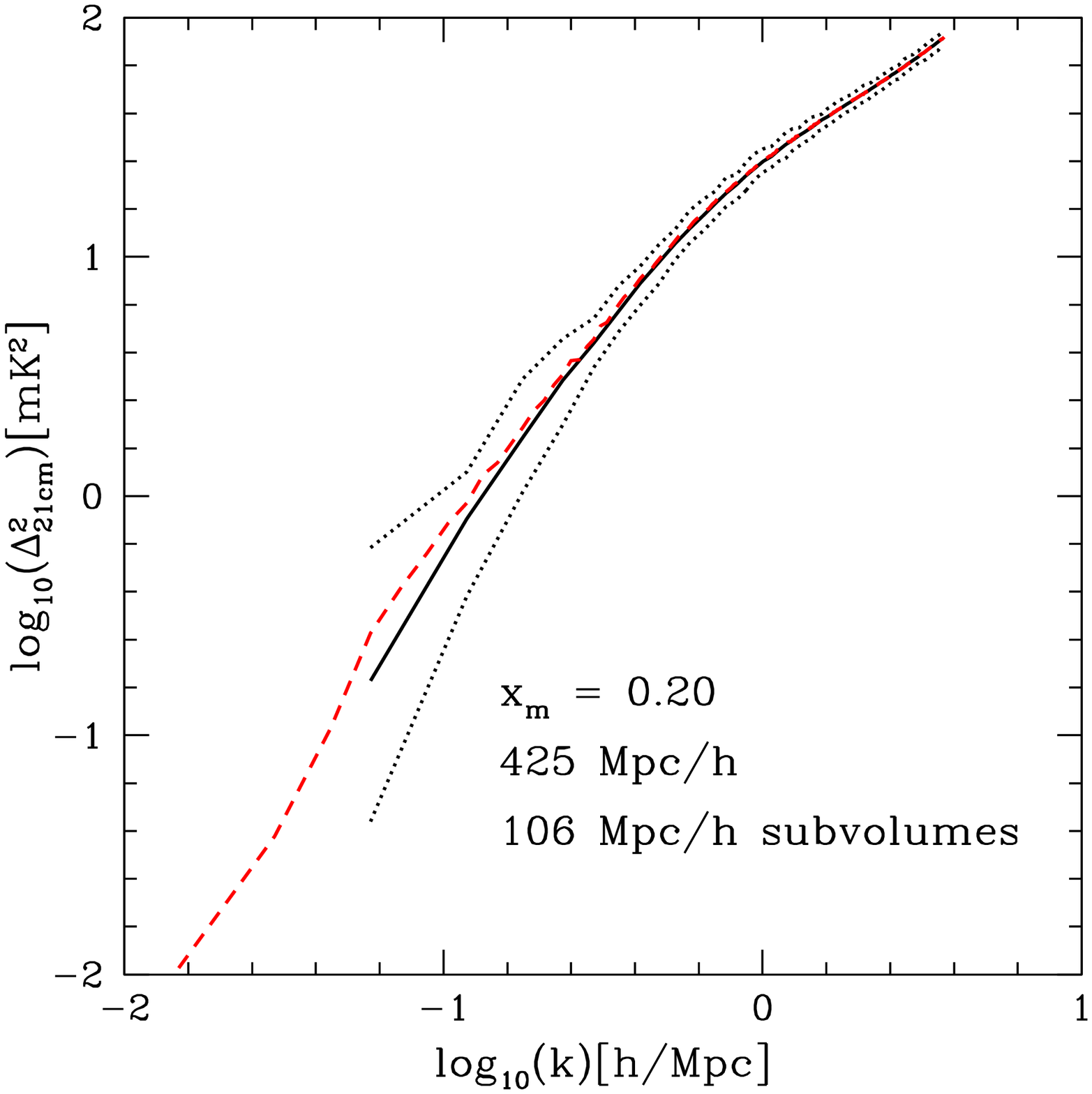}
    \includegraphics[width=2.2in]{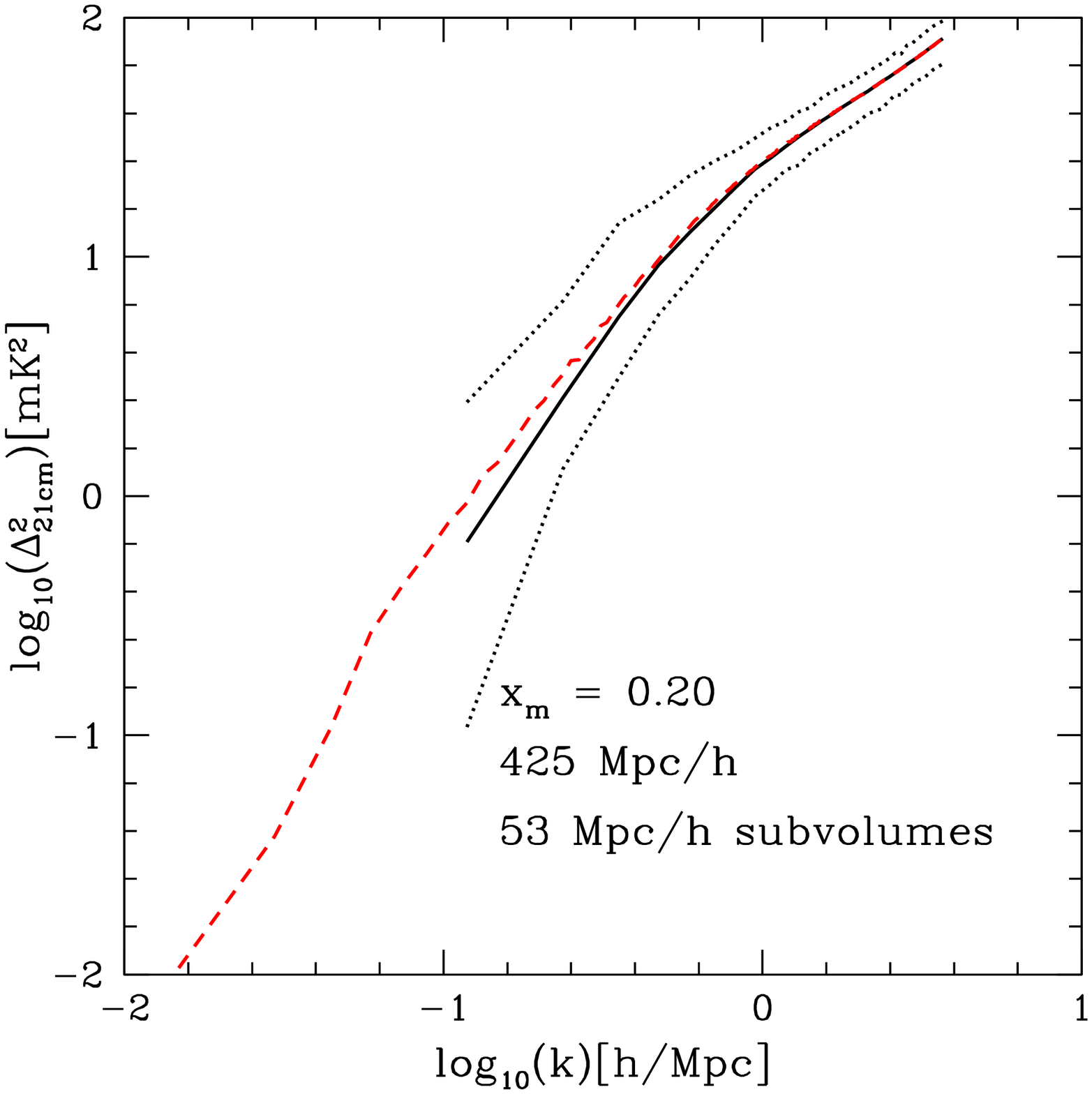}
    \includegraphics[width=2.2in]{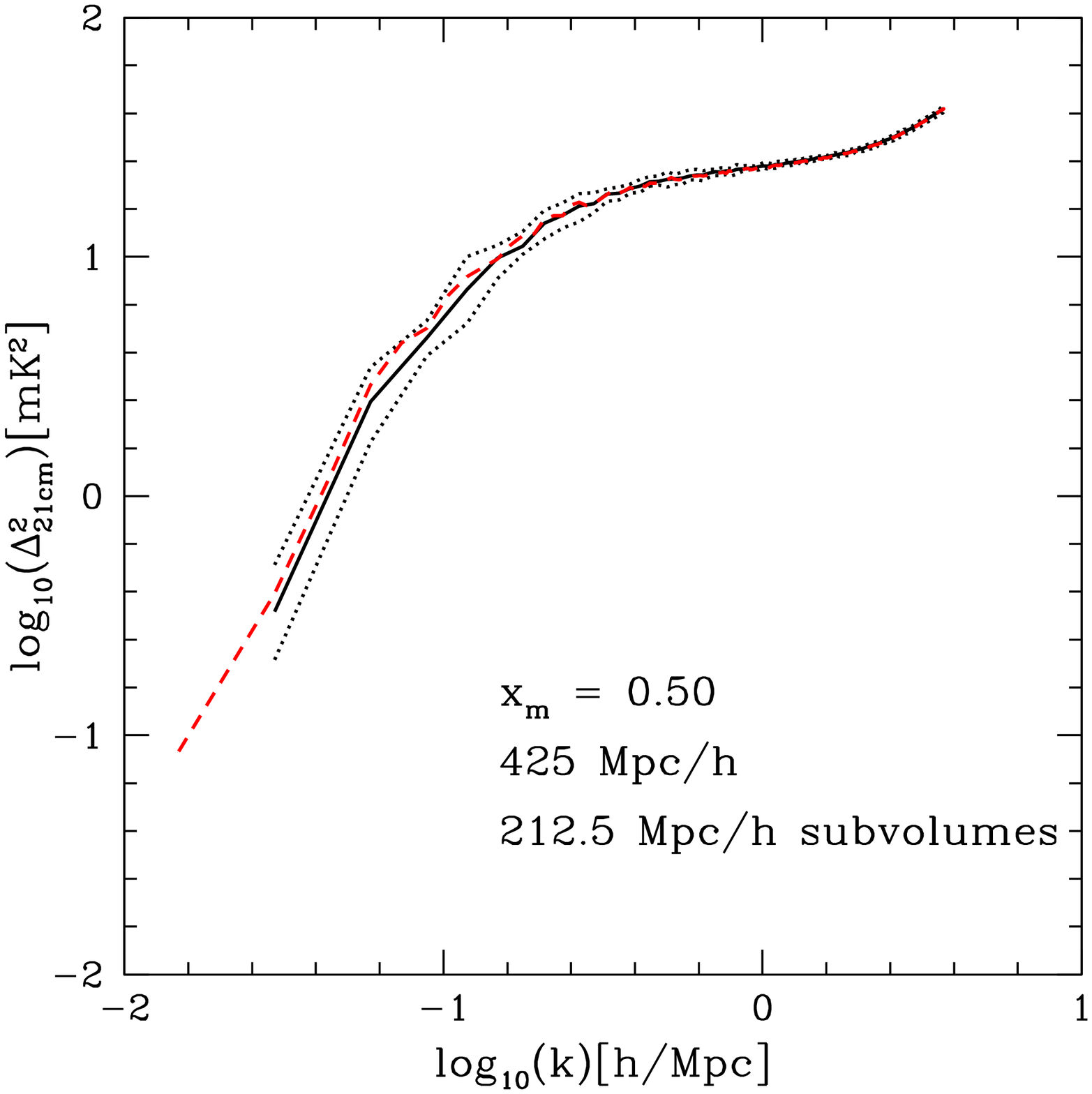}
    \includegraphics[width=2.2in]{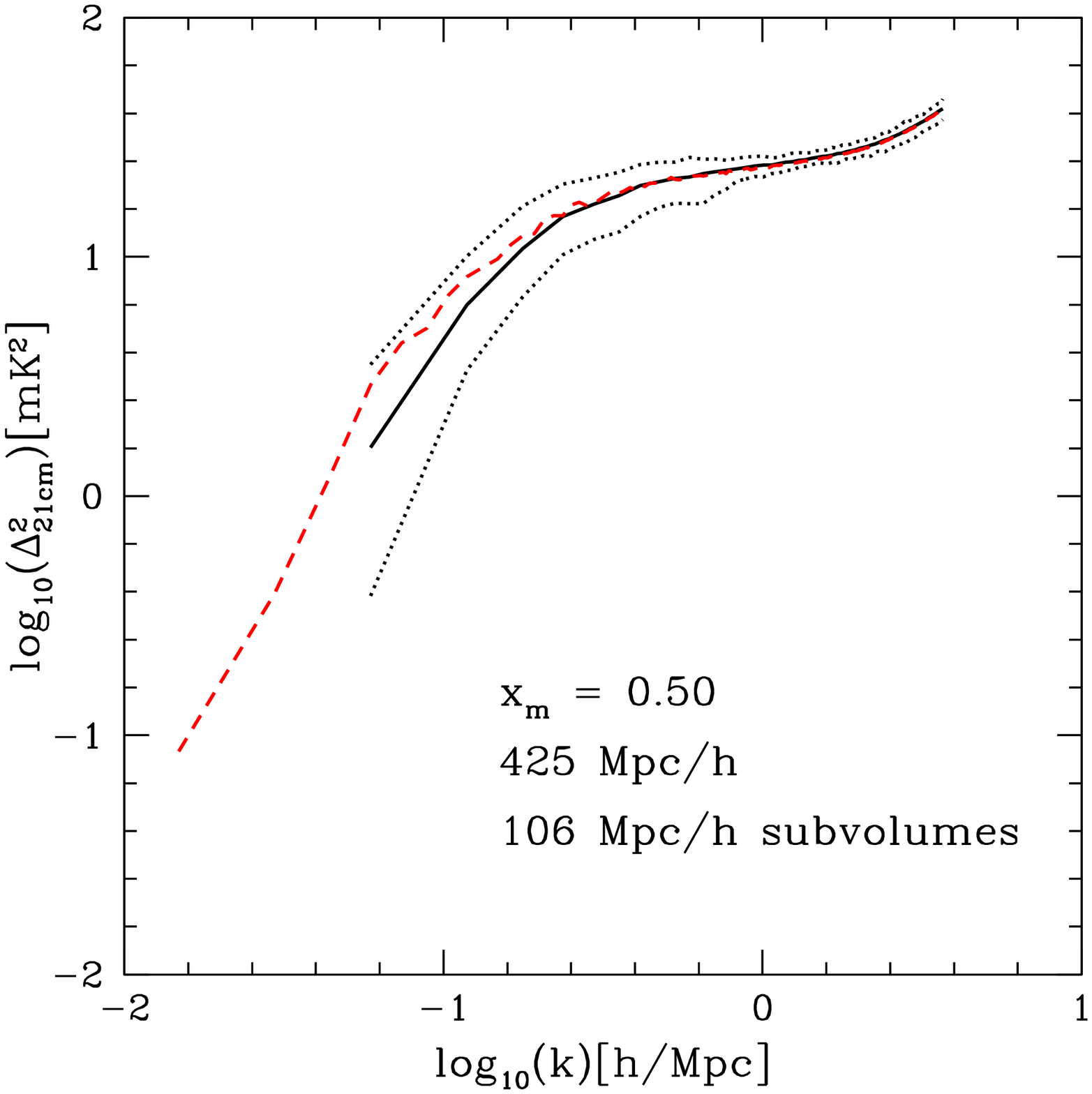}
    \includegraphics[width=2.2in]{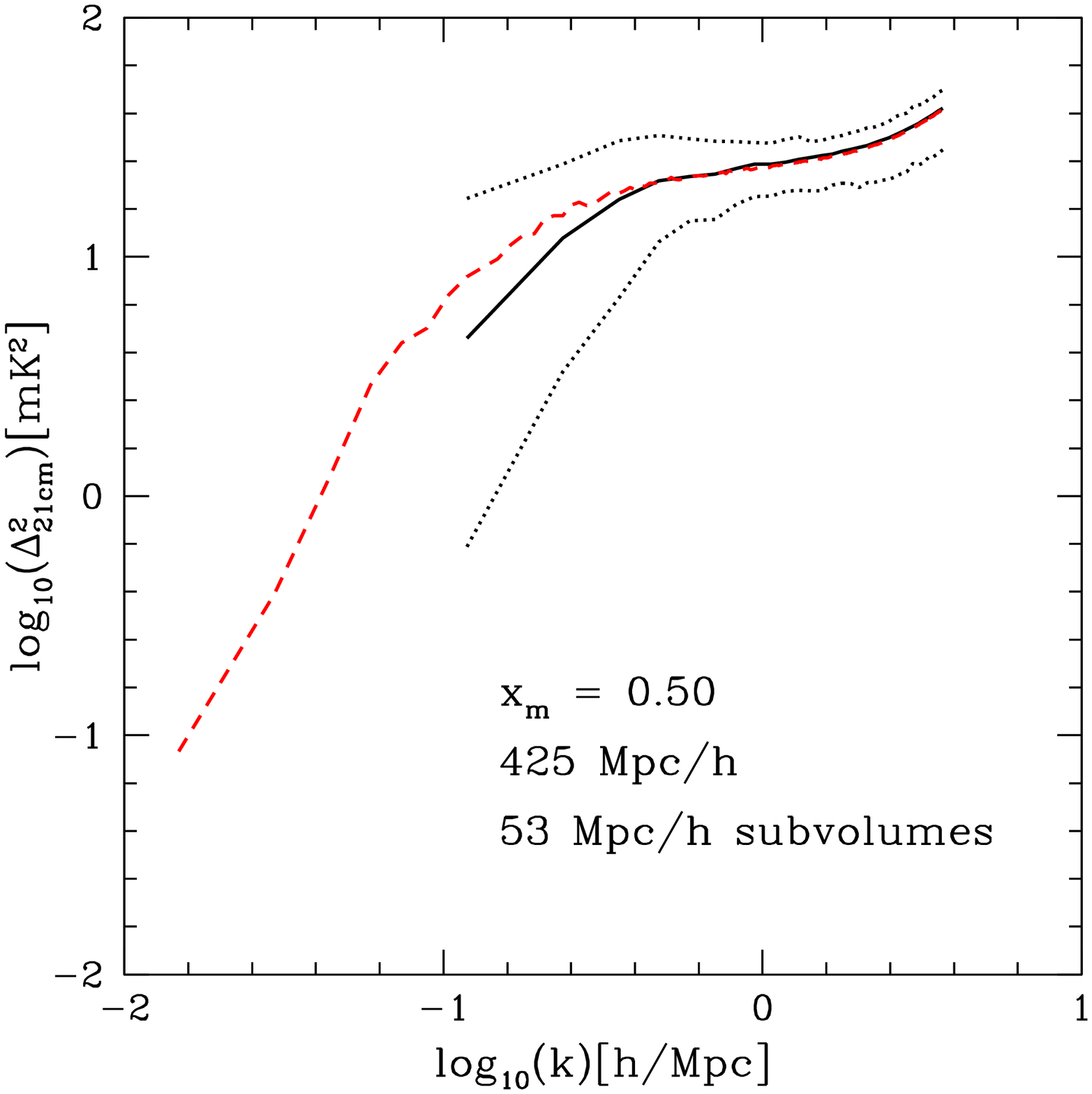}
    \includegraphics[width=2.2in]{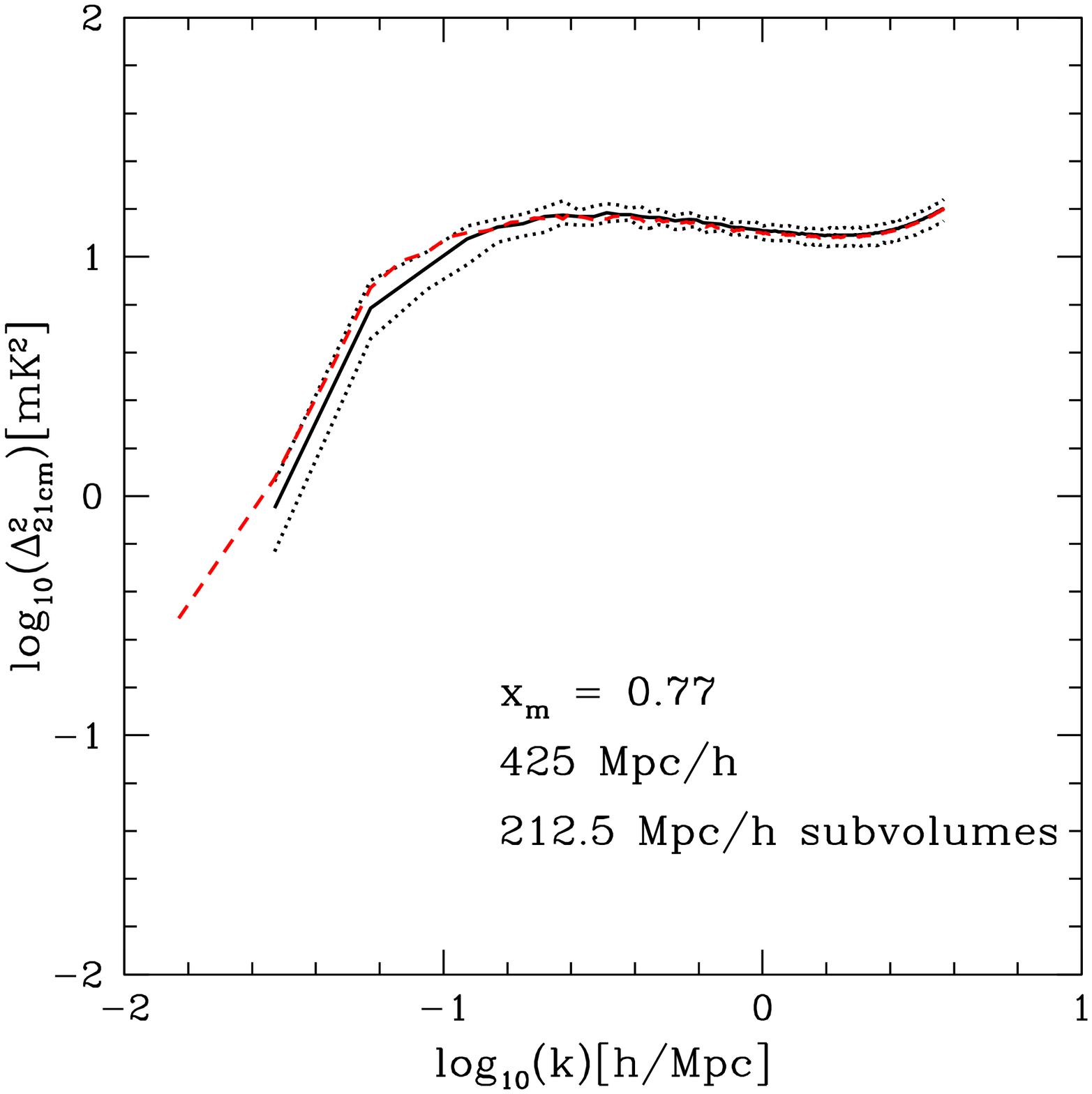}
    \includegraphics[width=2.2in]{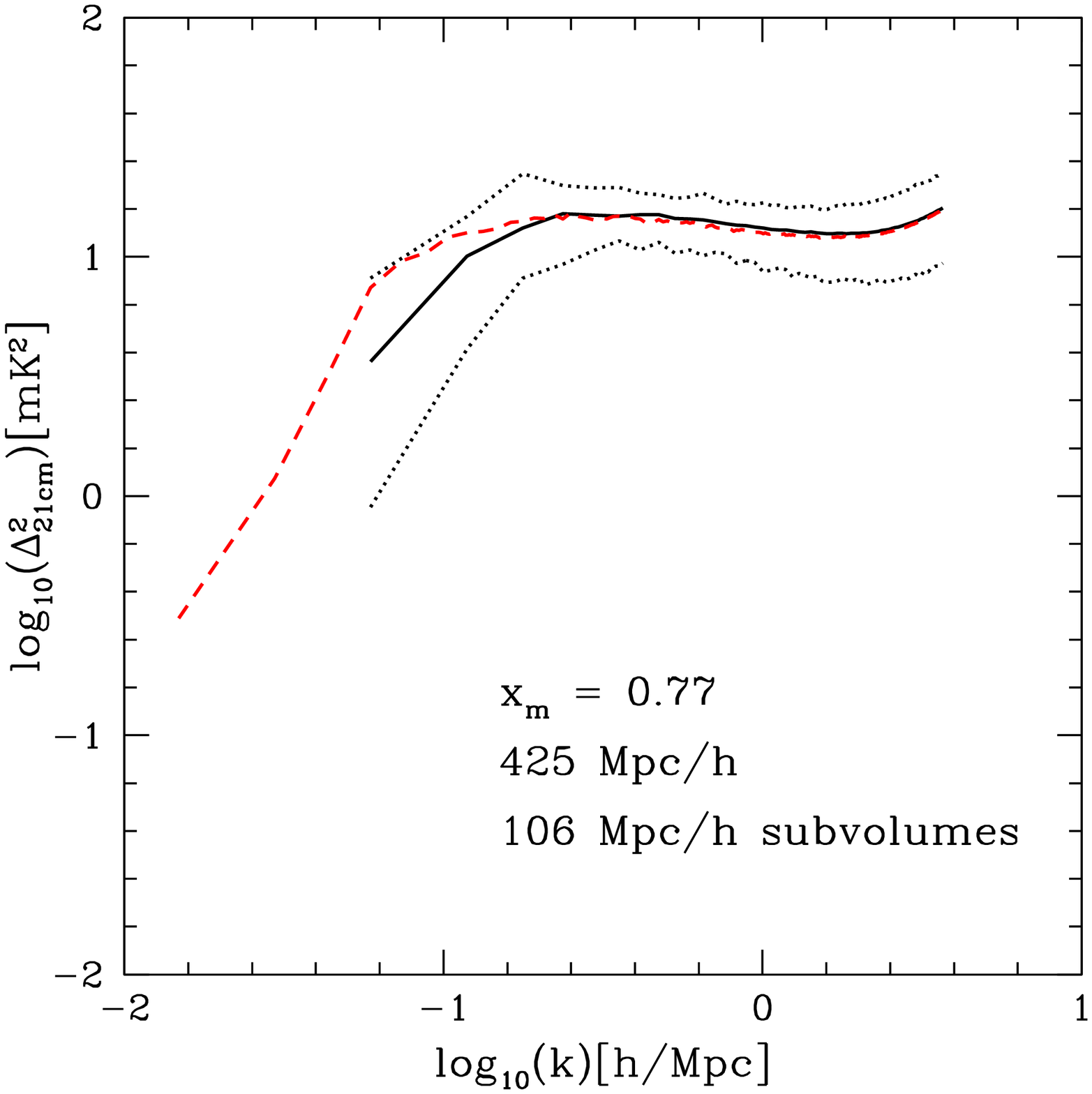}
    \includegraphics[width=2.2in]{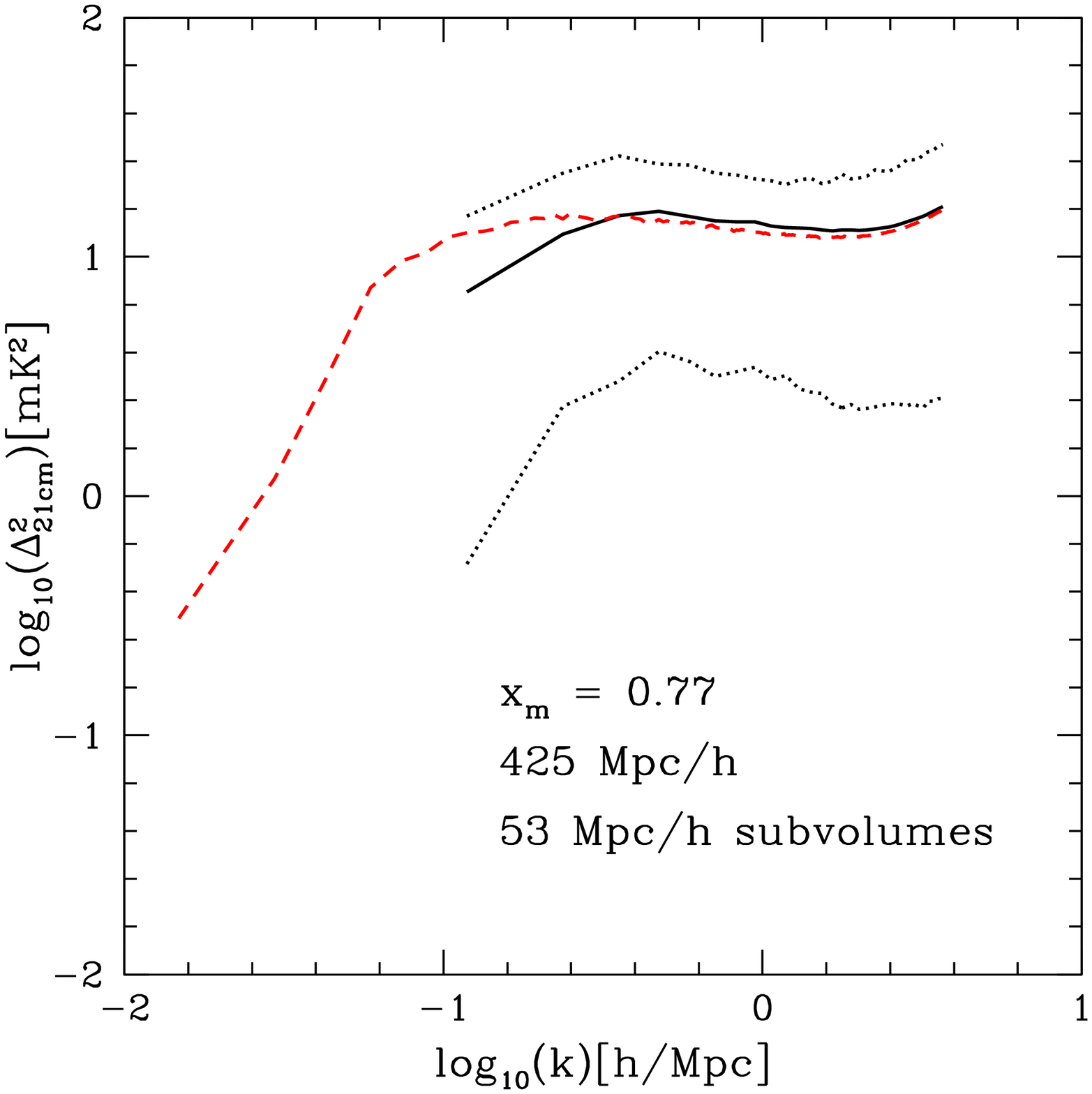}
  \end{center}
  \caption{21-cm differential brightness temperature fluctuation power 
    spectra, in sub-regions of XL2 with a given size 
    and at different stages of reionization, as labelled. 
    Shown are the mean of all sub-volumes (solid, black) and the minimum 
    and maximum lines enveloping the variations between all sub-regions 
    of that size. The average over the full volume is also indicated 
    (dashed, red). 
    \label{21cm-power-subvol:fig}}
\end{figure*}

Finally, we consider the variations of the 21-cm observables between
sub-volumes of a given size extracted from the XL2 simulation. Results 
are shown in Figures~\ref{dtrms_scatter:fig}--\ref{dt_kurt_scatter:fig}
for 21-cm rms, skewness and kurtosis, and in 
Figure~\ref{21cm-power-subvol:fig} for 21-cm power spectra.
  
For 21-cm rms, skewness and kurtosis, the mean over all subvolumes 
closely tracks the full volume evolution of the same quantities, regardless 
of the size of the subvolumes. All three 
quantities are essentially converged for $212.5\,h^{-1}\rm Mpc$ sub-volumes, 
but there are notable variations for smaller-size sub-volumes. For example, 
in terms of the peak differential brightness temperature rms value, these 
variations could be as high as 20\% for the $106.25\,h^{-1}\rm Mpc$ sub-volumes 
and reach over 60\% for the $53.125\,h^{-1}\rm Mpc$ sub-volumes. These variations 
grow even larger at late times. For example, at $190\,$MHz some sub-volumes 
are already ionized (thus have zero rms), while the mean for all sub-regions 
is still above 0.5\,mK and the maximum rms value is as large as $\sim2$~mK
($\sim4$~mK) for the $106.25\,h^{-1}\rm Mpc$ ($53.125\,h^{-1}\rm Mpc$) 
sub-volumes, respectively. Even the lower, pre-reionization peak at early times 
($\nu\sim80$~MHz) can differ by as much as 75\% for the smallest-size 
sub-regions, ranging between 2~mK and over 3~mK.

The higher moments of the 21-cm brightness PDF distributions, skewness 
and kurtosis, also show significant variations among sub-regions, as shown in 
Figures~\ref{dt_skew_scatter:fig} and \ref{dt_kurt_scatter:fig}. The average 
skewness of all sub-regions is also very close to the full-volume result, 
for all sub-region sizes. However, once again only the large, $212.5\,h^{-1}$Mpc
 volumes are converged and virtually indistinguishable from 
the full-box results. Even for $106\,h^{-1}$Mpc volumes there is a difference 
of up to 50\% in the depth of the skewness feature, which vary between 
$\sim2$~mK and $\sim3$~mK. For the smaller sub-volumes this variation becomes 
as large as a factor of 3, varying between $\sim1.5$~mK (i.e. very weak dip 
feature) and $\sim4.5$~mK. The frequency at which this feature occurs also 
varies, by as much as $\sim10$~MHz for the smallest sub-regions case. This 
behaviour is mirrored also in the kurtosis peak, found at the same frequencies 
as the skewness minimum, as shown in Figure~\ref{dt_kurt_scatter:fig}. The 
kurtosis is essentially zero throughout most of the evolution except for this 
broad peak feature, which in this particular simulation occurs around 150~MHz. 
The height of that peak varies by about a factor of 2 for $106\,h^{-1}$Mpc 
sub-volumes and by up to factor of $\sim7$ for the smaller volumes.
 
However, for 21-cm power spectra, the mean over subvolumes appears to lose 
powers on large scales but converge on small scales, relative to the power 
spectra calculated from the full volume, as shown in 
Figure~\ref{21cm-power-subvol:fig}. The loss depends on the stage of 
reionization: the larger the global ionization fraction is, the worse the 
mean over subvolumes tracks the values from the full volume. This is simply 
because of the growth of the ionization bubbles as the reionization proceeds. 
When the typical bubble size is comparable to the size of the subvolume, the 
information of the correlation of the fluctuations across subvolumes is lost 
when only fluctuations within subvolumes are counted for power spectrum 
calculation. On the other hand, the variations of power spectra for the 
106~$h^{-1}{\rm Mpc}$ and 53~$h^{-1}{\rm Mpc}$ subvolumes grow very large at late 
times ($x_{\rm m} \ge 0.50$). For example, when $x_{\rm m} = 0.50$, the variation 
for 106~$h^{-1}{\rm Mpc}$ subvolumes can be $\sim 60\%$ at 
$k=0.1\,h\,{\rm Mpc^{-1}}$. These variations can be even larger for smaller 
$k$, smaller subvolumes, and later times. 

\section{Summary and Conclusions}
\label{sect:summary}

We presented the first very large-scale, full radiative transfer 
simulation of Cosmic Reionization. We compared the results from 
this to a more typical-size, $\sim100\,h^{-1}$Mpc simulation with 
otherwise the same parameters. The large-scale density fluctuations 
at scales of tens to hundreds of comoving Mpc, which are damped or
not present at all in the smaller volume, strongly modulate the local 
number of ionizing sources and introduce significant additional 
patchiness in the ionization distribution. By all different size 
distribution measures - friends-of-friends, spherical average and 
power spectra - the ionized patches reach considerably larger sizes 
in the larger simulation volume. This in turn introduces increased 
rms fluctuations in the redshifted 21-cm emission, with the peak up 
by $\sim10$\% for our reionization model and LOFAR-like beamsize 
and bandwidth. The 21-cm rms signal boost is larger still, reaching 
more than a factor of 2, at late times, when extensive neutral patches 
still remain in the large volume, in contrast to the smaller one. 
Therefore, taking into account the long-wavelength perturbations 
should considerably facilitate the detection of the late-time 21-cm 
signal. 

We used jackknife approach to evaluate the variation of the reionization 
history in sub-volumes of sizes ranging from $212.5\,h^{-1}$Mpc to 
$53\,h^{-1}$Mpc. The reionization history is largely converged for the
larger subvolumes, but for $106\,h^{-1}$Mpc it shows significant 
variations at very high redshift, $z>20$, which persist, albeit at 
a lower level, all the way to the end of reionization. Finally, the 
$53\,h^{-1}$Mpc subvolumes vary in their degree of ionization by factors 
of a few or more, indicating that studying such volumes does not yield 
a reliable estimate of the mean global reionization history. 
The subvolumes of a certain size also reach each stage of the reionization 
history at a range of redshifts, correlated with the mean density of that
region. However, there is a significant scatter in that relation, which 
increases considerably for the smaller-size subvolumes. Consequently, even 
regions of mean density can reionize at quite different times, varying by
up to $\Delta z\sim1$ for $\sim50\,h^{-1}$Mpc volumes.


Recently, \citet{2013ApJ...776...81B} proposed a new semi-analytical 
approach for modelling reionization at large scales based on the
correlation between the density and ionized fraction fields observed 
in simulations. Using volumes of $100\,h^{-1}$Mpc per side, they showed 
that this correlation is reasonably tight for regions of size $\sim10$~Mpc
and larger ($k<0.6\,h\rm Mpc^{-1}$), but rapidly worsens for smaller sizes.
While this trend is in qualitative agreement with our results, we find 
quite large scatter in the reionization milestones even for regions as
large as $50-100\,h^{-1}$Mpc, indicating that the close density-ionized 
fraction correlation found by \citet{2013ApJ...776...81B} is likely 
over-estimated due to their relatively small simulation volume.
  
The same sub-volume variations of reionization result also in corresponding
variation in the 21-cm rms, skewness and kurtosis, which for the smaller 
sub-volumes can vary at peak by factors of $\sim2$,  $\sim2.5$, and 
$\sim10$, respectively for the three quantities. The variations are 
smaller, but still noticeable for $106\,h^{-1}$Mpc subvolumes, at $\sim15\%$,  
$\sim25\%$, and $\sim2.5$, respectively for rms, skewness and kurtosis
extreme values, and only converge for subvolumes as large as $212.5\,h^{-1}$Mpc. 
On the other hand, the 21-cm 3D power spectra prove more sensitive to the 
simulation volume, generally with larger variations between sub-volumes than
other 21-cm observables, and at large scales ($k\lesssim0.25\,h\,\rm Mpc^{-1}$) 
do not converge even for volumes $\gtrsim200\,h^{-1}$Mpc per side. For 
smaller, $\sim50\,h^{-1}$Mpc sub-volumes the 21-cm power spectra do not 
converge well at any scale probed here, varying by factor of $\sim2-10$ 
depending on the reionization stage. However, the average power of the 
sub-volumes does converge to the full-volume result at small scales. 


This work is just a proof of concept and an initial evaluation of the effects
of large scales for a particular reionization scenario. Clearly, the specific 
numbers derived here depend on the assumptions made about the reionization 
parameters like source efficiencies. Nonetheless, our conclusions about the 
convergence with simulation volume are general, at least as long as the 
reionization process proceeds along the lines assumed here, namely it is 
dominated by soft-UV stellar sources, which is currently the reionization
scenario best supported by the observational data. The source efficiencies 
assumed in this work were chosen so as to satisfy the current observational 
constraints like the EoR duration, overlap epoch and IGM photoionization rate
at late times. However, a full evaluation in terms of observations at 
redshifted 21-cm or other wavelengths requires running suites of simulations 
exploring the available parameter space. This large-scale simulation data has 
already been used for deriving the kSZ signal from reionization 
\citep{2013ApJ...769...93P}, signatures of luminous QSOs 
\citep{2012MNRAS.424..762D} and 21-cm redshift-space distortions 
\citep{2013PhRvL.110o1301S,Jensen2013}.



\section*{Acknowledgements} This work was supported by the Science and 
Technology Facilities Council [grant numbers ST/F002858/1 and ST/I000976/1]; 
and The Southeast Physics Network (SEPNet). PRS was supported in part by 
U.S. NSF grants AST-0708176 and AST-1009799, NASA grants NNX07AH09G, 
NNG04G177G and NNX11AE09G, and Chandra grant SAO TM8-9009X. GM was supported 
in part by Swedish Research Council grant 60336701. YM was supported by French 
state funds managed by the ANR within the Investissements d'Avenir programme 
under reference ANR-11-IDEX-0004-02. KA is supported by NRF grant funded by 
the Korean government MEST No. 2012R1A1A1014646 and by research funds from 
Chosun University, 2010. The authors acknowledge the Texas Advanced 
Computing Center (TACC) at The University of Texas at Austin for providing 
HPC resources that have contributed to the research results reported within 
this paper. This research was supported in part by an allocation of advanced 
computing resources provided by the National Science Foundation through TACC 
and the National Institute for Computational Sciences (NICS), with part of the 
computations performed on Lonestar at TACC (http://www.tacc.utexas.edu) and 
Kraken at NICS (http://www.nics.tennessee.edu/). Some of the numerical 
computations were done on the Apollo cluster at The University of Sussex and 
the Sciama High Performance Compute (HPC) cluster which is supported by the ICG, 
SEPNet and the University of Portsmouth. Part of the computations were performed 
on the GPC supercomputer at the SciNet HPC Consortium. SciNet is funded by: the 
Canada Foundation for Innovation under the auspices of Compute Canada; the 
Government of Ontario; Ontario Research Fund - Research Excellence; and the 
University of Toronto.



\begin{thebibliography}{51}
\expandafter\ifx\csname natexlab\endcsname\relax\def\natexlab#1{#1}\fi

\bibitem[{{Ahn} {et~al.}(2012){Ahn}, {Iliev}, {Shapiro}, {Mellema}, {Koda}, \&
  {Mao}}]{2012ApJ...756L..16A}
{Ahn} K., {Iliev} I.~T., {Shapiro} P.~R., {Mellema} G., {Koda} J., {Mao} Y.,
  2012, \apjl, 756, L16

\bibitem[{{Barkana} \& {Loeb}(2004)}]{2004ApJ...609..474B}
{Barkana} R., {Loeb} A., 2004, \apj, 609, 474

\bibitem[{{Battaglia} {et~al.}(2013){Battaglia}, {Trac}, {Cen}, \&
  {Loeb}}]{2013ApJ...776...81B}
{Battaglia} N., {Trac} H., {Cen} R., {Loeb} A., 2013, \apj, 776, 81

\bibitem[{{Ciardi} {et~al.}(2000){Ciardi}, {Ferrara}, {Governato}, \&
  {Jenkins}}]{2000MNRAS.314..611C}
{Ciardi} B., {Ferrara} A., {Governato} F., {Jenkins} A., 2000, \mnras, 314, 611

\bibitem[{{Ciardi} {et~al.}(2003){Ciardi}, {Ferrara}, \&
  {White}}]{2003MNRAS.344L...7C}
{Ciardi} B., {Ferrara} A., {White} S.~D.~M., 2003, \mnras, 344, L7

\bibitem[{{Crocce} {et~al.}(2006){Crocce}, {Pueblas}, \&
  {Scoccimarro}}]{2006MNRAS.373..369C}
{Crocce} M., {Pueblas} S., {Scoccimarro} R., 2006, \mnras, 373, 369

\bibitem[{{Datta} {et~al.}(2012){Datta}, {Friedrich}, {Mellema}, {Iliev}, \&
  {Shapiro}}]{2012MNRAS.424..762D}
{Datta} K.~K., {Friedrich} M.~M., {Mellema} G., {Iliev} I.~T., {Shapiro} P.~R.,
  2012, \mnras, 424, 762

\bibitem[{{Dor{\'e}} {et~al.}(2007){Dor{\'e}}, {Holder}, {Alvarez}, {Iliev},
  {Mellema}, {Pen}, \& {Shapiro}}]{cmbpol}
{Dor{\'e}} O., {Holder} G., {Alvarez} M.~A., {Iliev} I.~T., {Mellema} G., {Pen}
  U.-L., {Shapiro} P.~R., 2007, \prd, 76, 043002

\bibitem[{{Fan} \& {et al.}(2001)}]{2001AJ....122.2833F}
{Fan} X., {et al.}, 2001, \aj, 122, 2833

\bibitem[{{Field}(1959)}]{1959ApJ...129..536F}
{Field} G.~B., 1959, \apj, 129, 536

\bibitem[{{Friedrich} {et~al.}(2011){Friedrich}, {Mellema}, {Alvarez},
  {Shapiro}, \& {Iliev}}]{2011MNRAS.413.1353F}
{Friedrich} M.~M., {Mellema} G., {Alvarez} M.~A., {Shapiro} P.~R., {Iliev}
  I.~T., 2011, \mnras, 413, 1353

\bibitem[{{Furlanetto} {et~al.}(2006){Furlanetto}, {Oh}, \&
  {Briggs}}]{2006PhR...433..181F}
{Furlanetto} S.~R., {Oh} S.~P., {Briggs} F.~H., 2006, \physrep, 433, 181

\bibitem[{{Gnedin}(2000)}]{2000ApJ...535..530G}
{Gnedin} N.~Y., 2000, \apj, 535, 530

\bibitem[{{Gnedin} \& {Ostriker}(1997)}]{1997ApJ...486..581G}
{Gnedin} N.~Y., {Ostriker} J.~P., 1997, \apj, 486, 581

\bibitem[{{Harker} {et~al.}(2010){Harker}, {Zaroubi}, {Bernardi}, {Brentjens},
  {de Bruyn}, {Ciardi}, {Jeli{\'c}}, {Koopmans}, {Labropoulos}, {Mellema},
  {Offringa}, {Pandey}, {Pawlik}, {Schaye}, {Thomas}, \&
  {Yatawatta}}]{2010MNRAS.405.2492H}
{Harker} G., {Zaroubi} S., {Bernardi} G., {Brentjens} M.~A., {de Bruyn} A.~G.,
  {Ciardi} B., {Jeli{\'c}} V., {Koopmans} L.~V.~E., {Labropoulos} P., {Mellema}
  G., {Offringa} A., {Pandey} V.~N., {Pawlik} A.~H., {Schaye} J., {Thomas}
  R.~M., {Yatawatta} S., 2010, \mnras, 405, 2492

\bibitem[{{Harker} {et~al.}(2009){Harker}, {Zaroubi}, {Thomas}, {Jeli{\'c}},
  {Labropoulos}, {Mellema}, {Iliev}, {Bernardi}, {Brentjens}, {de Bruyn},
  {Ciardi}, {Koopmans}, {Pandey}, {Pawlik}, {Schaye}, \&
  {Yatawatta}}]{2009MNRAS.393.1449H}
{Harker} G.~J.~A., {Zaroubi} S., {Thomas} R.~M., {Jeli{\'c}} V., {Labropoulos}
  P., {Mellema} G., {Iliev} I.~T., {Bernardi} G., {Brentjens} M.~A., {de Bruyn}
  A.~G., {Ciardi} B., {Koopmans} L.~V.~E., {Pandey} V.~N., {Pawlik} A.~H.,
  {Schaye} J., {Yatawatta} S., 2009, \mnras, 393, 1449

\bibitem[{{Harnois-Deraps} {et~al.}(2013){Harnois-Deraps}, {Pen}, {Iliev},
  {Merz}, {Emberson}, \& {Desjacques}}]{2012arXiv1208.5098H}
{Harnois-Deraps} J., {Pen} U.-L., {Iliev} I.~T., {Merz} H., {Emberson} J.~D.,
  {Desjacques} V., 2013, MNRAS, in press

\bibitem[{{Holder} {et~al.}(2007){Holder}, {Iliev}, \& {Mellema}}]{pol21}
{Holder} G.~P., {Iliev} I.~T., {Mellema} G., 2007, \apjl, 663, L1

\bibitem[{{Iliev} {et~al.}(2006){Iliev}, {Mellema}, {Pen}, {Merz}, {Shapiro},
  \& {Alvarez}}]{2006MNRAS.369.1625I}
{Iliev} I.~T., {Mellema} G., {Pen} U.-L., {Merz} H., {Shapiro} P.~R., {Alvarez}
  M.~A., 2006, \mnras, 369, 1625

\bibitem[{{Iliev} {et~al.}(2007){Iliev}, {Mellema}, {Shapiro}, \&
  {Pen}}]{2007MNRAS.376..534I}
{Iliev} I.~T., {Mellema} G., {Shapiro} P.~R., {Pen} U.-L., 2007, \mnras, 376,
  534

\bibitem[{{Iliev} {et~al.}(2012){Iliev}, {Mellema}, {Shapiro}, {Pen}, {Mao},
  {Koda}, \& {Ahn}}]{2012MNRAS.423.2222I}
{Iliev} I.~T., {Mellema} G., {Shapiro} P.~R., {Pen} U.-L., {Mao} Y., {Koda} J.,
  {Ahn} K., 2012, \mnras, 423, 2222

\bibitem[{{Iliev} {et~al.}(2008){Iliev}, {Shapiro}, {McDonald}, {Mellema}, \&
  {Pen}}]{2008MNRAS.391...63I}
{Iliev} I.~T., {Shapiro} P.~R., {McDonald} P., {Mellema} G., {Pen} U.-L., 2008,
  \mnras, 391, 63

\bibitem[{{Jensen} {et~al.}(2013){Jensen}, {Datta}, {Mellema}, {Chapman},
  {Abdalla}, {Iliev}, {Mao}, {Santos}, {Shapiro}, {Zaroubi}, {Bernardi},
  {Brentjens}, {de Bruyn}, {Ciardi}, {Harker}, {Jeli{\'c}}, {Kazemi},
  {Koopmans}, {Labropoulos}, {Martinez}, {Offringa}, {Pandey}, {Schaye},
  {Thomas}, {Veligatla}, {Vedantham}, \& {Yatawatta}}]{Jensen2013}
{Jensen} H., {Datta} K.~K., {Mellema} G., {Chapman} E., {Abdalla} F.~B.,
  {Iliev} I.~T., {Mao} Y., {Santos} M.~G., {Shapiro} P.~R., {Zaroubi} S.,
  {Bernardi} G., {Brentjens} M.~A., {de Bruyn} A.~G., {Ciardi} B., {Harker}
  G.~J.~A., {Jeli{\'c}} V., {Kazemi} S., {Koopmans} L.~V.~E., {Labropoulos} P.,
  {Martinez} O., {Offringa} A.~R., {Pandey} V.~N., {Schaye} J., {Thomas} R.~M.,
  {Veligatla} V., {Vedantham} H., {Yatawatta} S., 2013, \mnras, 435, 460

\bibitem[{{Kohler} {et~al.}(2007){Kohler}, {Gnedin}, \&
  {Hamilton}}]{2007ApJ...657...15K}
{Kohler} K., {Gnedin} N.~Y., {Hamilton} A.~J.~S., 2007, \apj, 657, 15

\bibitem[{{Komatsu} {et~al.}(2009){Komatsu}, {Dunkley}, {Nolta}, {Bennett},
  {Gold}, {Hinshaw}, {Jarosik}, {Larson}, {Limon}, {Page}, {Spergel},
  {Halpern}, {Hill}, {Kogut}, {Meyer}, {Tucker}, {Weiland}, {Wollack}, \&
  {Wright}}]{2009ApJS..180..330K}
{Komatsu} E., {Dunkley} J., {Nolta} M.~R., {Bennett} C.~L., {Gold} B.,
  {Hinshaw} G., {Jarosik} N., {Larson} D., {Limon} M., {Page} L., {Spergel}
  D.~N., {Halpern} M., {Hill} R.~S., {Kogut} A., {Meyer} S.~S., {Tucker} G.~S.,
  {Weiland} J.~L., {Wollack} E., {Wright} E.~L., 2009, \apjs, 180, 330

\bibitem[{{Krug} {et~al.}(2012){Krug}, {Veilleux}, {Tilvi}, {Malhotra},
  {Rhoads}, {Hibon}, {Swaters}, {Probst}, {Dey}, {Dickinson}, \&
  {Jannuzi}}]{2012ApJ...745..122K}
{Krug} H.~B., {Veilleux} S., {Tilvi} V., {Malhotra} S., {Rhoads} J., {Hibon}
  P., {Swaters} R., {Probst} R., {Dey} A., {Dickinson} M., {Jannuzi} B.~T.,
  2012, \apj, 745, 122

\bibitem[{Lewis {et~al.}(2000)Lewis, Challinor, \& Lasenby}]{Lewis:1999bs}
Lewis A., Challinor A., Lasenby A., 2000, Astrophys. J., 538, 473

\bibitem[{{Lonsdale} \& et~al.(2009)}]{2009IEEEP..97.1497L}
{Lonsdale} C.~J., et~al., 2009, IEEE Proceedings, 97, 1497

\bibitem[{{Luki{\'c}} {et~al.}(2007){Luki{\'c}}, {Heitmann}, {Habib},
  {Bashinsky}, \& {Ricker}}]{2007ApJ...671.1160L}
{Luki{\'c}} Z., {Heitmann} K., {Habib} S., {Bashinsky} S., {Ricker} P.~M.,
  2007, \apj, 671, 1160

\bibitem[{{Mao} {et~al.}(2012){Mao}, {Shapiro}, {Mellema}, {Iliev}, {Koda}, \&
  {Ahn}}]{Mao12}
{Mao} Y., {Shapiro} P.~R., {Mellema} G., {Iliev} I.~T., {Koda} J., {Ahn} K.,
  2012, \mnras, 422, 926

\bibitem[{{McQuinn} {et~al.}(2007){McQuinn}, {Lidz}, {Zahn}, {Dutta},
  {Hernquist}, \& {Zaldarriaga}}]{2007MNRAS.377.1043M}
{McQuinn} M., {Lidz} A., {Zahn} O., {Dutta} S., {Hernquist} L., {Zaldarriaga}
  M., 2007, \mnras, 377, 1043

\bibitem[{{Mesinger} {et~al.}(2011){Mesinger}, {Furlanetto}, \&
  {Cen}}]{2011MNRAS.411..955M}
{Mesinger} A., {Furlanetto} S., {Cen} R., 2011, \mnras, 411, 955

\bibitem[{{Park} {et~al.}(2013){Park}, {Shapiro}, {Komatsu}, {Iliev}, {Ahn}, \&
  {Mellema}}]{2013ApJ...769...93P}
{Park} H., {Shapiro} P.~R., {Komatsu} E., {Iliev} I.~T., {Ahn} K., {Mellema}
  G., 2013, \apj, 769, 93

\bibitem[{{Parsons} \& et~al.(2010)}]{2010AJ....139.1468P}
{Parsons} A.~R., et~al., 2010, \aj, 139, 1468

\bibitem[{{Planck Collaboration} {et~al.}(2013){Planck Collaboration}, {Ade},
  {Aghanim}, {Armitage-Caplan}, {Arnaud}, {Ashdown}, {Atrio-Barandela},
  {Aumont}, {Baccigalupi}, {Banday}, \& et~al.}]{2013arXiv1303.5076P}
{Planck Collaboration}, {Ade} P.~A.~R., {Aghanim} N., {Armitage-Caplan} C.,
  {Arnaud} M., {Ashdown} M., {Atrio-Barandela} F., {Aumont} J., {Baccigalupi}
  C., {Banday} A.~J., et~al., 2013, ArXiv e-prints

\bibitem[{{Press} \& {Schechter}(1974)}]{1974ApJ...187..425P}
{Press} W.~H., {Schechter} P., 1974, \apj, 187, 425

\bibitem[{{Reed} {et~al.}(2007){Reed}, {Bower}, {Frenk}, {Jenkins}, \&
  {Theuns}}]{2007MNRAS.374....2R}
{Reed} D.~S., {Bower} R., {Frenk} C.~S., {Jenkins} A., {Theuns} T., 2007,
  \mnras, 374, 2

\bibitem[{{Ricotti} {et~al.}(2002){Ricotti}, {Gnedin}, \&
  {Shull}}]{2002ApJ...575...33R}
{Ricotti} M., {Gnedin} N.~Y., {Shull} J.~M., 2002, \apj, 575, 33

\bibitem[{{Santos} {et~al.}(2010){Santos}, {Ferramacho}, {Silva}, {Amblard}, \&
  {Cooray}}]{2010MNRAS.406.2421S}
{Santos} M.~G., {Ferramacho} L., {Silva} M.~B., {Amblard} A., {Cooray} A.,
  2010, \mnras, 406, 2421

\bibitem[{{Shapiro} {et~al.}(2013){Shapiro}, {Mao}, {Iliev}, {Mellema},
  {Datta}, {Ahn}, \& {Koda}}]{2013PhRvL.110o1301S}
{Shapiro} P.~R., {Mao} Y., {Iliev} I.~T., {Mellema} G., {Datta} K.~K., {Ahn}
  K., {Koda} J., 2013, Physical Review Letters, 110, 151301

\bibitem[{{Sheth} {et~al.}(2001){Sheth}, {Mo}, \&
  {Tormen}}]{2001MNRAS.323....1S}
{Sheth} R.~K., {Mo} H.~J., {Tormen} G., 2001, \mnras, 323, 1

\bibitem[{{Sheth} \& {Tormen}(2002)}]{2002MNRAS.329...61S}
{Sheth} R.~K., {Tormen} G., 2002, \mnras, 329, 61

\bibitem[{{Sokasian} {et~al.}(2003){Sokasian}, {Abel}, {Hernquist}, \&
  {Springel}}]{2003MNRAS.344..607S}
{Sokasian} A., {Abel} T., {Hernquist} L., {Springel} V., 2003, \mnras, 344, 607

\bibitem[{{Songaila} \& {Cowie}(2010)}]{2010ApJ...721.1448S}
{Songaila} A., {Cowie} L.~L., 2010, \apj, 721, 1448

\bibitem[{{Spergel} \& {et al.}(2003)}]{2003ApJS..148..175S}
{Spergel} D.~N., {et al.}, 2003, \apjs, 148, 175

\bibitem[{{Trac} \& {Gnedin}(2011)}]{2011ASL.....4..228T}
{Trac} H.~Y., {Gnedin} N.~Y., 2011, Advanced Science Letters, 4, 228

\bibitem[{{Watson} {et~al.}(2013{\natexlab{a}}){Watson}, {Iliev}, {D'Aloisio},
  {Knebe}, {Shapiro}, \& {Yepes}}]{will2013}
{Watson} W.~A., {Iliev} I.~T., {D'Aloisio} A., {Knebe} A., {Shapiro} P.~R.,
  {Yepes} G., 2013{\natexlab{a}}, \mnras, 433, 1230

\bibitem[{{Watson} {et~al.}(2013{\natexlab{b}}){Watson}, {Iliev}, {Diego},
  {Gottl{\"o}ber}, {Knebe}, {Mart{\'{\i}}nez-Gonz{\'a}lez}, \&
  {Yepes}}]{2013arXiv1305.1976W}
{Watson} W.~A., {Iliev} I.~T., {Diego} J.~M., {Gottl{\"o}ber} S., {Knebe} A.,
  {Mart{\'{\i}}nez-Gonz{\'a}lez} E., {Yepes} G., 2013{\natexlab{b}}, ArXiv
  e-prints

\bibitem[{{Zahn} \& {et al.}(2012)}]{2012ApJ...756...65Z}
{Zahn} O., {et al.}, 2012, \apj, 756, 65

\bibitem[{{Zahn} {et~al.}(2007){Zahn}, {Lidz}, {McQuinn}, {Dutta}, {Hernquist},
  {Zaldarriaga}, \& {Furlanetto}}]{2007ApJ...654...12Z}
{Zahn} O., {Lidz} A., {McQuinn} M., {Dutta} S., {Hernquist} L., {Zaldarriaga}
  M., {Furlanetto} S.~R., 2007, \apj, 654, 12

\bibitem[{{Zaroubi} {et~al.}(2012){Zaroubi}, {de Bruyn}, {Harker}, {Thomas},
  {Labropolous}, {Jeli{\'c}}, {Koopmans}, {Brentjens}, {Bernardi}, {Ciardi},
  {Daiboo}, {Kazemi}, {Martinez-Rubi}, {Mellema}, {Offringa}, {Pandey},
  {Schaye}, {Veligatla}, {Vedantham}, \& {Yatawatta}}]{2012MNRAS.425.2964Z}
{Zaroubi} S., {de Bruyn} A.~G., {Harker} G., {Thomas} R.~M., {Labropolous} P.,
  {Jeli{\'c}} V., {Koopmans} L.~V.~E., {Brentjens} M.~A., {Bernardi} G.,
  {Ciardi} B., {Daiboo} S., {Kazemi} S., {Martinez-Rubi} O., {Mellema} G.,
  {Offringa} A.~R., {Pandey} V.~N., {Schaye} J., {Veligatla} V., {Vedantham}
  H., {Yatawatta} S., 2012, \mnras, 425, 2964

\end{thebibliography}


\end{document}